\documentclass[twocolumn]{aastex63}
\usepackage{amsmath,lipsum}
\usepackage{amssymb}
\usepackage{makecell}
\usepackage{threeparttable}

\newcommand{\tabincell}[2]{\begin{tabular}{@{}#1@{}}#2\end{tabular}}

\makeatletter

\newcommand{\Rmnum}[1]{\expandafter\@slowromancap\romannumeral #1@}
\makeatother

\submitjournal{ApJ}

\shorttitle{M Subdwarf Research. II}
\shortauthors{Zhang et al.}

\begin{document}

\title{M Subdwarf Research. \Rmnum{2}. Atmospheric Parameters and Kinematics}

\correspondingauthor{A-Li Luo \& Georges Comte}
\email{lal@nao.cas.cn; georges.comte@lam.fr}

\author[0000-0003-1454-1636]{Shuo Zhang}
\affiliation{CAS Key Laboratory of Optical Astronomy, National Astronomical Observatories, Beijing 100101, China}
\affiliation{University of Chinese Academy of Sciences, Beijing 100049, China}
\affiliation{Department of Astronomy, School of Physics, Peking University, Beijing 100871, P. R. China}
\affiliation{Kavli institute of Astronomy and Astrophysics, Peking University, Beijing 100871, P. R. China}

\author[0000-0001-7865-2648]{A-Li Luo}
\affiliation{CAS Key Laboratory of Optical Astronomy, National Astronomical Observatories, Beijing 100101, China}
\affiliation{University of Chinese Academy of Sciences, Beijing 100049, China}
\affiliation{School of Information Management \& Institute for Astronomical Science, Dezhou University, Dezhou 253023, China}
\affiliation{Department of Physics and Astronomy, University of Delaware, Newark, DE 19716, USA}

\author{Georges Comte}
\affiliation{Aix-Marseille Univ, CNRS, CNES, LAM, Laboratoire d'Astrophysique de Marseille, Marseille, France}

\author[0000-0001-6767-2395]{Rui Wang}
\affiliation{CAS Key Laboratory of Optical Astronomy, National Astronomical Observatories, Beijing 100101, China}
\affiliation{University of Chinese Academy of Sciences, Beijing 100049, China}

\author{Yinbi Li}
\affiliation{CAS Key Laboratory of Optical Astronomy, National Astronomical Observatories, Beijing 100101, China}

\author{Bing Du}
\affiliation{CAS Key Laboratory of Optical Astronomy, National Astronomical Observatories, Beijing 100101, China}

\author{Wen Hou}
\affiliation{CAS Key Laboratory of Optical Astronomy, National Astronomical Observatories, Beijing 100101, China}

\author{Li Qin}
\affiliation{CAS Key Laboratory of Optical Astronomy, National Astronomical Observatories, Beijing 100101, China}
\affiliation{School of Information Management \& Institute for Astronomical Science, Dezhou University, Dezhou 253023, China}

\author{John Gizis}
\affiliation{Department of Physics and Astronomy, University of Delaware, Newark, DE 19716, USA}

\author{Jian-Jun Chen}
\affiliation{CAS Key Laboratory of Optical Astronomy, National Astronomical Observatories, Beijing 100101, China}

\author{Xiang-Lei Chen}
\affiliation{CAS Key Laboratory of Optical Astronomy, National Astronomical Observatories, Beijing 100101, China}
\affiliation{University of Chinese Academy of Sciences, Beijing 100049, China}

\author{Yan Lu}
\affiliation{CAS Key Laboratory of Optical Astronomy, National Astronomical Observatories, Beijing 100101, China}
\affiliation{School of Information Management \& Institute for Astronomical Science, Dezhou University, Dezhou 253023, China}

\author{Yi-Han Song}
\affiliation{CAS Key Laboratory of Optical Astronomy, National Astronomical Observatories, Beijing 100101, China}

\author{Hua-Wei Zhang}
\affiliation{Department of Astronomy, School of Physics, Peking University, Beijing 100871, P. R. China}
\affiliation{Kavli institute of Astronomy and Astrophysics, Peking University, Beijing 100871, P. R. China}

\author{Fang Zuo}
\affiliation{CAS Key Laboratory of Optical Astronomy, National Astronomical Observatories, Beijing 100101, China}
\affiliation{University of Chinese Academy of Sciences, Beijing 100049, China}

\begin{abstract}
Applying the revised M subdwarf classification criteria discussed in Paper \Rmnum{1} to LAMOST DR7, combining the M subdwarf sample from Savcheva et al, a new M subdwarf sample was constructed for further study. The atmospheric parameters for each object were derived fitting with the PHOENIX grid, combining with Gaia DR2, the relationship between the gravity and metallicity were explored according to the locus both in the color-absolute magnitude diagram and the reduced proper motion diagram. Objects that have both the largest gravity and the lowest metallicity are located away from the main-sequence cloud and may be considered as the intrinsic M subdwarfs, which can be classified as luminosity class \Rmnum{6}. Another group of objects whose spectra show typical M subdwarf characters have lower gravity and relatively moderate metal deficiency and occupy part of the ordinary M dwarf region in both diagrams. The Galactic $U$, $V$, $W$ space velocity components and their dispersion show that the local Galactic halo population sampled in the solar neighborhood is represented by objects of high gravity and an inconspicuous bimodal metallicity distribution, with a fraction of prograde orbits. The other M subdwarfs seem to partly belong to the thick disk component with a significant fraction of thin disk moderately metal-poor objects intricately mixed with them. However, the selection effects, especially the favored anti-center direction of investigation in the LAMOST sub-sample, but also contamination by multiplicity and parameter coupling could play important roles and need to be further investigated. 

\end{abstract}


\section{Introduction}
M subdwarfs are usually recognized as the oldest members of the low-mass stellar populations and metal-poor counterparts of the late-type M dwarfs. With their extremely long nuclear-burning lifetimes, subdwarfs are supposed to represent early generations of star formation and are thus potential tracers of Galactic dynamic and chemical evolution history \citep{2014ApJ...794..145S}. Despite their scarcity in the solar neighborhood, M subdwarfs are supposed to comprise the majority of stars in the Milky Way stellar halo \citep{2013AJ....145...40B}, while a number of them also populate the thick disk, whose population is supposed to be older than that of the thin disk. Additionally, the cool atmospheres of M subdwarfs provide conditions for studying molecule and dust formation as well as radiative transfer in metal-poor environments \citep{1997ARA&A..35..137A}.

In many early investigations, M subdwarfs are considered as metal-poor main-sequence dwarfs (e.g. \citealt{2007ApJ...669.1235L,2014A&A...564A..90R}) solely with some noteworthy kinematic and spectral properties: red subdwarfs, exhibiting large space velocities relative to the Sun, kinematically associated with the Galactic halo \citep{1997AJ....113..806G} and thick disk, display shallower TiO absorption bands in the optical spectra than the ordinary red dwarfs dominating the Galactic thin disk. Historically, they were usually searched based on their low luminosity combined with large proper motion \citep{1969ApJ...158.1115S,1972ApJ...173..671J,1984ApJ...286..269H,1986AJ.....91.1140S,1997AJ....113..806G,2003AJ....125.1598L,2007ApJ...669.1235L,2009MNRAS.399.1223S,2011AJ....141..117J} and characteristic spectral features \citep{1976ApJ...210..402M,1978ApJ...220..935M,1980ApJ...240..859A,2008ApJ...681L..33L,2014ApJ...794..145S,2016RAA....16..107B,2019ApJS..240...31Z,2020AJ....159...30H}. The research of metal-poor subdwarfs also extends to the late M-type and L-type in recent years (e.g. \citealt{2017MNRAS.464.3040Z,2017MNRAS.468..261Z,2018MNRAS.480.5447Z,2019MNRAS.489.1423Z}).

On the other hand, some studies have suggested that subdwarfs should have a luminosity classification independent of dwarfs---\Rmnum{6} because of 1$\sim$2 magnitude less luminous than the main sequence dwarfs in the H-R diagram, forming an independent sequence, and more or less showing larger surface gravity than the dwarfs \citep{2008AJ....136..840J}. Results from \citet{2019AJ....157...63K} show that ultra subdwarfs can be as much as five times smaller than their solar-metallicity counterparts for a given temperature. However, subdwarf identification in most of the researches depends much more on the metallicity rather than the gravity, and whether gravity constrain for subdwarfs is necessary still remains an open question.

Characterizing M subdwarfs in terms of atmospheric parameters has long been a challenge. Like M dwarfs, the opacity sources are dominated by the molecular absorption bands such as TiO, VO, MgH, FeH and CaH in the optical, as well as H$_\text{2}$O and CO in the infrared, each of which has millions of spectral lines. These complex band structures leave no window to access the true continuum and only allow the strongest atomic lines such as Ca\Rmnum{1}, Na\Rmnum{1}, K\Rmnum{1} to show out at low spectral resolution. In addition, dust clouds form with decreasing temperature, the stellar atmosphere becomes more complicated, and the spectra show increasingly diatomic and triatomic molecules \citep{2016A&A...596A..33R}. Thus, the traditional technique of estimating the effective temperature ($T_{\text{eff}}$) of a star based on black body approximation and broadband photometry is not applicable to cool M subdwarfs in which the true continuum is covered by a wide variety of molecular absorbers\citep{1997ARA&A..35..137A}. It is also difficult to disentangle the effect of reduced metal abundance from that of the increasing surface gravity or the temperature decrease because these three parameters affect the pressure structure of the photosphere in a similar way \citep{2008AJ....136..840J}.

Since accurate metallicities for most M subdwarfs are difficult to measure, proxies for metallicity have been introduced. The most widely used metallicity-tracer parameter $\zeta$, (a parameter derived from CaH2, CaH3, and TiO5 molecular bandhead indices originally used by \citealt{1997AJ....113..806G}), was introduced by \citet{2007ApJ...669.1235L} and revised by \citet{2012AJ....143...67D} and \citet{2019ApJS..240...31Z}. With the parameter $\zeta$ or revised definition, two large samples of M subdwarfs are spectroscopically identified by \citet{2014ApJ...794..145S} and \citet{2019ApJS..240...31Z} using Sloan Digital Sky Survey (SDSS; \citealt{2000AJ....120.1579Y}) and the Large Sky Area Multi-Object Fiber Spectroscopic Telescope (LAMOST; \citealt{2012RAA....12.1197C}) respectively. According to the decreasing order of $\zeta$, the M subdwarfs can be divided into three metallicity subclasses: M subdwarfs (sdM), extreme metal-poor M subdwarfs (esdM), and ultra metal-poor M subdwarfs (usdM). 

The first model grid based on the PHOENIX radiative transfer code \citep{1995ApJ...445..433A} has been used by several authors to compare synthetic spectra to observed ones and to the broad-band colors: \citet{1995bmsb.conf..239D,1996ApJS..104..117L,1996MNRAS.280...77J}. \citet{1997AJ....113..806G} was thus able to constrain the metallicity scale of subdwarfs from $-$2 dex for esdM to $-$1 to $-$1.5 for sdM, while the stars not classified as subdwarfs were spanning the range $-$1 dex to 0. However, this pioneering model grid was far from being optimized because of the lack of theoretical molecular opacities, especially for the very important species CaH. 

The NextGen model grid \citep{1999ApJ...512..377H} brought a number of improvements and was in turn used by e.g. \citet{2005MNRAS.356..963W} who demonstrated the $\alpha$ enrichment of halo metal-poor M stars. Several papers using NextGen models to interpret spectra of cool M dwarfs and subdwarfs were subsequently published (e.g.\citealt{2000ApJ...535..965L,2004ApJ...602L.125L}) while \citet{2006ApJ...645.1485B} compared the performances of NextGen models and AmesCond models on the spectrum of an esdM star. Progress in modeling led the PHOENIX-based computations to the BT-Settl series of grids, the last generation of which proved highly successful in reproducing the features of the observed spectra in detail. Comparisons of low mass stars between a few high-resolution spectroscopic spectra and theoretical synthetic ones show that the latest PHOENIX models have already reached a milestone.\citep{2014A&A...564A..90R,2018A&A...620A.180R,2018A&A...610A..19R}.  

Gaia Data Release 2 \citep{2018A&A...616A...1G} has provided astrometry and photometry on more than one billion stars. As a result, a large number of subdwarfs that were out of the reach of parallax studies now have accurate available trigonometric parallaxes permitting distance determination and hence their placement in the Hertzsprung-Russell diagram, while their 3D kinematics can be studied thanks to high accuracy proper motions from Gaia and radial velocity from spectroscopy.

M subdwarfs have been found both at high Galactic latitudes, as expected for halo population members, and at much lower latitudes in fields located towards the Galactic anti-center, suggesting that one could use them to study the mix of various population components in the Solar neighborhood. It becomes thus especially interesting to explore the links between their atmospheric parameters, especially gravity and metallicity, and their kinematic properties. For this, we need a reasonably accurate estimation of the atmospheric parameters of a large sample of subdwarfs for which we have accurate photometry and astrometry and the same data on a control sample of ordinary M dwarfs. The present work, based only on low-resolution spectroscopy, has its limitations and must be considered as a preliminary attempt towards this goal.  

The paper is organized as follows. In Section \ref{sec:data}, we briefly summarize the observational material that will be used in Sections \ref{sec:sample} and \ref{sec:property}. In Section \ref{sec:model}, we analyze the properties of the PHOENIX-based BT-Settl CIFIST model grid to evaluate its performance by studying spectral indices of BT-Settl synthetic spectra compared with those measured on real-world template spectra of M dwarfs and subdwarfs. Then, we fit the template spectra with the synthetic grid to get the atmospheric parameter ranges for each template spectra. In section \ref{sec:sample}, we derive the atmospheric parameters for spectra of a sample of M dwarf and subdwarf from the best-fit PHOENIX BT-Settl synthetic spectra. This leads to the construction of a cleaned genuine subdwarf sample, a normal dwarf sample and a third population whose properties point towards non-genuine subdwarfs but rather metal-deficient dwarfs. In section \ref{sec:property}, we explore the links between the distribution of the atmospheric parameters of the stars belonging to these three samples and their photometric and kinematic properties as provided by the Gaia Data Release 2, using color-absolute magnitude diagram, reduced proper motion diagram and 3D Galactic motion distributions. Finally, conclusions and summary are provided in Section \ref{sec:con}. \\

\section{Observational data} \label{sec:data}

The low-resolution spectra of M dwarfs and subdwarfs are from LAMOST DR7 and the SDSS DR7, and the astrometric data is from Gaia DR2.

LAMOST is a reflecting Schmidt telescope located in Xinglong Station of National Astronomical Observatory, China (40$^\circ$N, 105$^\circ$E) with a mean aperture of 4.3 meters and a field of view of 5$^\circ$. The observable sky area extends from $-$10$^\circ$ to +90$^\circ$ declination. 4000 optical fibers positioned on the focal plane yield a high spectrum acquisition rate per night. Until March 2020, DR7 published more than 10 million low-resolution spectra (R$\sim$1800) covering 3800$-$9000$\rm\AA$ \citep{2015RAA....15.1095L} and 2.17 million medium resolution spectra (R$\sim$7500) covering 3700$-$5900$\rm\AA$ and 5700$-$9000$\rm\AA$ \citep{2020arXiv200507210L}.

The SDSS began regular survey operations in 2000, and has progressed through several phases until now: SDSS-\Rmnum{1} (2000$-$2005), SDSS-\Rmnum{2} (2005$-$2008), SDSS-\Rmnum{3} (2008$-$2014), and SDSS-\Rmnum{4} (2014$-$2020). The SDSS DR7 contains over 1.6 million low-resolution spectra in total, including 930,000 galaxies, 120,000 quasars, and 460,000 stars. It was collected by a dedicated wide-field 2.5 m telescope at Apache Point Observatory \citep{2006AJ....131.2332G}. The telescope employed a drift-scan technique, imaging the sky in $u$,$g$,$r$,$i$,$z$ wide bands along five camera columns \citep{1998AJ....116.3040G}. The fiber-fed spectrographs acquired 640 spectra (R$\sim$2000) simultaneously, and the spectral range is 3800$-$9200 $\rm\AA$.

The proper motions, trigonometric parallaxes, and photometry used in this work come from Gaia Data Release 2 made available in April 2018 by European Space Agency. Gaia DR2 contains the position and brightness information of more than 1.69 billion stars, the parallax and proper-motion measurements for 1.33 billion stars \citep{2018A&A...616A..10G,2018A&A...616A...2L,2018A&A...616A...9L}, the color information for 1.38 billion stars \citep{2018A&A...616A..10G}, the radial velocities for more than 7 million stars \citep{2019A&A...622A.205K}, etc. The mean parallax error is up to 0.7 mas for sources having G=20 mag, and the proper motion uncertainty is up to 0.2 and 1.2 mas yr$^{-1}$ for G = 17 mag and G = 20 mag respectively.

\section{PHOENIX BT-Settl Model Analysis}\label{sec:model}

\subsection{PHOENIX BT-Settl Model Grid}

Classical theoretical model calculation codes include ATLAS code from \citet{1973PhDT.........4K} and \citet{2004A&A...419..725C}, MARCS code from \citet{1975A&A....42..407G,2008A&A...486..951G}, PHOENIX code from \citet{1995ApJ...445..433A}, and so on. In the atmosphere of very low mass stars, various molecular absorption (each with hundreds of thousands to millions of spectral lines) and the presence of numerous condensates make accurate modeling extremely complicated; the convection zone extends to the most outer photosphere layer, which means that the evolution of the model strictly depends on the precise treatment of surface boundary \citep{2013MmSAI..84.1053A}. It is worth celebrating that in recent years, the PHOENIX code differs from previous methods by computing the opacities during the model execution (or on-the-fly), considering non-local thermodynamic equilibrium effects, etc., and combining with the latest solar abundance \citep{2009ARA&A..47..481A,2011SoPh..268..255C}. The atmospheric model, especially for modeling very low mass stars such as M-type stars, brown dwarfs, and even Jupiter-like planets, has made breakthrough progress.

The latest version PHOENIX BT-Settl models represent decisive progress compared to the previous model grids \citep{2001ApJ...556..357A} in different aspects, including updated molecular line lists, better solar abundance estimates and treatment of convection, though residual incompleteness of opacities and a limitation of the mixing length theory formalism still exist. Further studies on a complete, comprehensive and uniform grid of models are addressing these shortcomings and will better reproduce observational constraints \citep{2016sf2a.conf..223A}.

The synthetic spectra generated by BT-Settl model atmospheres \citep{2012RSPTA.370.2765A,2013MmSAI..84.1053A,2015A&A...577A..42B} have been compared with observed spectra and their consistency verified in a number of recent M dwarf/subdwarf studies (e.g., \citealt{2014A&A...564A..90R,2016A&A...596A..33R,2018A&A...620A.180R,2018A&A...610A..19R}). 

In this work, we use the version BT-Settl CIFIST2011 model atmosphere which is valid across the entire parameter range \citep{2011ASPC..448...91A,2012RSPTA.370.2765A}. A part of pre-calculated grid as list in \ref{tab:param_grid} is used in the analysis and the following parameter measurement process.

The models are calculated with effective temperatures from $T_{\text{eff}}$ = 400 to 8000 K in 100 K steps, surface grativity ranging from log $g$ = 0.0 to 6.0 in steps of 0.5 dex, and metallicity from [M/H]=$-$2.5 to +0.5 in steps of 0.5 dex (M stands for all elements heavier than H and He; assuming [X/H] = [Fe/H] for most elements and so [Fe/H] denotes the overall metallicity). Alpha element (O, Ne, Mg, Si, S, Ar, Ca, and Ti) enhancement is taken into account as follows: for [M/H] = 0.0, no enhancement; for [M/H]=$-$0.5, a value of [$\alpha$/M]=+0.2 has been used; for [M/H] $\leq-$1.0, [$\alpha$/M]=+0.4. Note that in PHOENIX the [$\alpha$/M] is similar to [$\alpha$/Fe] since it is computed as the change in dex of the abundance of alpha elements relative to the non-alpha average, and the heaviest is Fe.

\begin{deluxetable}{ccc}[htb!]
\tablecaption{Parameter Space of the Grid Used in this Work \label{tab:param_grid}}
\tablecolumns{3}
\tablewidth{0pt}
\tablehead{
\colhead{Variable} &
\colhead{Range} &
\colhead{Step size}
}
\startdata
\textit{T}$_{\text{eff}}$& 2700 $-$ 4500 	& 100 K 			\\                 
log \textit{g}  				& 0 $-$ 6.0    		&  0.5 dex      	\\
$[\text{M/H}]$  			& $-$2.5 $-$ +0.5  	&  0.5 dex  		\\
$[\alpha/\text{M}]$  	& 0 $-$ +0.4  			&  0.2 dex      	\\
\enddata
\end{deluxetable}

\subsection{Comparison of Models with Real-world Template Spectra in the ``canonical'' diagram}\label{subsec:diagrams_and_templates}

In this section, we firstly compare M-type template spectra from \cite{2015AJ....150...42Z} with BT-Settl models. The templates, assembled from SDSS low-resolution (R$\sim$2000) spectra and classified using \citet{2007ApJ...669.1235L,2013AJ....145..102L} calibration, are spanning the spectral subtypes K7.0 to M8.5 and covering every half-subtype across the range.

The relevant pre-calculated synthetic spectra\footnote{\url{https://phoenix.ens-lyon.fr/Grids/BT-Settl/CIFIST2011/SPECTRA/}} generated by PHOENIX code have been convolved down to the spectral resolution of SDSS, using a series of Gaussian kernels with parameters sigma and corresponding FWHM listed in Table \ref{tab:kernels}. To determine the sigma value of a convolution kernel, we fit gaussians profiles to night sky lines (assuming they have a natural width much smaller than the spectral psf). In the red range, clean lines of OH molecule are numerous, in the blue the 5577$\rm \AA$ is also good, as well as some mercury lines, except 5461$\rm \AA$ which has fine structure. Na D lines of city lights are avoided because they have a very large width (high pressure lamps).

\begin{deluxetable}{cccc}[htb!]
\tablecaption{The gaussian kernels with the following sigmas 
and corresponding fwhm used for convolving synthetic spectra. \label{tab:kernels}}
\tablecolumns{4}
\tablewidth{0pt}
\tablehead{
\colhead{Central $\rm \lambda (\rm \AA)$} &
\colhead{$\rm \sigma $} &
\colhead{Pixels of 0.5 $\rm \AA$} &
\colhead{FWHM}
}
\startdata
4200 & 1.05 & 2.10 & 4.935 \\
5100 & 1.08 & 2.16 & 5.075 \\
6000 & 1.25 & 2.50 & 5.875 \\
6900 & 1.43 & 2.86 & 6.720 \\
7800 & 1.40 & 2.80 & 6.580 \\
8700 & 1.39 & 2.78 & 6.535 \\
\enddata
\end{deluxetable}

 \begin{figure*}[htb!]
\center
\includegraphics[width=150mm]{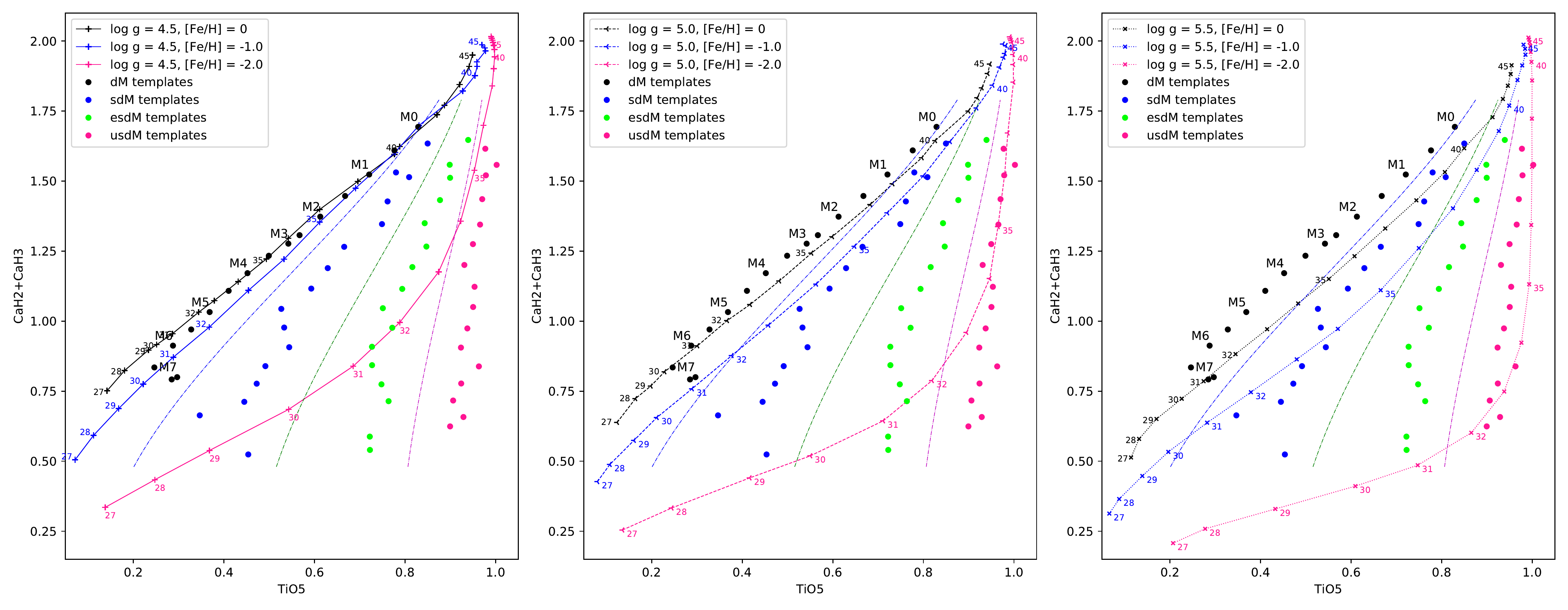}
\caption{The ``canonical'' diagram classically used to identify subdwarfs from dwarfs: spectral composite index CaH2+CaH3 versus TiO5. Models are plotted with the following codes : continuous lines: log $g$ = 4.5 models, dashed lines: log $g$ = 5.0 models, and dotted lines: log $g$ = 5.5 models. Black : solar metallicity, blue: $-$1.0 dex, and magenta: $-$2.0 dex. The small 2-digit labels indicate $T_{\text{eff}}$ points (e.g. ``32'' means 3200 K). 
Observed template spectra: black dots represent normal main-sequence dwarfs, subdwarf templates colored in blue are for ``sdM'' subclass, green for ``esdM'' and magenta for ``usdM'' subclass. The M0\Rmnum{5} to M7\Rmnum{5} ``standard dwarf'' spectral type sequence is labeled on all graphs.\label{fig:3-TiO5-CaH}}
\end{figure*}

Figure \ref{fig:3-TiO5-CaH} is the ``canonical'' [TiO5, CaH2+CaH3] diagram classically used by many authors to identify and spectroscopically separate the subdwarfs from the normal dwarfs. The spectral indices were defined initially by \citet{1995AJ....110.1838R}. The boundaries of the d/sd/esd/usd metallicity classification scheme are also plotted. This diagram may represent more or less a metallicity-gravity diagram, since TiO5 is a metallicity (and alpha) indicator while CaH2+CaH3 is rather a gravity indicator, and both are naturally strongly $T_{\text{eff}}$ dependent. The subdwarfs are supposed to order along temperature sequences, while the fan-shape of their distribution reflects their atmospheric metal deficiencies.

The normal dwarf template sequence is well-ordered and fits quite well with the model family \{log $g$ = 4.5, [Fe/H] = 0.0\} track, although with the coolest types (M6$\sim$M7) being a bit offset, as it would be expected from an insufficient value of the model gravity.This is an important point and needs to be further discussed. All late-type M dwarfs are expected to have higher surface gravity than early M dwarfs due to the near 1-to-1 mass radius relation for low-mass stars. Note that between metallicities of 0.0 and $-$1.0, the displacement of the model track is quite small even at the lowest $T_{\text{eff}}$, making the estimates of moderate metal deficiencies rather difficult on this diagram.

The subdwarf templates also draw quite well-defined and well separated sequences with increasing slopes from sd to usd, the latter one being almost vertical at a constant TiO5 index of almost null absorption. The coolest types of dwarfs and sd are somehow confused, this questions the validity of the templates themselves which are built from spectra either in insufficient number or too noisy to be really representative of the genuine subtype definition.

The metallicity of the sd subclass does not comply with a $-$1.0 value, it must be more deficient, or affected by a higher gravity, or both. The model family \{log $g$ = 5.0, [Fe/H] = $-$2.0\} could provide a reasonable fit for usdM0$\sim$3, but the track moves towards significant values of the TiO5 opacity for later subtypes, thus the real stars must be more deficient. The model family \{log $g$ = 5.5, [Fe/H] = $-$2.0\} track yields more or less an outer envelope for the subdwarfs, implying that most of the right-hand distribution of their template representative points is driven more by large metal deficiencies than by high gravity.

\subsection{Fitting the dM and sdM Templates with the Model} \label{subsec:temp_fitting}

Using template spectra rather than hundreds of individual spectra to compare with the models enables to smooth accidental differences due to noise and uncertainties in individual classification, and in principle gives access to high signal-to-noise data. However, the final quality of a template depends on the number of individual spectra co-added to build it. We shall hereafter use the COMPLETE collection of templates of \citet{2015AJ....150...42Z}, i.e. the interpolated 221 spectra in total spanning 12 subclasses (from dMr to usdMp) and 18 subtypes (from M0 to M8.5).

The parameter grid of the synthetic spectra is shown in Table \ref{tab:param_grid}. Note that the grid is not uniform: the value log $g$ = 6.0 only pertains for $T_{\text{eff}} \geq$ 3000 K, and the [$\alpha$/M] ratio is dependent on the metallicity.

The spectral fitting region extends from 6000$\rm\AA$ to 8800$\rm\AA$ with a step size of 1$\rm\AA$ which covers most prominent spectral features used for classification. Before the fitting process, the synthetic spectra were smoothed at the resolution of the observed spectra.

In the fitting process, we set the spectral flux as a function of $T_{\text{eff}}$, log $g$, and [M/H], and assume: 

\begin{equation}
F_{\text{o},\lambda} = F_{\text{m},\lambda} \times P_{\text{n},\lambda} + \epsilon_{\lambda}, 
\end{equation}

\noindent where $F_{\text{o},\lambda}$ is the flux of the observed spectrum at wavelength $\lambda$, $F_{\text{m},\lambda}$ is the flux of the synthetic spectrum generated by PHOENIX model wavelength $\lambda$, P$_{\text{n},\lambda} $ is an n-order polynomial factor used to correct the inaccuracies of the continuum of the observed spectrum (here we set n = 4 based on the optimal experimental results), $\epsilon_{\lambda}$ is the difference between the synthetic spectrum.

The equation above can be solved by minimizing the loss function L=$\sum_{\lambda=1}^{\text{N}}(F_{\text{m},\lambda} \times$ P$_{\text{n},\lambda} - F_{\text{o},\lambda})^2$ using the least square method to obtain the coefficient matrix solution of P$_{\text{n},\lambda} $. Then we obtain the best-fit synthetic spectrum by minimizing the $\chi^{2}$ distance, 

\begin{equation}
\chi^2\ =\ \frac{\sum_{i\text{=1}}^N[(F_{\text{o}_{i}}-F'_{\text{m}_{i}})^2/ \epsilon_{i}^2]}{N} 
\end{equation}

\noindent where $F'_{\text{m}_{i}}$ is the flux of the $i^{\text{th}}$ pixel of the synthetic spectrum corrected by polynomial, $\epsilon_i$ is the flux difference of the $i^{\text{th}}$ pixel, and $N$ is the number of pixels involved in the calculation.

For each given template spectrum, we calculate the $\chi^{2}$ values when matching all the synthetic spectra in the grid and then interpolate between the parameters by synthesizing linear combinations of top-five best-fit synthetic spectra, as the same way in \citet{2017ApJ...836...77Y}.
\begin{figure}[htb!]
\center
\includegraphics[width=90mm]{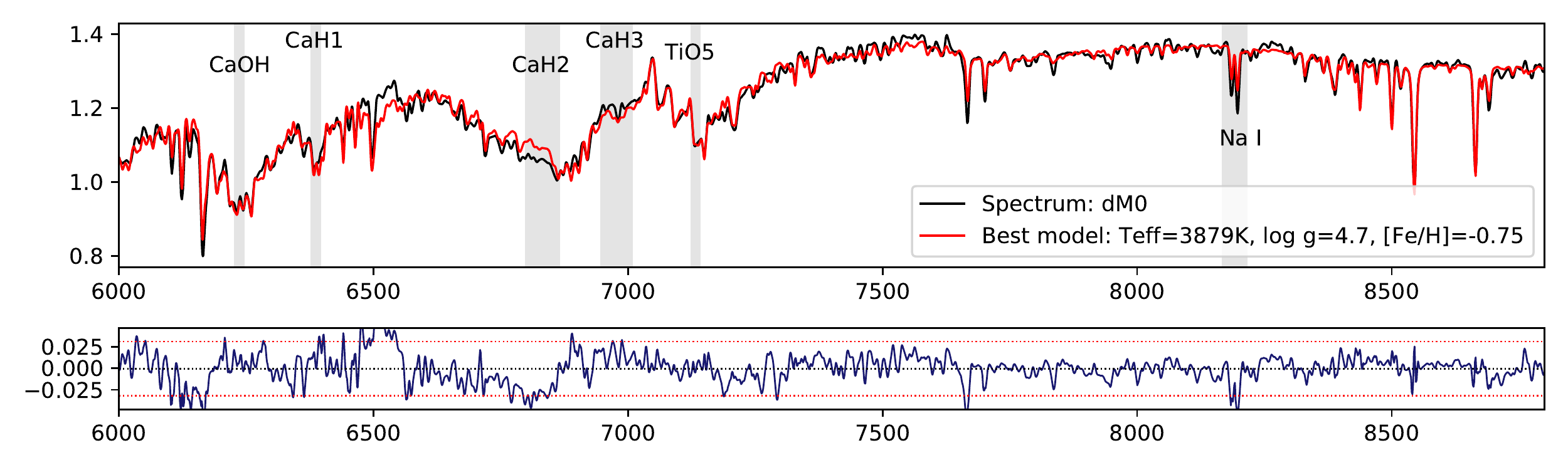}
\includegraphics[width=90mm]{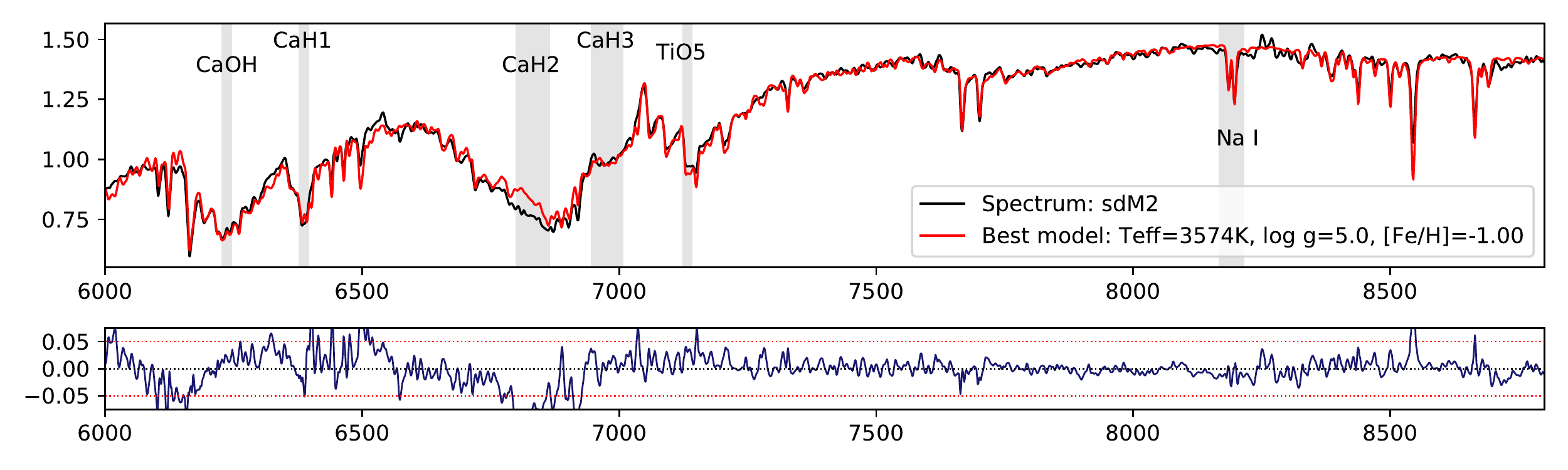}
\includegraphics[width=90mm]{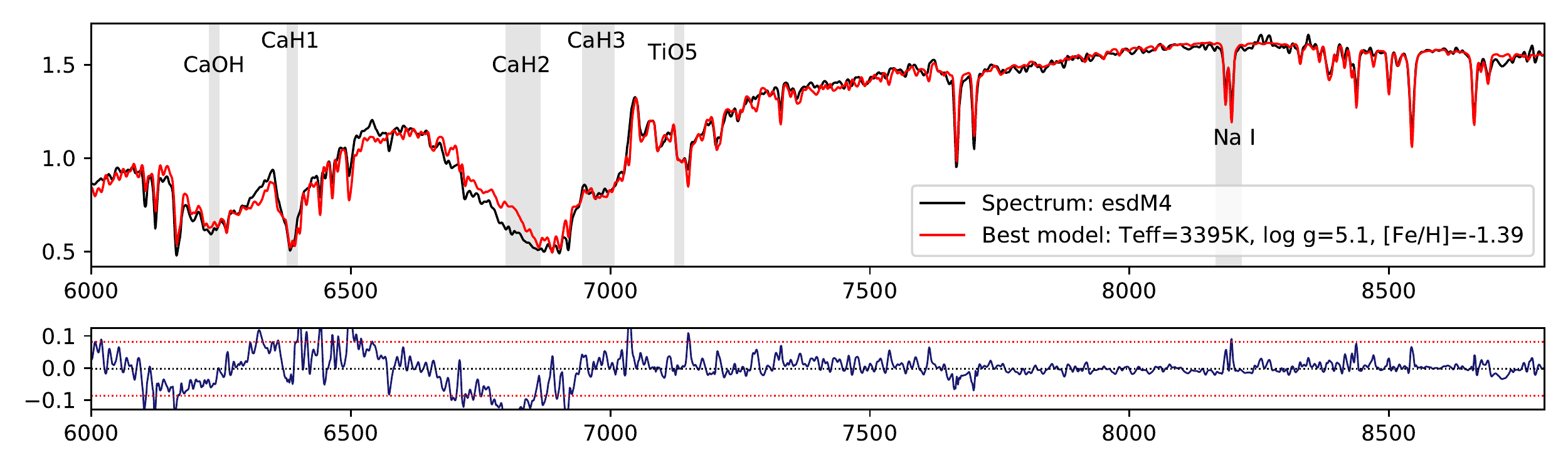}
\includegraphics[width=90mm]{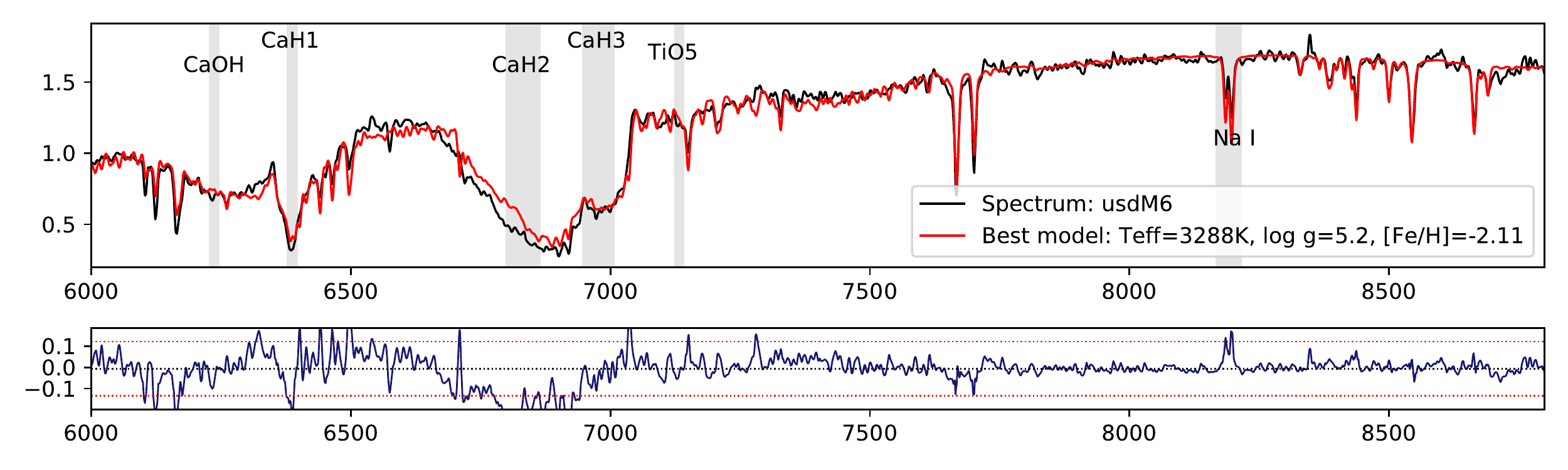}
\caption{Template spectra of subtype dM0, sdM2, esdM4 and usdM6 with their corresponding synthetic spectra and fitting residuals. \label{fig:3-templates}}
\end{figure}

\begin{figure}[htb!]
\center
\includegraphics[height=63mm, width=90mm]{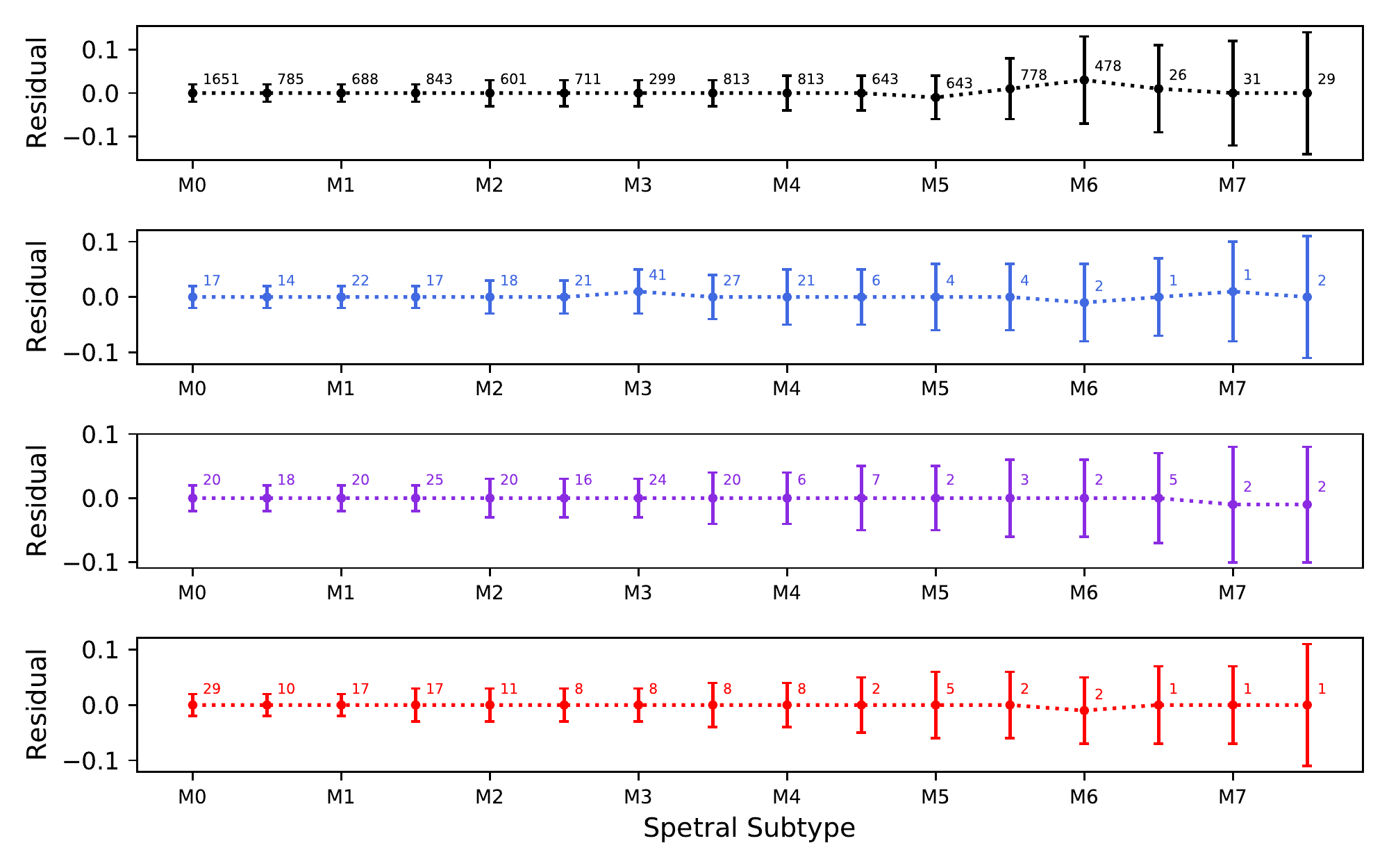}
\caption{The residuals between template spectra of different spectral subtypes and their best-fit synthetic spectra. Black, blue, purple, and red represent subclass dM, sdM, esdM, and usdM respectively. The error bar is the standard deviation of the difference between the flux of the template spectrum and the flux of the synthetic spectrum. The number beside each node represents the number of spectra used to construct the template spectrum. \label{fig:3-template-all}}
\end{figure}

\begin{figure}[htb!]
\center
\includegraphics[width=90mm]{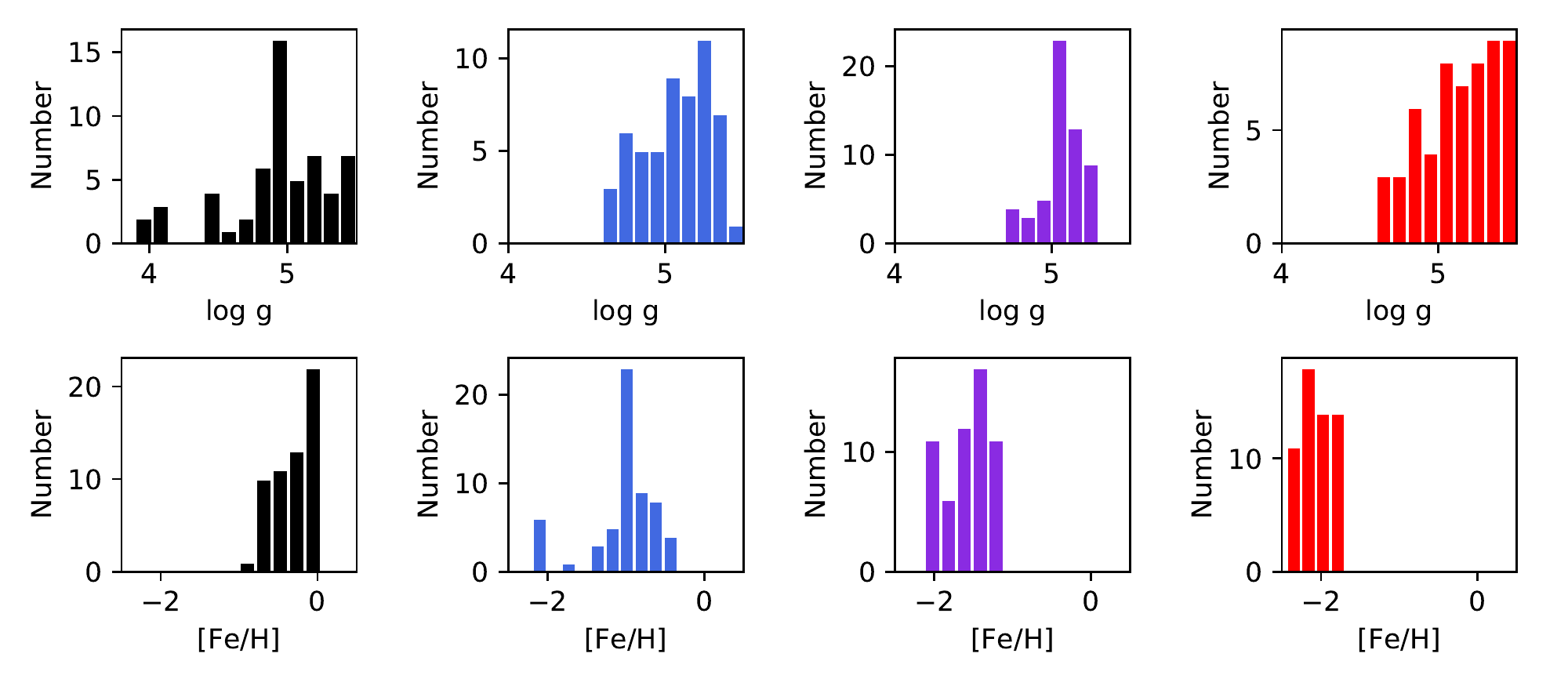}
\caption{The distributions of estimated atmospheric parameters of the templates spectra of different subclasses. The colors represent the same as in Figure \ref{fig:3-template-all}. 
\label{fig:3-template-dis}}
\end{figure}

The fitting results of four template spectra of different spectral subtypes and metallicity subclasses (dM0, sdM2, esdM4 and usdM6) with their corresponding estimated parameters are shown in Figure \ref{fig:3-templates} as examples. The shape of the residual spectra indicate that the errors mainly come from the region of 6000$-$7000\AA, where there remain some incompleteness in line lists and/or missing/incomplete opacities on some molecules used in the modeling (CaOH, CN, CaH, etc.). Besides, the far red region of the observed spectra could be contaminated by telluric absorption. 

Figure \ref{fig:3-template-all} shows the residual levels of the fitting results of all the templates. The residual level of the fitting results gradually increases from $\sim$5\% for early type M stars to 10\% for the late types: both spectral quality and the number of objects for template building decrease dramatically for the latest types. 

Finally, the distributions of the estimated results of log $g$ and [Fe/H] of the templates above are shown in Figure \ref{fig:3-template-dis}. The dispersions are large for both the two parameters, with the basic trend being a clear decrease of the average metallicity from sd to usd accompanied by a more confuse increase of gravity. Note that when using the complete set of templates (221 in total), we can meet spectra of slightly atypical characteristics, as for instance illustrated in the top panel of Figure \ref{fig:3-templates} in which a dM0 is metal-poor and 100 K cooler than a standard dM0 of solar metallicity.

\section{Sample Construction and Atmospheric Parameter Estimates} \label{sec:sample}
To estimate the atmospheric parameters $T_{\text{eff}}$, log $g$ and [M/H] of M subdwarfs from low-resolution spectra and investigate the trends of these parameters with the kinematic properties of the stars, we need a sufficiently large sample of subdwarfs spanning the broadest possible range of parameters. 

In Paper \Rmnum{1}, we had explored LAMOST DR4 in which 2,791 M subdwarf were identified. Extending the search to LAMOST DR7, a total of 4,767 visually selected candidates are retrieved including the previously obtained DR4 ones. Combining this sample with SDSS objects published by \citet{2014ApJ...794..145S}, we submitted these spectra to a least-squares based fitting process with PHOENIX synthetic spectra. The determination of atmospheric parameters is not always possible with a reasonable accuracy because the error budget of the fitting process depends on the signal-to-noise (S/N) of the spectra. Thus, the final combined LAMOST + SDSS sample of stars with known atmospheric parameters will be reduced with respect to the initial numbers.

\subsection{Sample Construction and Individual Spectra Fitting}\label{subsec:construct}

Because of magnitude limitation, most M dwarfs of LAMOST range in spectral type before M3, and only a few cooler and fainter spectra of low S/N are available. The data of SDSS DR7 covers a more extended parameter space, with cooler and lower metallicity objects. The strategies of two surveys point to different fields, i.e. SDSS at high Galactic latitude favoring halo and thick disk objects detection, while LAMOST survey has in priority been targeted at regions of the Galactic anti-center. Therefore, in this study we combine the subdwarfs from SDSS DR7 (SEGUE-I/II) and LAMOST DR7 to construct a consistent sample of M subdwarfs, more representative of the total population of these stars.

First, it is useful to discuss three points on which the processing of the LAMOST and SDSS spectra sets will differ: (1) selecting criteria, (2) spectroscopic resolution, and (3) spectral data reduction.\\


(1) Selecting criteria. \\

\hspace*{1em}Several works have been conducted to identify M subdwarfs from SDSS and LAMOST low resolution spectral data sets \citep{2004AJ....128..426W,2008ApJ...681L..33L,2013AJ....145...40B,2014ApJ...794..145S,2016NewA...44...66Z,2016RAA....16..107B}, Paper \Rmnum{1}, and the two largest samples published were from \citet{2014ApJ...794..145S} and Paper \Rmnum{1} respectively. 

\hspace*{1em}The selecting conditions differ for these two latter samples: the principal separator used is the standard parameter $\zeta$ initially introduced by \citet{2007ApJ...669.1235L} based on a combination of CaH2, CaH3 and TiO5 spectral indices:

\begin{equation}
\zeta_{\text{CaH}/\text{TiO}}=\frac{1-\text{TiO5}}{1-[\text{TiO5}]_{Z_\odot}}
\end{equation}
in which the [TiO5]$_{Z_\odot}$ is a third order polynomial fit of the TiO5 spectral index as a function of the CaH2+CaH3 index as: 
\begin{equation} 
\begin{split}
[\text{TiO5}]_{Z_\odot}\ &=\ a\\+
&b\times (\text{CaH2}\ +\ \text{CaH3})\\+
&c\times (\text{CaH2}\ +\ \text{CaH3})^2\\+
&d\times (\text{CaH2}\ +\ \text{CaH3})^3
\end{split}
\end{equation}
so that $\zeta$ = 1 for solar metallicity and $\zeta$ = 0 for the most metal-poor objects where TiO5 absorption becomes undetectable. The coefficients $a$, $b$, $c$, $d$ derived for SDSS objects by \citet{2007ApJ...669.1235L} led to the calibration $\zeta_{\text{L07}}$ which was later revised by \citet{2013AJ....145..102L} (hereafter $\zeta_{\text{L13}}$). The condition for a star to belong to the subdwarf population was set as $\zeta_{\text{L07}}<$0.825 and $\zeta_{\text{L13}}<$0.825

Since $\zeta$ was found to be dependent on the M dwarfs reference sample, instrumental setup and various factors influential in the reduction process \citep{2013AJ....145..102L}, we proposed a new calibration adapted to LAMOST spectra in Paper \Rmnum{1} (hereafter $\zeta_{\text{Z19}}$), and we suggested the condition for LAMOST subdwarfs being set as $\zeta_{\text{Z19}}<$0.75 with an additional screening condition based on the CaH1 index.\\

(2) Spectroscopic resolution.\\

\hspace*{1em}Along the wavelength of a spectrum from SDSS or LAMOST, the physical resolution (depending on the grating$\times$focal length of camera$\times$pixel size of the detector) varies in a complicated way. For LAMOST, when spectra are extracted and rebinned in constant log-wavelength steps by the 1D pipeline, this variation is +/- ``dissolved'' into another variation. Additionally, though in theory, all spectra obtained from the same instrument are supposed to have an identical resolving power, in practice for a large-scale survey project, the resolution changes slightly at different observation dates, due to imperfect instrument stability.

\hspace*{1em}In the actual calculation, assuming a gaussian instrumental profile, the resolution R of a spectrum at wavelength $\lambda$ can be derived from $\lambda$/$\Delta\lambda$, where $\Delta\lambda$ is the full width at half maximum (FWHM) of the spectral line broadening at that wavelength. The Gaussian kernel $\sigma$ = FWHM/2.355 used to convolve the synthetic spectrum is derived from $\sqrt{{\sigma_{\text{o}}}^2 - {\sigma_{\text{m}}}^2}$, where $\sigma_{\text{m}}$ corresponds to the synthetic spectrum, and $\sigma_{\text{o}}$ corresponds to the observed spectrum. Since the resolution of the synthetic spectra is extremely high, we set $\sigma_{\text{m}}$ as 0 here. 

\hspace*{1em}Prior to the fitting process, the observed spectra are shifted to rest-frame based on the measured radial velocities. The fitting range was set to 6000$-$8800\AA, in which the pixels marked by ``bad'' flags and the spectral intervals heavily contaminated by telluric absorption (7210$-$7350\AA, 7560$-$7720\AA, and 8105$-$8240\AA) were masked. We divided the region 6000$-$8800$\rm\AA$ into 7 bands evenly and used a dedicated convolution kernel for each 400$\rm\AA$ band. 

\hspace*{1em}For SDSS and LAMOST spectra, we adopted different convolution kernels calculated as follows. For an SDSS spectrum, we computed the mean value of the $\sigma$ in each band by the ``dispersion'' value of each pixel stored in the fits file. For a LAMOST spectrum, we calculated the mean value of the FWHM of the arc lamp spectrum in each band, obtained on the same day for wavelength calibration purpose. \\

(3) Data reduction.\\

\begin{figure}[htb!]
\center
\includegraphics[width=80mm]{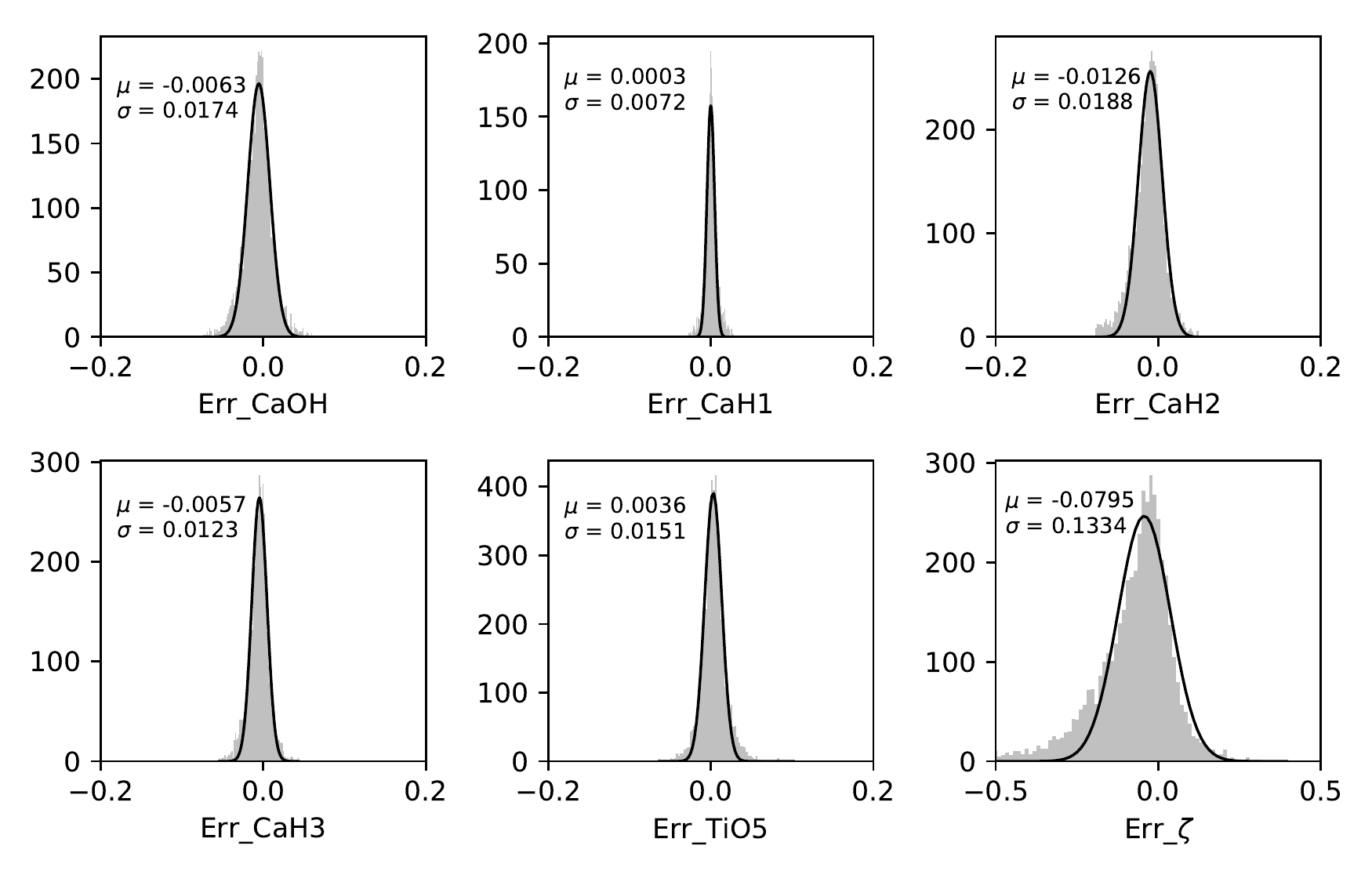}
\caption{Each subdiagram shows a spectral index error distribution for the LAMOST subdwarfs. The errors are the differences between the values measured from the original spectra minus the values measured from the continuum-corrected spectra. They reflect the deformation level of the local pseudocontinuum, due to imperfect flux calibration. The bottom right panel shows the distribution of the derived parameter $\zeta$ values with a scatter 0.133 although the scatters of each of the indices (CaH2, CaH3 and TiO5) used to calculate $\zeta$ are quite smaller.
\label{fig:4-index-errors}}
\end{figure}

\hspace*{1em}In the fitting process described above, we first correct the continuum of an observed spectrum using a fourth-order polynomial and then choose the best-fit synthetic spectrum based on the chi-square minimum principle. We find that the continuum slopes of some observed spectra of LAMOST set differ largely from their corresponding corrected spectra. This deformation is mainly caused by flux calibration problems \citep{2016ApJS..227...27D} and would affect the accuracy of the spectral indices. To control this unwanted bias, we measure the spectral indices of each polynomial-corrected observed spectrum, and calculate the difference between them and the original indices measured on the uncorrected spectrum. We then construct the distribution of each index differences, as shown in Figure~\ref{fig:4-index-errors}. Although the offset of a single index is generally small, the error on the $\zeta$ parameter composed of multiple indices may be considerably enlarged. Hence, targets with at least one spectral index difference beyond 3$\sigma$ of the corresponding difference distribution are removed from the final subdwarf sample.

We summarize the specific solution to build the combined LAMOST + SDSS subdwarf sample in seven steps as follows:

(1) Select 4,767 visually inspected subdwarfs from LAMOST DR7. 

(2) Recalculate the indices of the SDSS subdwarfs from \citet{2014ApJ...794..145S} with the radial velocities recorded in the catalog (see Paper \Rmnum{1} for more comments).

(3) Calculate the resolution of each spectrum using the profile of corresponding arc-lamp spectrum (LAMOST) or the ``dispersion'' value stored in the fits file (SDSS) and get the gaussian convolution kernels.

\begin{figure}[htb!]
\center
\includegraphics[width=90mm]{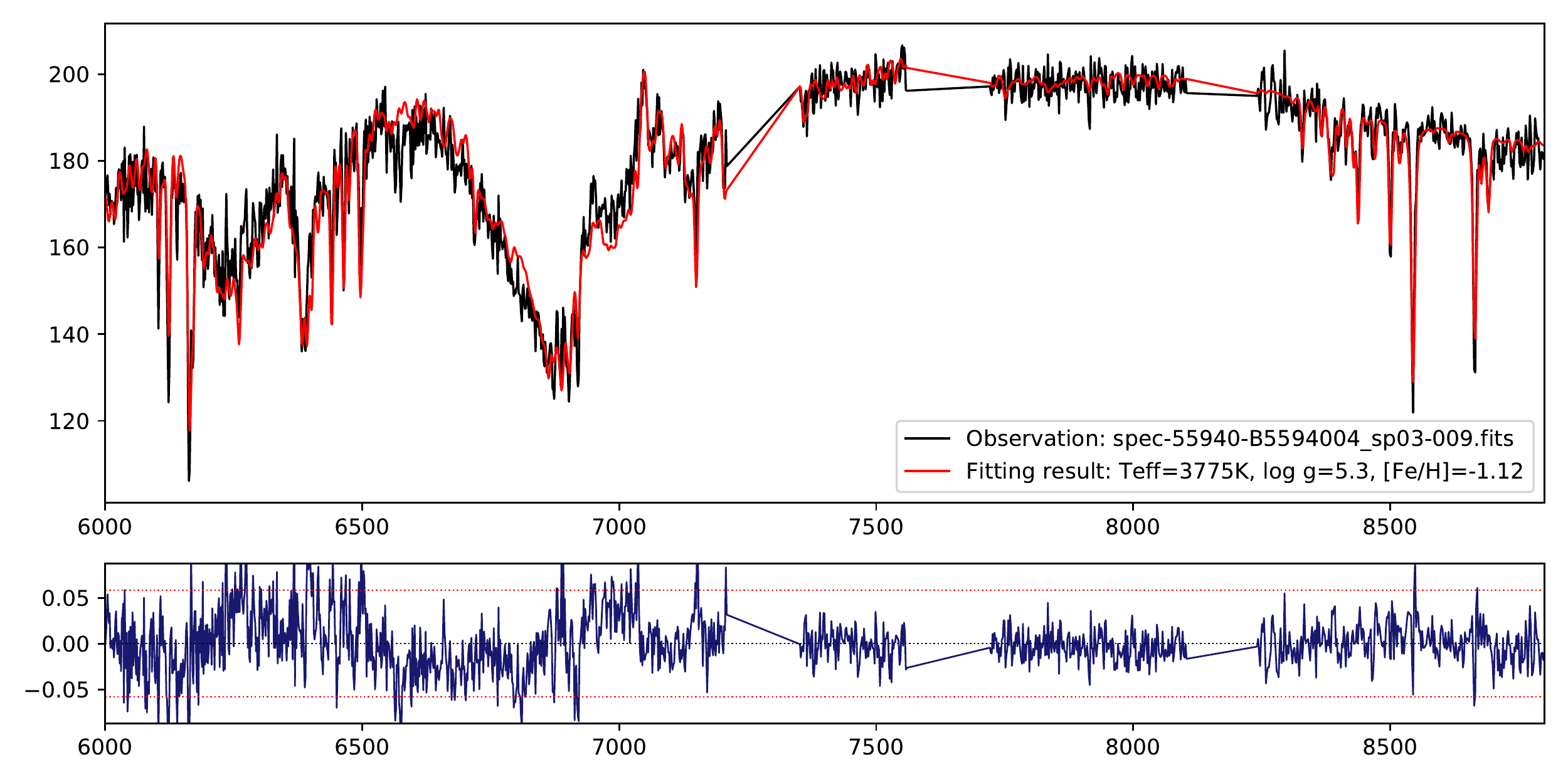}
\includegraphics[width=90mm]{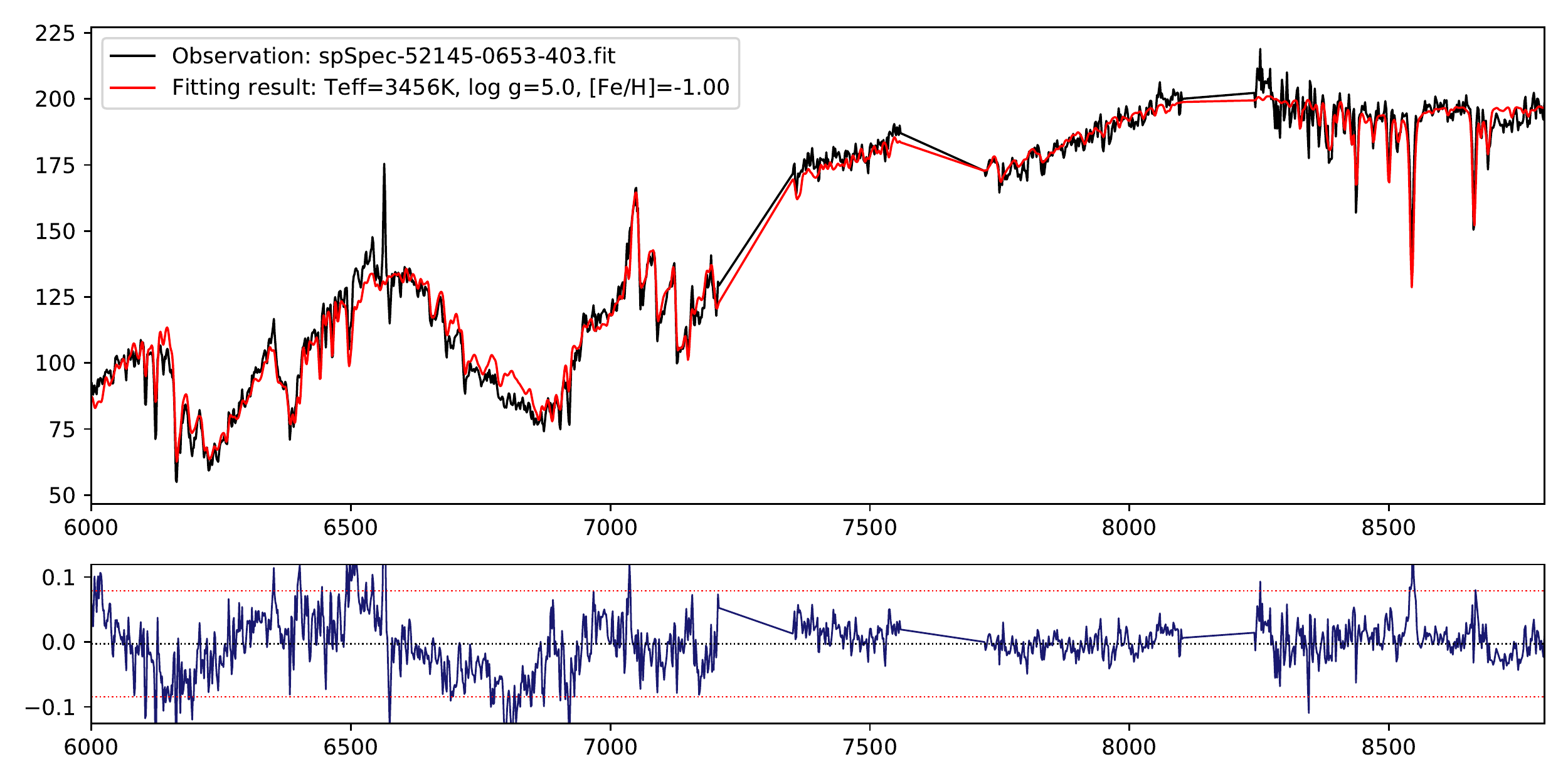}
\caption{Upper panel: a LAMOST extreme subdwarf spectrum and its best-fit synthetic spectrum. The spectral bands contaminated by telluric absorptions are masked. Lower panel: an SDSS subdwarf spectrum showing H$\alpha$ emission and its best-fit synthetic spectrum. \label{fig:4-fiting-example}}
\end{figure}

(4) Smooth the synthetic spectra to fit every observed spectrum and estimate the atmospheric parameters. Figure \ref{fig:4-fiting-example} shows the fitting result of a LAMOST extreme subdwarf in the upper panel and an SDSS subdwarf spectrum showing H$\alpha$ emission in the lower panel. The spectral bands contaminated by telluric absorptions are masked.

(5) For each index, measure its value on each observed spectrum and on its corresponding continuum-corrected spectrum, and get their difference. 

(6) Exclude the objects with one or more spectral index differences beyond 3$\sigma$ of the corresponding difference distribution.

(7) Combine the two samples above, and apply the selecting cuts by removing the objects of $\zeta_{\text{L07}}\geq$0.825, $\zeta_{\text{L13}}\geq$0.825, and $\zeta_{\text{Z19}}\geq$0.75.

Finally, a sample of 3,131 subdwarfs is assembled, 1,722 of which have reliable kinematic properties. To provide better understanding, we summarize in Table \ref{tab:subsamples} the various subsamples used along this work. 

\begin{table*}[htb!]
\setcounter{table}{2}
\caption{The various stellar samples used in the present work.} \label{tab:subsamples}
\begin{threeparttable}
\footnotesize
\begin{tabular*}{18.5cm}{p{0.6cm}p{1.3cm}<{\centering} p{1.8cm}<{\centering} p{2cm}<{\centering} p{3.1cm}<{\centering} p{2.7cm}<{\centering} p{2.8cm}<{\centering} p{1.3cm}<{\centering}}
	\hline\hline
	\tabincell{c}{\quad \\ \quad \\ \quad \\ \quad \\ (1)} &
	\tabincell{c}{LAMOST\\ M dwarfs\\ \quad \\ \quad \\ (2)} &
	\tabincell{c}{LAMOST\\ M subdwarfs\\ initial sample \\ \quad \\ \quad } &
	\tabincell{c}{SDSS\\ M subdwarfs\\ initial sample\\ \\ (3)} &
	\tabincell{c}{LAMOST+SDSS\\ combined subdwarf sample\\ with atmospheric param.\\ before CaOH/CaH1 cut\\ \quad} &
	\tabincell{c}{LAMOST+SDSS\\ \qquad green-marked \qquad \quad \\ ``subdwarfs'' \\ \quad\\ \quad} &
	\tabincell{c}{LAMOST+SDSS\\ ``genuine'' subdwarfs\\ after CaOH/CaH1 cut\\ \quad \\ (4)} &
	\tabincell{c}{LAMOST\\ M giants\\ \quad \\ \quad \\ (5)}\\
	\hline
N	& $\sim$70,000	& 4,767 &   2,780   &   \tabincell{c}{3,131\\ \tiny(LAMOST: 1,852)\\ \tiny(SDSS: 1,279)\\ \quad  }  & 855 & 2,276 & 7,200	\\                               
N$_{\text{Gaia}}$ & 35,382 &  - & - & \tabincell{c}{1,722\\  \tiny(LAMOST: 1,160)\\  \tiny (SDSS: 562)} & 526 & 1,196 & - \\
	\hline
\end{tabular*}
\begin{tablenotes}
        \footnotesize
        \item{}
        \item{Notes to Table \ref{tab:subsamples}:}
 		\item{\quad (1) The numbers corresponding to N$_{\text{Gaia}}$ means the objects (N) have available data from Gaia DR2 and can be used for kinematic analysis. The flag is set as 1 if all of the following conditions are true: \\
 		\hspace*{0.7cm} a. phot\_bp\_mean\_flux\_over\_error $>$ 10 \\
		\hspace*{0.7cm} b. phot\_rp\_mean\_flux\_over\_error $>$ 10 \\
		\hspace*{0.7cm} c. phot\_g\_mean\_flux\_over\_error $>$ 10 \\
		\hspace*{0.7cm} d. ResFlag = 1 $\&$ ModFlag = 1 (from \citealt{2018AJ....156...58B}) \\
		\hspace*{0.7cm} e. astrometric\_excess\_noise\_sig $\leq$ 2 \\
		\hspace*{0.7cm} f. ruwe $\leq$ 1.4}
        \item{\quad (2) A visually identified sample from \cite{2019ApJS..240...31Z} for comparison.}
        \item{\quad (3) \citet{2014ApJ...794..145S}, the spectral indices are recalculated using the radial velocities published in the catalog.}
       \item{\quad (4) i.e. using the Equation \ref{equa:cah1_caoh} in the present paper.}
       \item{\quad (5) A visually identified sample from \cite{2019ApJS..240...31Z} for comparison.}
\end{tablenotes}
\end{threeparttable}
\end{table*}

\begin{figure}[htb!]
\center
\includegraphics[width=82mm]{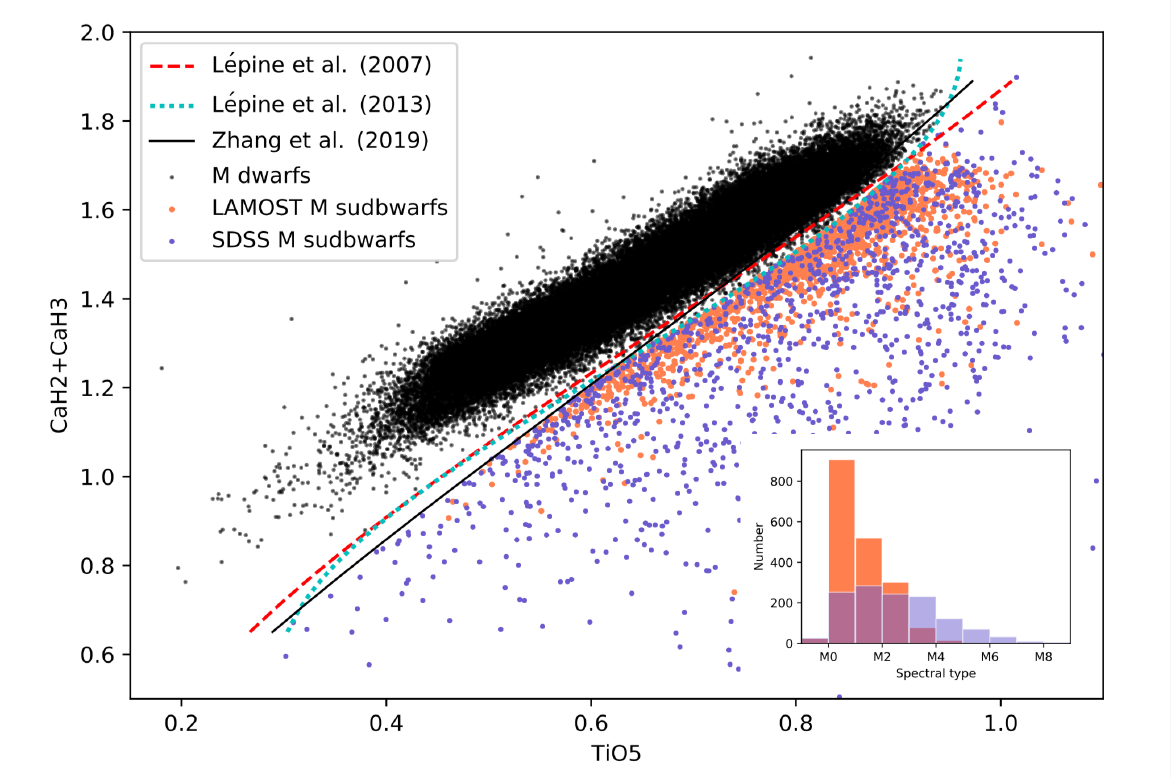}
\caption{The M subdwarf sample constructed in this work, combining the low-resolution spectra from SDSS (blue dots) and LAMOST (red dots) data sets. The M dwarfs used for comparison (black dots) are visually confirmed from LAMOST data \citep{2019ApJS..240...31Z}. The bottom right internal subdiagram shows the spectral type distributions of the two sub-samples. \label{fig:4-combine-two}}
\end{figure}

We must underline that we haven't used the CaH1 index as a filter until now. For comparison, we also estimate the atmospheric parameters of a M dwarf sample which contains around 70,000 visually inspected spectra from LAMOST \citep{2019ApJS..240...31Z}. Figure~\ref{fig:4-combine-two} shows the combined LAMOST+SDSS sample of subdwarfs together with the comparison sample of LAMOST M dwarfs. The LAMOST sample contains a large number of early-type and medium metal-poor stars, while the SDSS sample covers a broader range of spectral types and metal abundances.

\subsection{Results of Parameter Estimation} \label{subsec:result_param}

\subsubsection{Signal-to-Noise impact on quality}
\begin{figure}[htb!]
\center
\includegraphics[width=95mm]{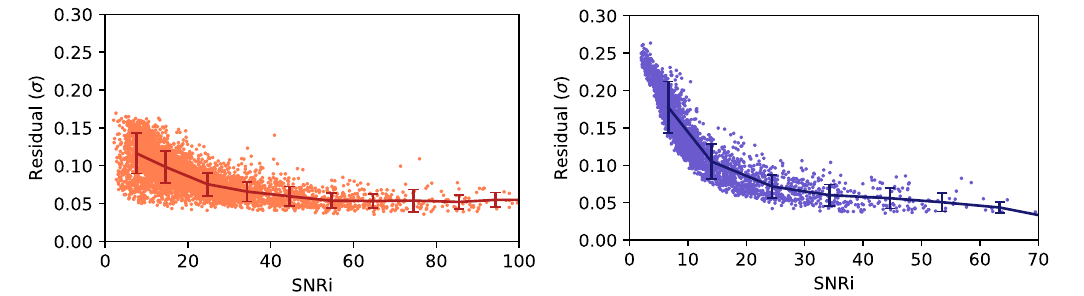}
\caption{The spectral fitting residual of subdwarfs as a function of the signal-to-noise in $i$ band. Left panel: LAMOST subdwarfs, right panel: SDSS subdwarfs. 
\label{fig:4-subdwarf-sn}}
\end{figure}

Statistically, the fit residuals between the observed spectra and the synthetic spectra vary with signal-to noise ratio (S/N). Figure \ref{fig:4-subdwarf-sn} shows the distribution of fitting residuals for subdwarfs from the two surveys. The left panel shows the LAMOST subdwarfs mainly covering M0$-$M3 while the SDSS subdwarfs on the right panel cover M0 to M8. The residual levels stabilize below 10\% after the S/N becomes larger than $\sim$30. The later-type subdwarfs from both samples show higher residual levels because of an average lower spectrum quality.

\subsubsection{Comparison with literature results}
\begin{figure*}[htb!]
\center
\includegraphics[width=140mm]{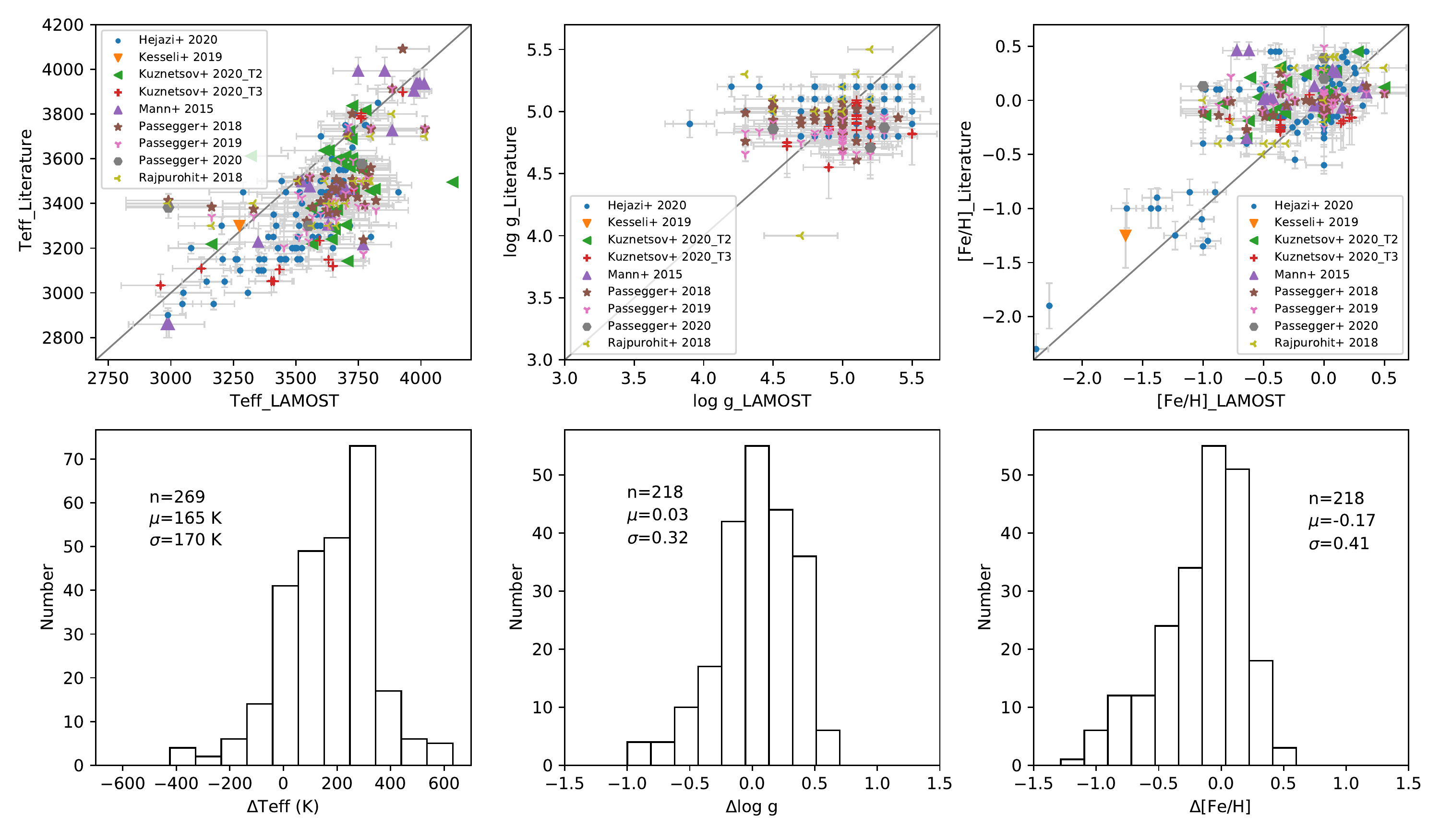}
\caption{The upper panels show the comparison of atmospheric parameters of 167 M stars estimated from 269 LAMOST spectra (including multi-observations) with the ones compiled from literature. Parameter values from different literature are shown in different colors as listed in legend. Error estimates provided are shown as corresponding error bars. The uncertainties for the values from LAMOST spectra are the upper values of internal errors determined from signal-to-ratio and $\sqrt{\chi^2}$. The bottom panels show the distribution of differences along with the corresponding mean value and standard deviation. \label{fig:4-labelM}}
\end{figure*}

In order to estimate the accuracy of the derived parameters, we cross-match the entire LAMOST DR7 M-type stars with objects having known parameters from literature. As a result, we find a total of 269 observations for 167 stars studied with high or medium spectral resolution in 9 papers, and the comparison is shown in Figure \ref{fig:4-labelM}. Although a systematic bias in the effective temperature appears from around 3700K to the cool end, it generally follows a 1-to-1 trend with the literature values. The metallicity also seems to follow a 1-to-1 trend, while the gravity values show a larger range than the literature ones. Note that we have masked several telluric absorption bands where gravity-sensative K and Na lines are located in the fitting process, the accuracy of our estimated gravity values is therefore affected somehow. The distributions of the differences (measurements from this work $-$ literature values) for the three parameters are shown in Figure \ref{fig:4-labelM}. The bias and scatter are 165 K and 170 K for $T_{\text{eff}}$, 0.03 dex and 0.32 dex for log $g$, and $-$0.17 dex and 0.41 dex for [Fe/H], respectively.

\section{Sample Properties Analysis} \label{sec:property}

In this section, we analyze the properties of the subdwarf sample by comparing them with the dwarf sample on spectral index diagrams and Gaia DR2 H-R diagram, and provide a preliminary account of their kinematic properties from the reduced proper motion diagram and in the 3D galactic motion system. 

In Paper \Rmnum{1}, we suggested a screening condition associated with the CaH1 index be combined with the classical $\zeta$ condition to identify a subdwarf via its low-resolution spectrum. To show the importance of gravity and gravity-dependent index---CaH1 for identifying a ``genuine'' subdwarf, we divide the subdwarf sample selected above into two groups by the separator curve (equation 9 in Paper \Rmnum{1}) on the [CaOH, CaH1] index diagram:
\begin{equation}{\label{equa:cah1_caoh}}
\begin{aligned}
\text{CaH1}&<0.4562\times\text{CaOH}^3-0.1977\times \text{CaOH}^2\\
&-0.01899\times\text{CaOH}-0.7631
\end{aligned}
\end{equation}
The objects that do not meet Equation (\ref{equa:cah1_caoh}) are marked in green in the following figures. 

For further analysis, we retrieve proper motions, and photometric magnitudes in $G$, $BP$, and $RP$ bands of these stars from Gaia Data Release 2, and estimated distance from \cite{2018AJ....156...58B}. To ensure reliability, we set a series of filter conditions as listed in the notes to Table \ref{tab:subsamples}, and finally 1,722 objects among the 3,131 selected in Section \ref{sec:sample} meet the conditions, 526 of which are green-marked ``subdwarfs''.

\begin{figure*}[htb!]
\center
\includegraphics[width=120mm]{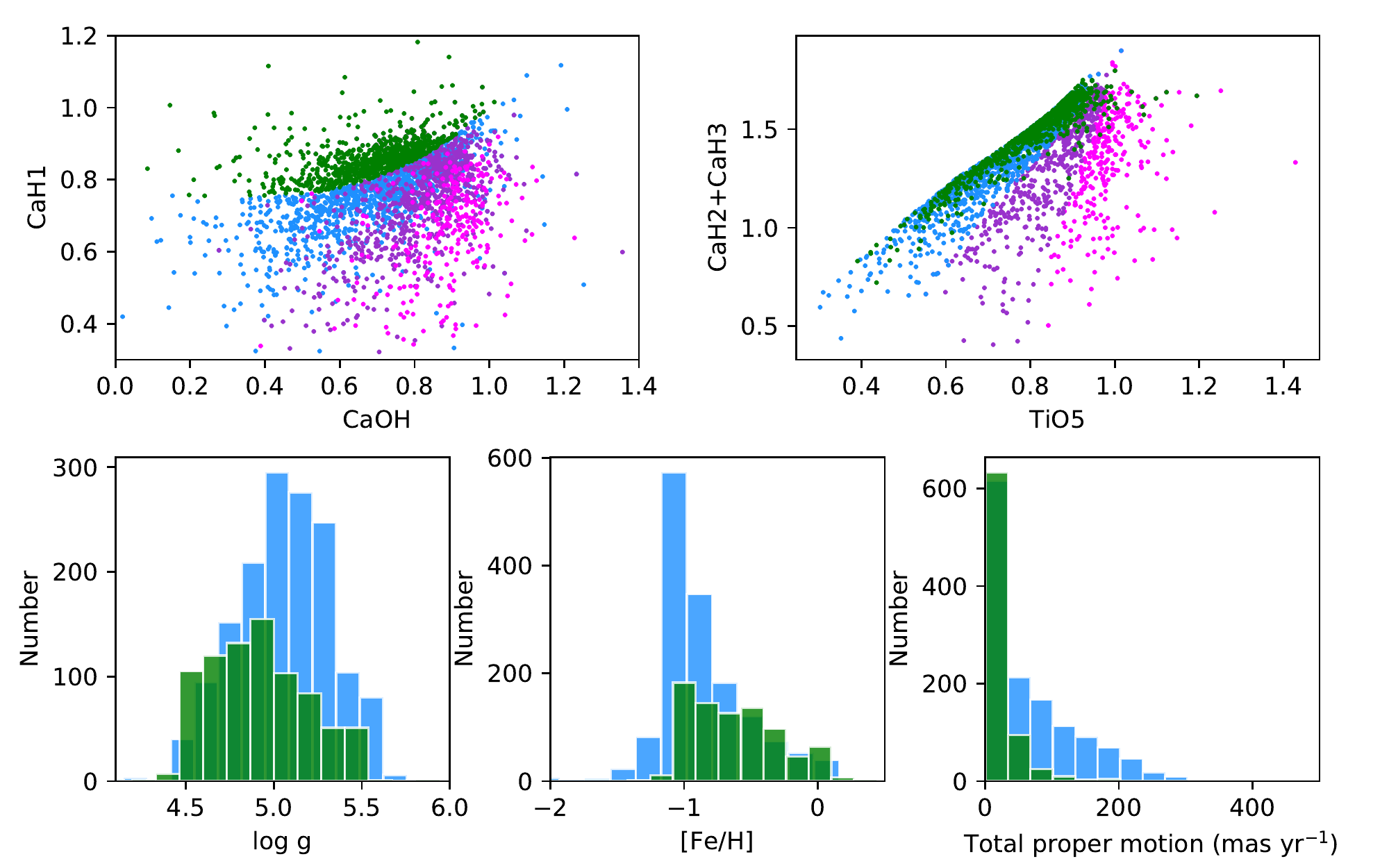}
\caption{The gravity and metallicity distributions of the green-marked ``subdwarfs'' vs. the other sdMs. The two upper panels show the distribution of spectral indices of four groups of subdwarfs: green-marked ``subdwarfs'' (the subdwarfs that don't meet Equation \ref{equa:cah1_caoh}), sdMs, esdMs, and usdMs (the remaining subdwarfs classified into to categories according to the $\zeta$ values). The sdMs, esdMs and usdMs are colored in blue, purple and magenta respectively. The bottom left and middle panels show the comparison of green-marked ``subdwarfs'' with the sdMs. The bottom right panel compares the total proper-motion distribution of the green-marked ``subdwarfs'' with that of the other sdMs. \label{fig:4-CaOH-CaH1-1}}
\end{figure*}

\subsection{Spectral Index Diagram} \label{subsec:index}

As shown in Figure \ref{fig:4-CaOH-CaH1-1}, the upper left panel shows the green marked ``subdwarfs''. According to the $\zeta$ values, we also divide the remaining subdwarfs into sdM/esdM/usdM subclasses on the [TiO5, CaH2+CaH3] diagram, as can be seen in the upper right panel, and the objects in the three subclasses are plotted as blue, purple and magenta points respectively. Most of the green marked ``subdwarfs'' appear located in the ``sdM'' region, which is colored in blue. Therefore, we compare the gravity and metallicity of the two groups and display those distributions in the bottom left and middle panels. As shown in the bottom middle panel, the green objects share similar metal abundance distribution with the blue labeled ones, however, their gravities are systematically lower than the latter ones by some 0.2 dex.

With the available proper-motions of these targets from Gaia DR2, we also compare the total proper motions of the two groups in bottom right panel of Figure \ref{fig:4-CaOH-CaH1-1}. The proper-motions of green marked ``subdwarfs'' are overall smaller than those of the sdMs. Generally speaking, large proper motion is a classical property of subdwarfs, though some objects may have relatively small proper motion due to the 3D motion direction. Proper motions of esdMs and usdMs are larger than sdMs (not shown in the figure), which means that the green ones do behave more like dwarfs than subdwarfs.

\begin{figure}[htb!]
\center
\includegraphics[width=80mm]{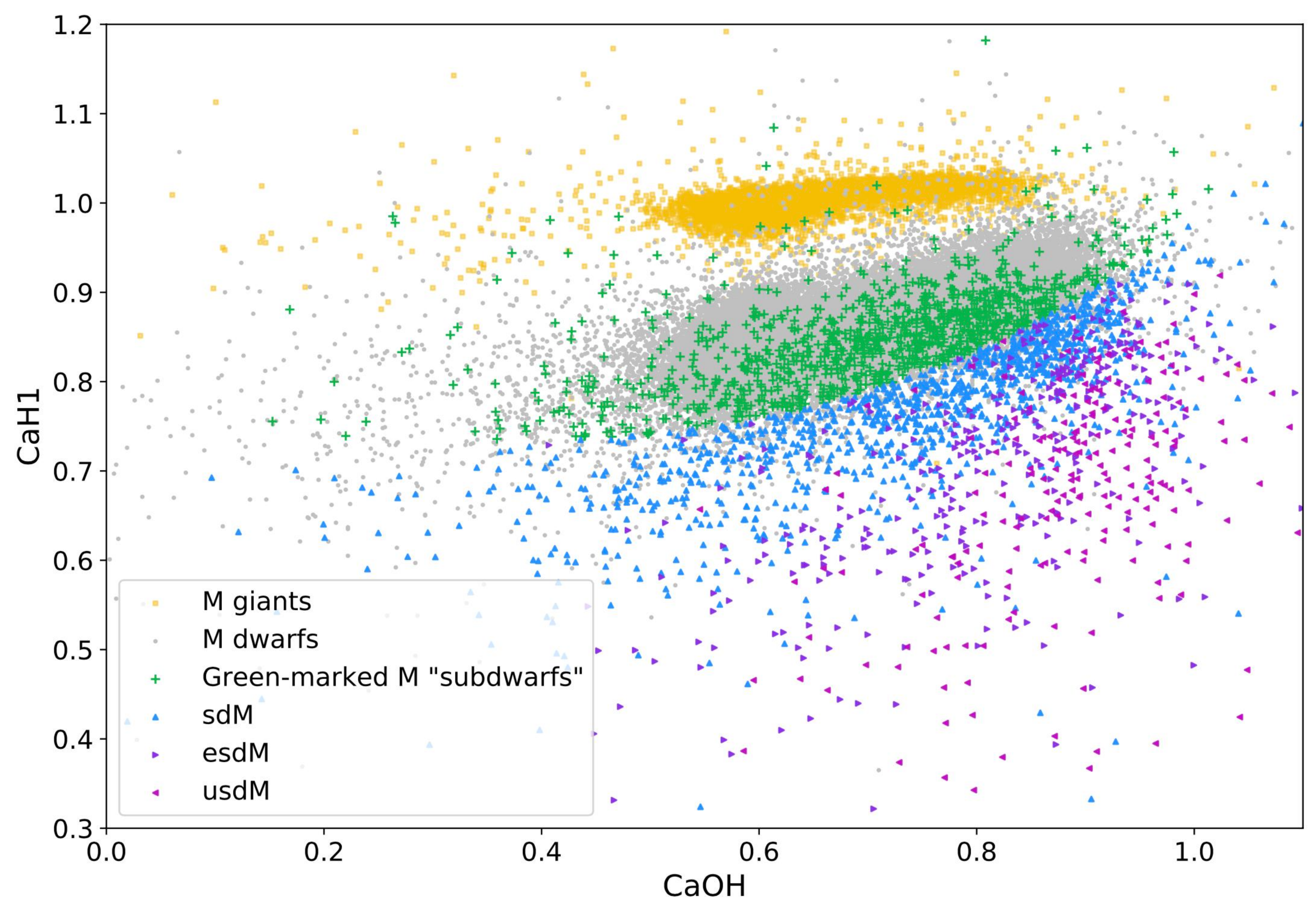}
\caption{M-type stars with different gravity shown on the spectral index diagram CaH1 vs. CaOH. Color coding is the same as in the upper left panel of Figure \ref{fig:4-CaOH-CaH1-1}. To illustrate the sensitivity to gravity of this diagram, an M giant sample (7,200 objects visually identified from \citealt{2019ApJS..240...31Z} and colored in yellow) and the M dwarf sample are added. 
\label{fig:4-CaOH-CaH1-2}}
\end{figure}

The gravity differences can be seen more clearly in Figure \ref{fig:4-CaOH-CaH1-2}, in which we add the LAMOST M dwarfs and a new sample of M giants onto the [CaOH, CaH1] index diagram. The giants were also spectroscopically identified by visual inspection \citep{2019ApJS..240...31Z}. Giants, and dwarfs subdwarfs are classified as luminosity class \Rmnum{3}, \Rmnum{5} respectively and \citet{2008AJ....136..840J} have proposed that subdwarfs be classified as \Rmnum{6}, this sequence tracing the gravity variation trend. Therefore, Figure \ref{fig:4-CaOH-CaH1-2} clearly illustrates the gravity segregation between the luminosity classes of giants/dwarfs/subdwarfs thanks to the sensitivity of CaH1 index to log $g$ especially when compared to CaOH index.

\subsection{H-R Diagram} \label{subsec:hrd}

\begin{figure}[htb!]
\center
\includegraphics[width=80mm]{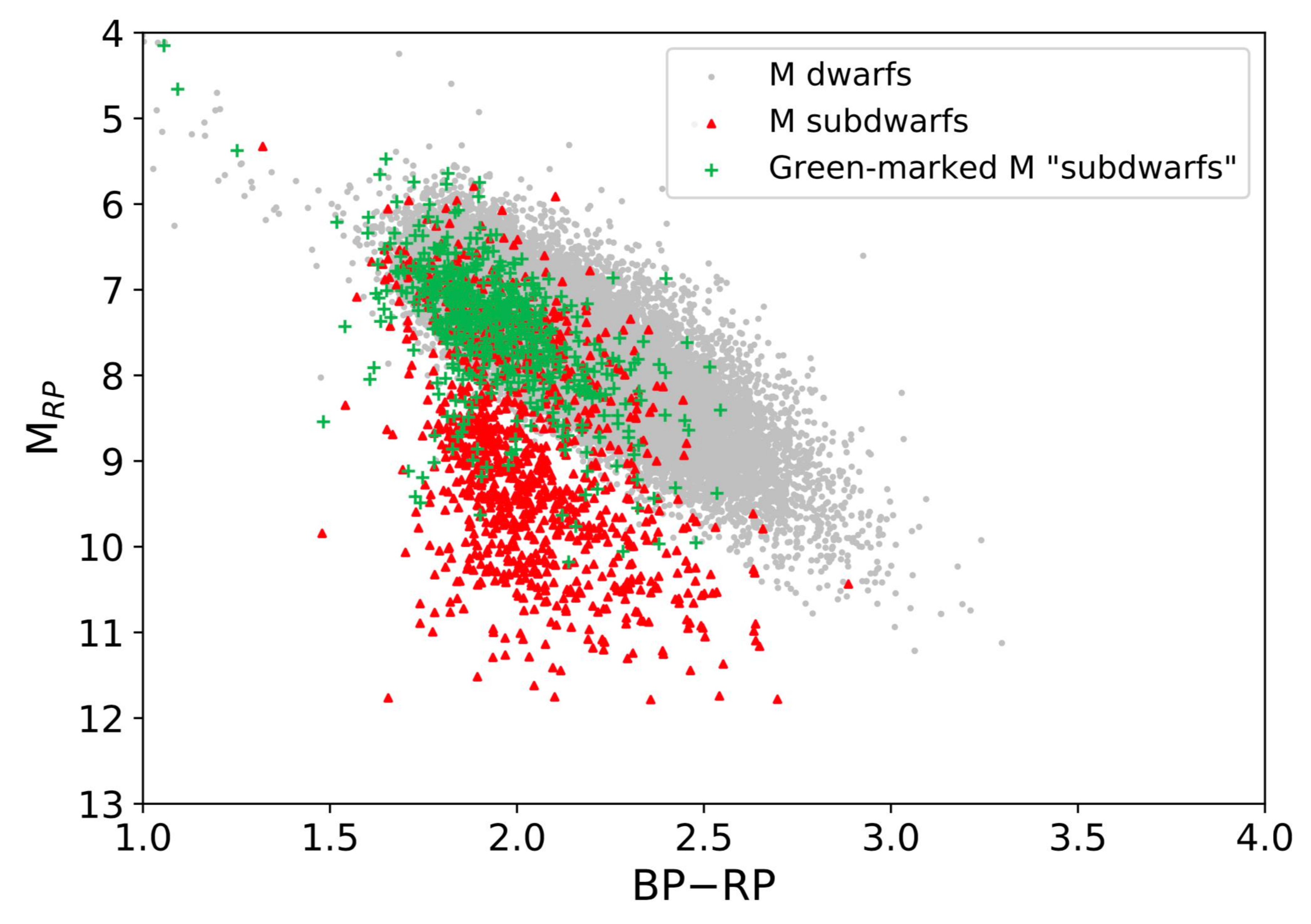}
\caption{All the subdwarfs and dwarfs on the Gaia color-magnitude diagram. The absolute magnitudes in RP band are derived from the Gaia DR2 apparent magnitudes and estimated distances from \citet{2018AJ....156...58B}. Most of the green-marked ``subdwarfs'' fall into the main sequence region populated by the M dwarfs plotted as black dots, and the remaining subdwarfs plotted as red dots lie in part below the main sequence with a large dispersion, but some are also distributed across the main sequence region. \label{fig:5-BP-RP_Mrp_all}}
\end{figure}

Using estimated distance by \citet{2018AJ....156...58B}, we derive the absolute magnitudes M$_{\text{G}}$, M$_{\text{BP}}$ and M$_{\text{RP}}$ of the dwarfs and subdwarfs. Since most of the M-type stars we analyze here are located within 1 kpc, we choose to ignore the effect of extinction in the calculation. Figure \ref{fig:5-BP-RP_Mrp_all} shows the stars on Gaia color-absolute magnitude diagram [BP$-$RP, M$_{\text{RP}}$]. On this diagram, which is an observational Hertzsprung-Russel diagram, the gray dots represent the dwarf sample, the green dots represent the ``subdwarfs'' that do not meet Equation (\ref{equa:cah1_caoh}), and the red dots represent the remaining subdwarfs (all subclasses together). The green-marked ``subdwarfs'' fall almost completely in the dwarf region, and thus once again verify the conclusion that they behave more like dwarfs than subdwarfs. 

\begin{figure}[htb!]
\center
\includegraphics[width=90mm]{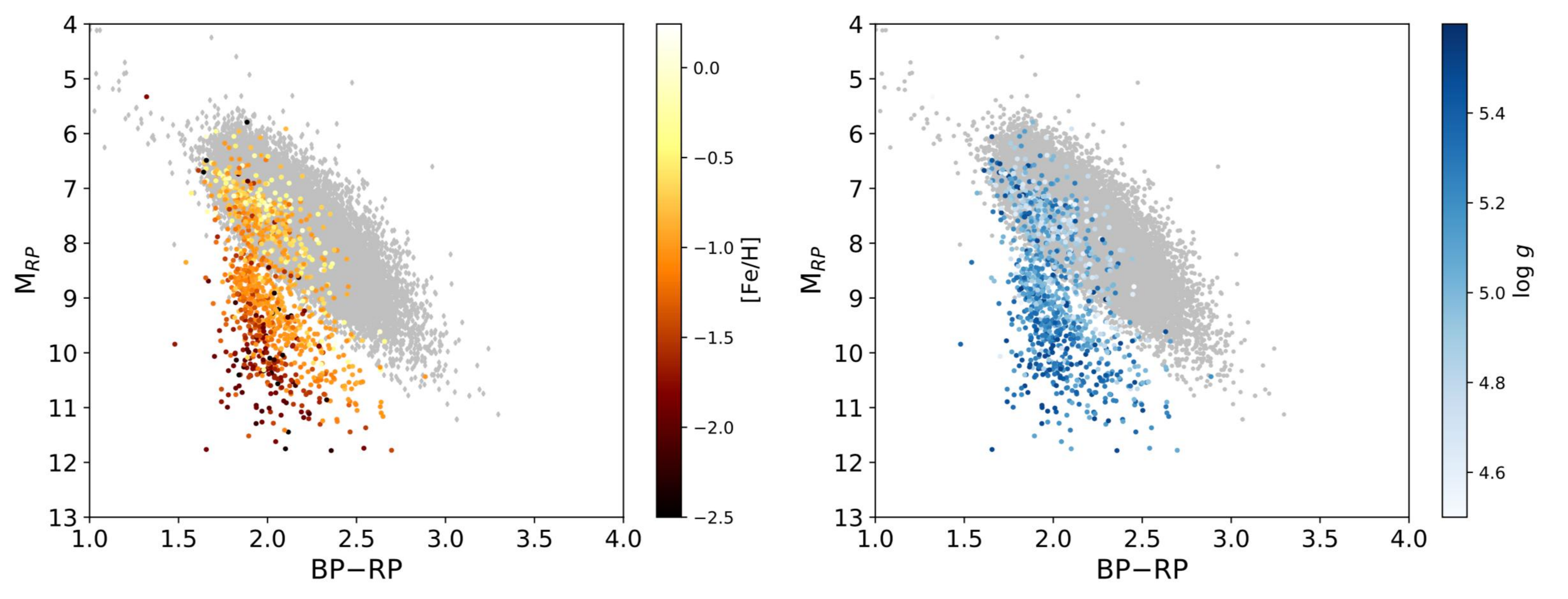}
\caption{After the removal of the green-marked ``subdwarfs'', the remaining samples are shown on the Gaia color-absolute magnitude diagram again. On the left panel, subdwarf are color-coded for different metallicities, while the same ones on the right panel are color-coded for different gravities. Note that the subdwarfs with log $g>$ 5.5 or log $g<$ 4.5 (38 objects in total) are not plotted on the right panel to make the color bar more tight.\label{fig:5-BP-RP_Mrp}}
\end{figure}

Besides, we find that a number of Equation (\ref{equa:cah1_caoh}) selected subdwarfs (red dots) also overlap with dwarfs on this diagram. This point will be developed below. 

Figure \ref{fig:5-BP-RP_Mrp} shows the metal abundance and gravity variation of these stars on the Gaia color-magnitude diagram. The subdwarfs falling in the dwarf region are the ones with the highest metal abundances or the smallest surface gravities. In other words, stars that are away from the main sequence have both large gravity and low metal abundance, and they should be the ``genuine'' subdwarfs.

\subsection{Kinematics} \label{subsec:kinematic}

\subsubsection{Reduced Proper Motion diagram} \label{subsubsec:rpm}

Aside from the H-R diagram, another useful tool for effectively separating different stellar populations is the reduced proper motion (RPM; \citealt{1972ApJ...173..671J}) diagram (e.g. \citealt{2002ApJ...575L..83S,2005AJ....130.1658S,2007ApJ...669.1235L,2009AJ....137....1F}), which uses photometric information (color) and proper motions of the stars. Even without knowledge of the distance to the stars, the RPM diagram can be used to constrain kinematic properties of large stellar samples, because a given stellar population whose members are orbiting in the Galactic potential has a characteristic space velocity distribution reflected in a combination of luminosity and transverse motion. 
The definition of the reduced proper motion $H$ is
\begin{equation}
H = m + 5log \ \mu + 5 = M + 5log \ T
\end{equation}
where $m$ is the apparent magnitude of a star in a given photometric band, $\mu$ is the annual total proper motion (in arcsec yr$^{-1}$), $M$ is the absolute magnitude, and $T$ is the tangential velocity (in a.u. yr$^{-1}$). 

Since the dwarfs belong to the thin disk population and the subdwarfs are claimed to belong to the thick disk and halo populations, the comparison between blue to red or optical to near-infrared colors and the reduced proper motions can in principle relatively clearly distinguish the Population \Rmnum{1} dwarfs and Population \Rmnum{2} subdwarfs : (1) given an absolute magnitude, subdwarfs are bluer in color; (2) subdwarfs being members of a population spanning a spatial volume more extended out of the Galactic plane with more eccentric orbits, they have higher average tangential velocities \citep{2002ApJ...575L..83S,2007ApJ...669.1235L}. The RPM diagram combines these two features so that the subdwarfs and main sequence stars fall into two separate regions. This method has long been used to separate high-velocity subdwarfs from dwarfs \citep{1972ApJ...173..671J,2003PASP..115...22Y,2003PASP..115..796Y}.

\begin{figure}[htb!]
\center
\includegraphics[width=90mm]{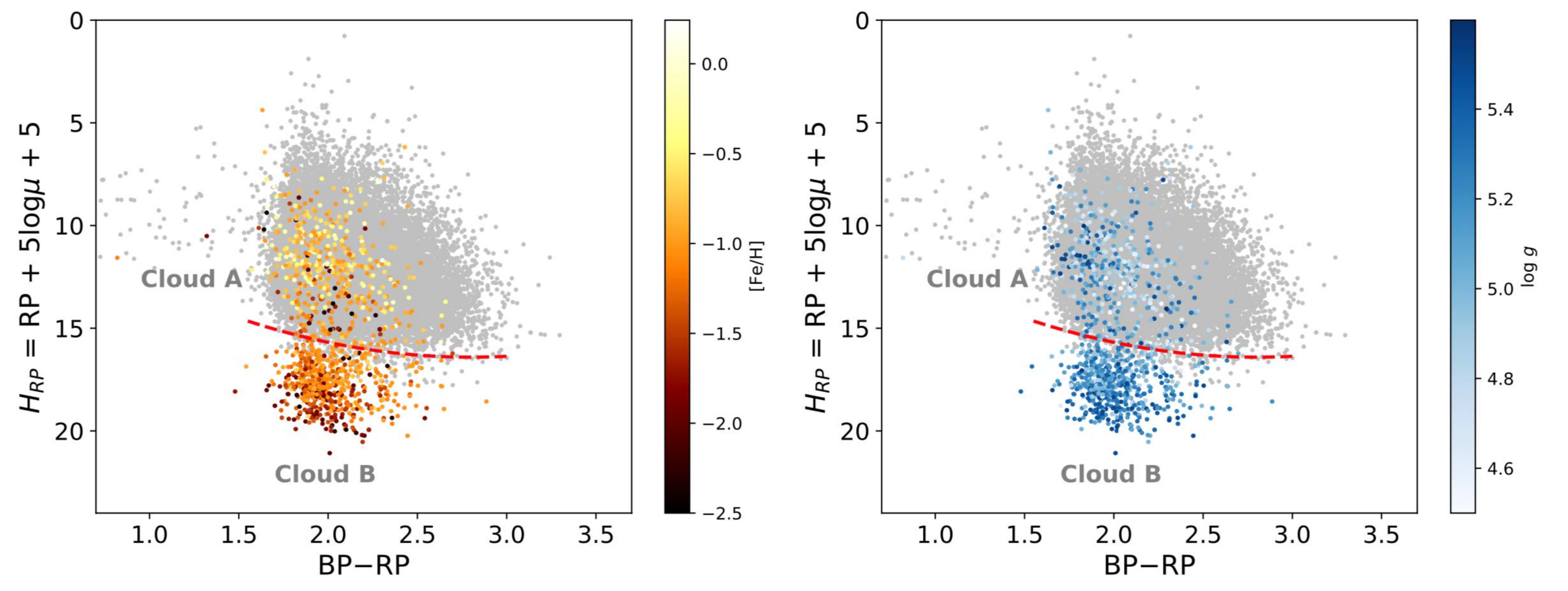}
\caption{After the removal of the green-marked ``subdwarfs'', the remaining samples are shown on the reduced proper motion diagram. The color coding is the same as in Figure \ref{fig:5-BP-RP_Mrp_all}. The fiducial separation line between Cloud A and Cloud B objects is drawn (see text).\label{fig:5-BP-RP_Hrp}}
\end{figure}

Figure \ref{fig:5-BP-RP_Hrp} shows the RPM diagram for our stars, using Gaia DR2 color and proper motions. A clear similarity with the information contained in the H-R diagram of Figure \ref{fig:5-BP-RP_Hrp} can be noted. Although the selection biases of the combined LAMOST+SDSS subdwarf sample are poorly known (mix of pointing directions, different limiting depth of surveys, velocity ellipsoid sampling, density distribution sampling, spatial inhomogeneities, etc.), basic trends can be outlined. First, the subdwarfs falling below the region occupied by the main-sequence dwarfs have low metal abundances and higher gravities. Second, remaining subdwarfs appear clearly divided in two clouds (noted A and B in the following) with a quite clean separation: Cloud A is not distinct from the lower part of the main-sequence dwarf cloud, while Cloud B is centered around a half magnitude below the center of Cloud A. The fact that Cloud A and B are especially distinct in the RPM diagram imply that these two clouds sample two distinct populations whose kinematic behaviors are different.

\subsubsection{Galactic velocity components} \label{subsubsec:uvw}

Using the usual textbook transformation formula, we compute the three-dimensional space velocity components $U$, $V$, $W$ in the right-handed Galactic frame: the $U$ and $V$ are contained in the Galactic plane, $U$ points toward the Galactic center, while $V$ is the velocity component in the direction of the solar motion. The $W$ component is perpendicular to the Galactic plane and points towards the North Galactic pole. $U$, $V$, $W$ for our targets are determined using the radial velocities, coordinates, proper motions and distances derived from the parallaxes. The components are reduced to the Local Standard of Rest (LSR; \citealt{2009AJ....137.4149F}) by correcting from the solar motion assumed to be ($U_\odot $, $V_\odot $, $W_\odot $)=[11.1, 12.24, 7.25] km s$^{-1}$ \citep{2010MNRAS.403.1829S}. 

\begin{figure*}[htb!]
\center
\includegraphics[width=140mm]{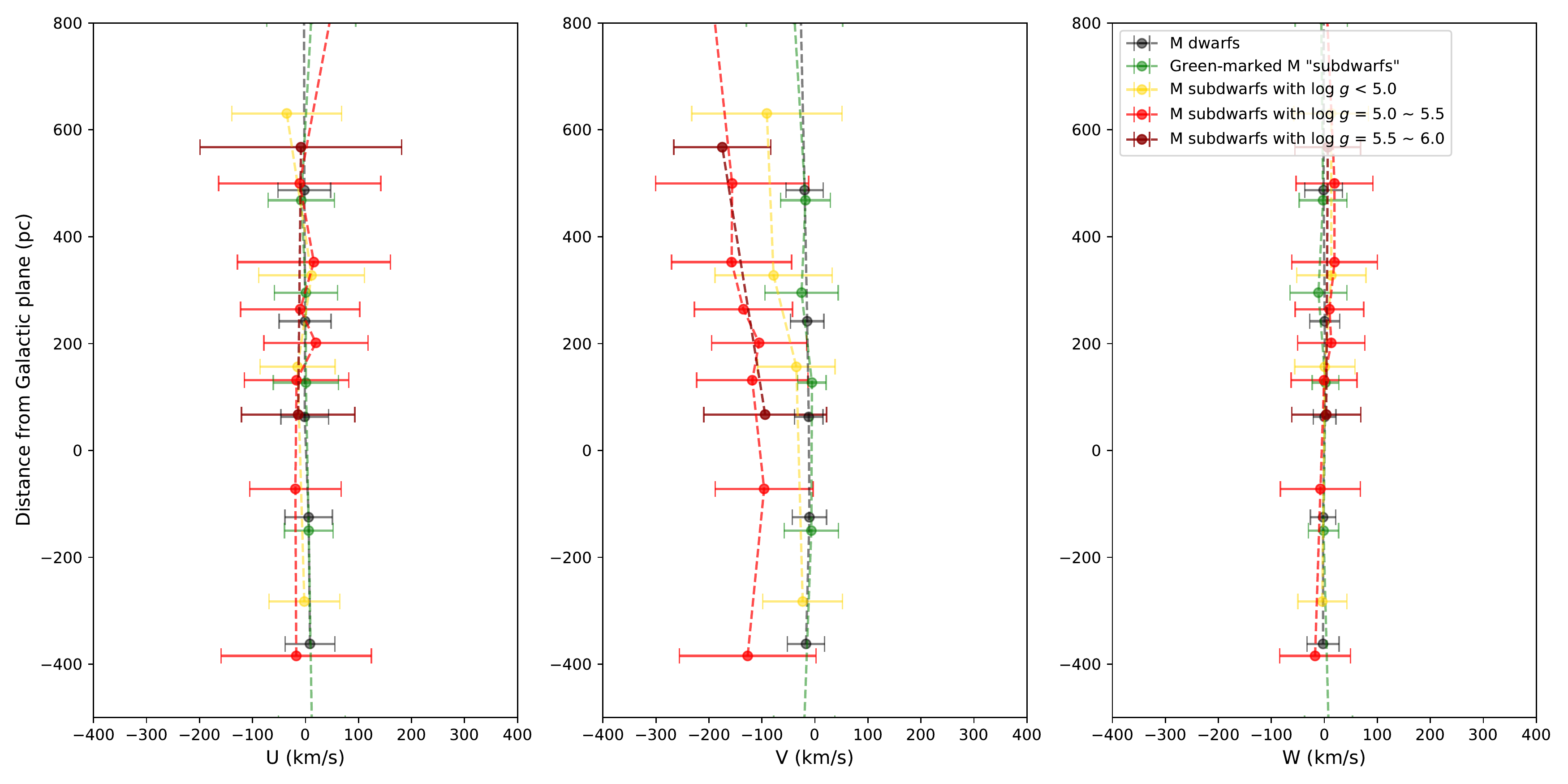}
\caption{Velocity component distribution of dwarfs and subdwarfs with different gravities in the Galactic 3D motion system ($U$, $V$, $W$). From left to right, the abscissa represents the velocity component of $U$, $V$, and $W$ respectively. The ordinate represents the distance from Galactic plane. Except for the green-marked ``subdwarfs'', the remaining subdwarfs are divided into groups with different gravities and distinguished by different colors. Each dwarf bin contains 7,000 stars and each subdwarf bin contains 100 stars. The node and the error bar show the mean value and the standard deviation of each bin respectively. 
\label{fig:5-uvw-1}}
\end{figure*}

\begin{figure*}[htb!]
\center
\includegraphics[width=140mm]{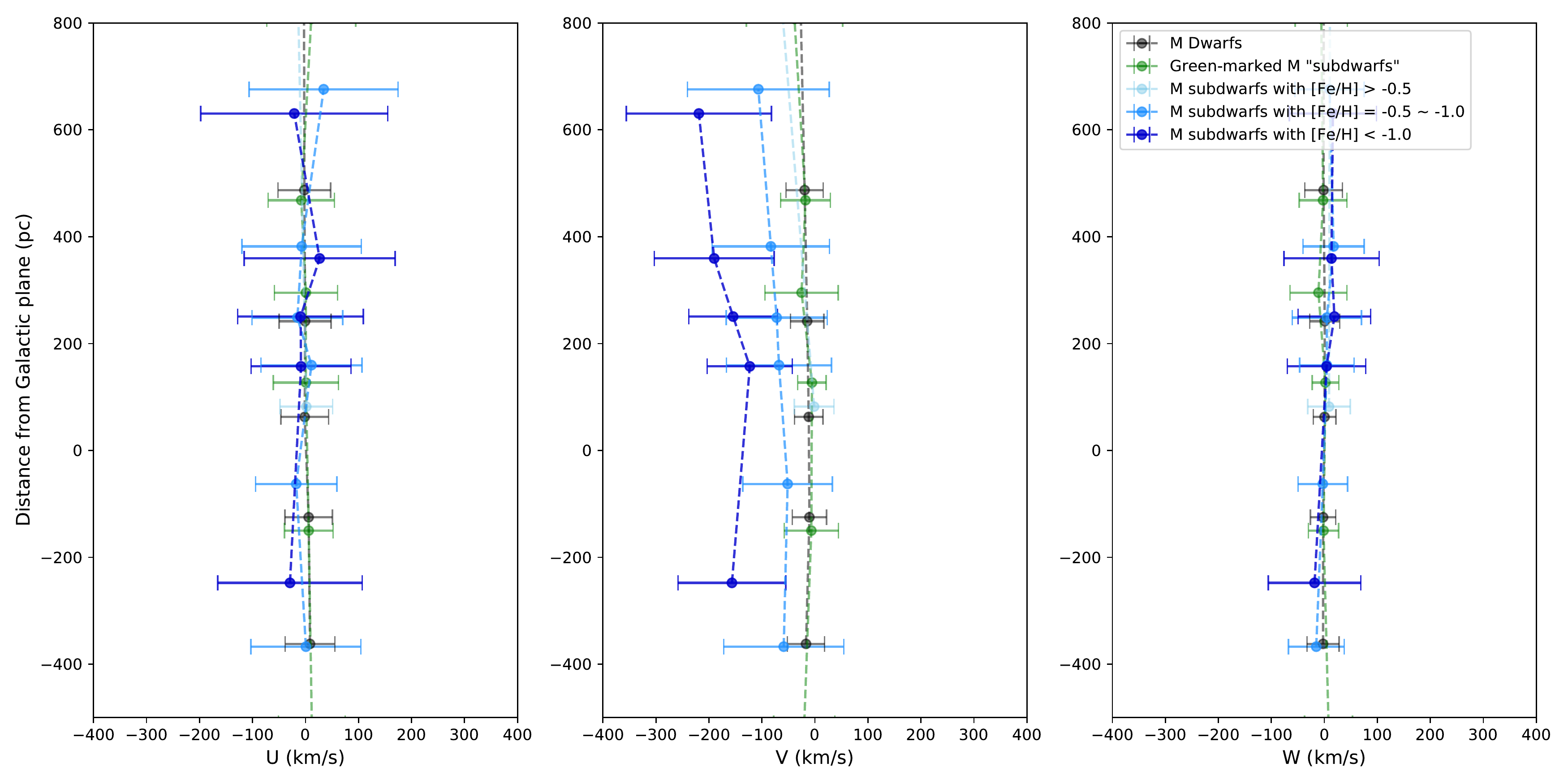}
\caption{Velocity component distribution of dwarfs and subdwarfs with different metallicities in the Galactic 3D motion system ($U$, $V$, $W$). Except for the green-marked ``subdwarfs'', the remaining subdwarfs are divided into groups with different metallicities and distinguished by different colors. The meaning of the axes and the number of stars contained in each bin are the same as in Figure \ref{fig:5-uvw-1}.
\label{fig:5-uvw-2}}
\end{figure*}

In Figures \ref{fig:5-uvw-1} and \ref{fig:5-uvw-2}, we divide our remaining subdwarfs into different gravity and metallicity groups and compare them with the green-marked ``subdwarfs'' and dwarfs. Each dwarf bin contains 7,000 objects and each subdwarf bin contains 100 objects. As expected, the green-marked ``subdwarfs'' behave mostly like field dwarfs with only a slightly larger dispersion in all the velocity components. In addition, the yellow bins in Figure \ref{fig:5-uvw-1} representing the subdwarf group with the lowest gravity and the light blue bins in Figure \ref{fig:5-uvw-2} representing the subdwarf group with the highest metallicity have nearly the same behaviors in $V$ as the dwarfs and the green-marked ``subdwarfs''. 

To further explore the kinematic behavior of the full subsamples, we draw several histograms in Figure \ref{fig:5-histos}, in which the dwarfs, green-marked ``subdwarfs'', Cloud A and Cloud B subdwarfs are plotted separately. The mean values and standard deviations are also shown on each panel, except for $|$z$|$, the absolute value of the distance to the Galactic plane, for which the quartiles are computed. The separation between Cloud A and Cloud B objects is defined by the fiducial line drawn on Figure \ref{fig:5-BP-RP_Hrp}. Several points emerge:

\begin{figure*}[htb!]
\center
\includegraphics[width=130mm]{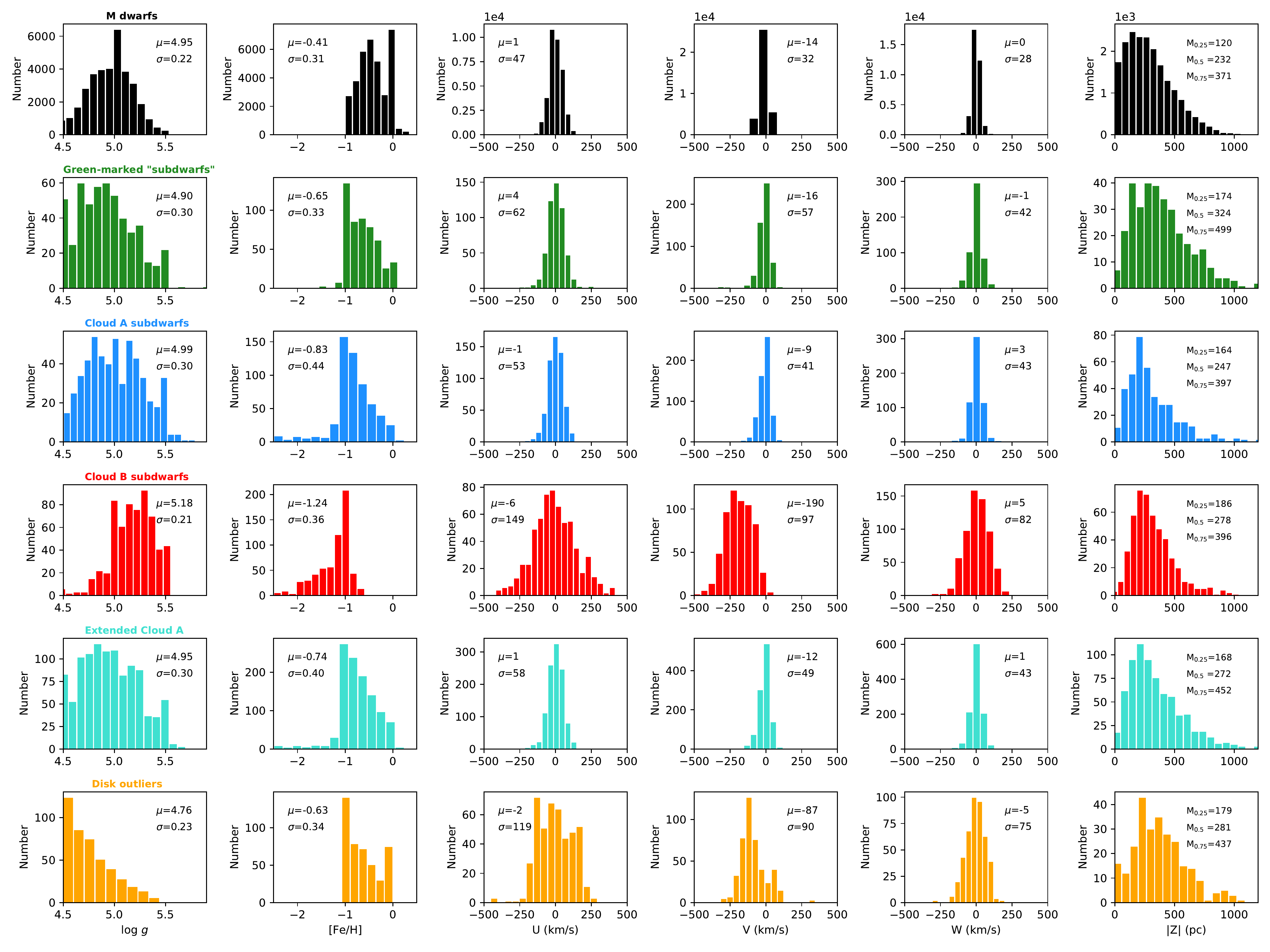}
\caption{Histograms of some characteristics for the various sub-populations of dwarfs and subdwarfs for which Gaia astrometry is available. From left to right, log $g$, [Fe/H], $U$, $V$, $W$ space velocity components and $|$Z$|$. From top to bottom: (a) all spectroscopically identified M dwarfs, (b) green-marked ``subdwarfs'', (c) Cloud A subdwarfs (see Figure \ref{fig:5-BP-RP_Hrp} for definition), (d) Cloud B subdwarfs, (e) Extended Cloud A (see text for definition), (f) Disk outliers (spectroscopically identified M dwarfs whose total space velocity w.r.t. the LSR is larger than that of 99 \% of the population, see Figure \ref{fig:5-Toomre} for their location on Toomre diagram). The $\mu$ and $\sigma$ on each panel are indicated for the mean value and standard deviation respectively. For the $|$Z$|$ distributions, the quartiles are indicated. 
\label{fig:5-histos}}
\end{figure*}

(a) The gravities increase systematically from dwarfs (mean close to 5.0, in agreement with the results from Section \ref{sec:model}) to Cloud B subdwarfs (mean around 5.2).

(b) Metallicity decreases systematically from dwarfs (mean around $-$0.4 dex) to Cloud B subdwarfs. The metallicity of the dwarfs, quite surprisingly, appears systematically poorer than solar, with a distribution tail down to $-$1. dex. Although the error assessment and control is difficult in the spectra fitting process, we recall that the comparison in Subsection \ref{subsec:result_param} with the sample of 167 stars having a metallicity measurement in the literature did not reveal any major systematic deviation.

\begin{figure}[htb!]
\includegraphics[width=80mm]{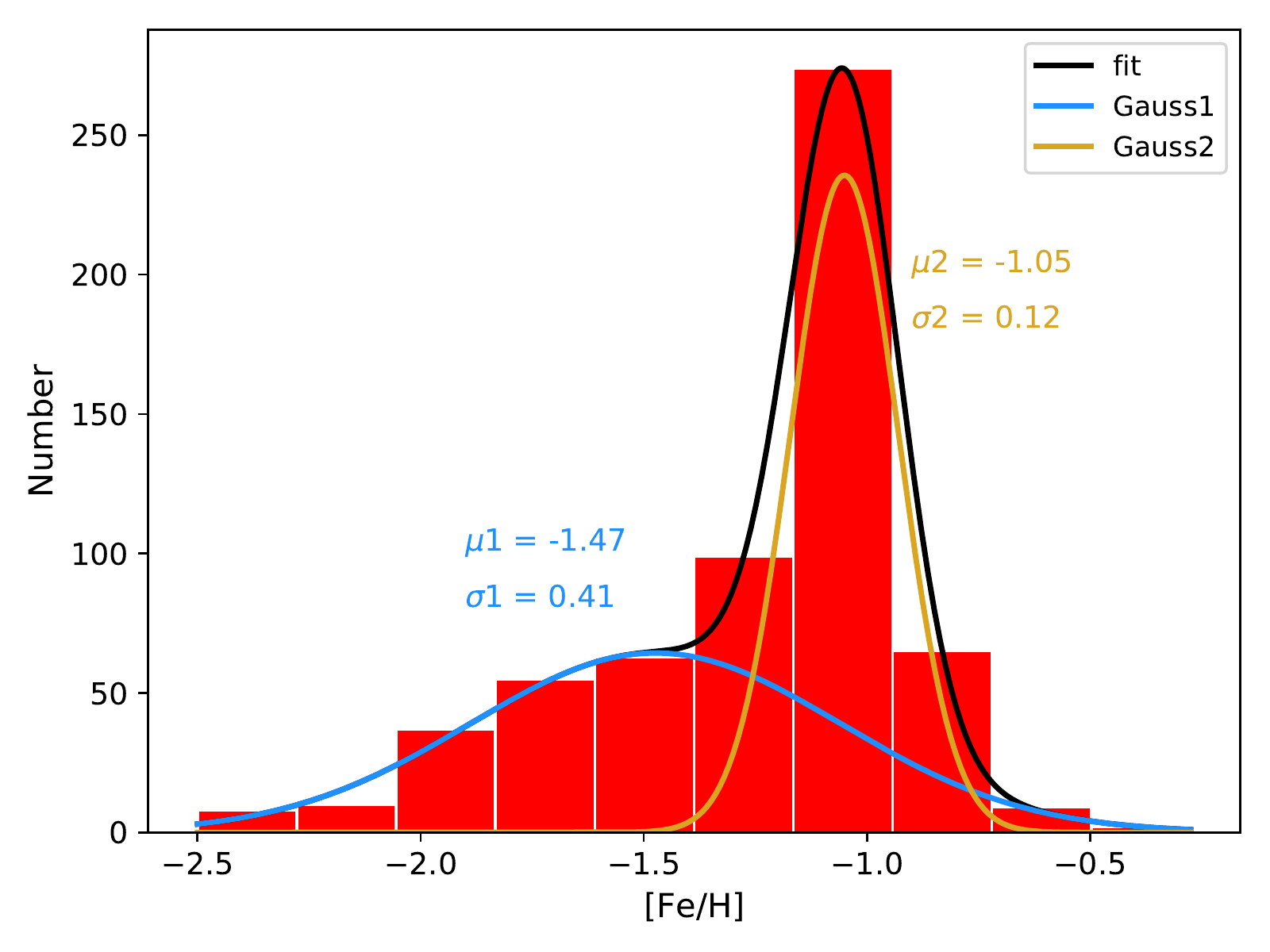}
\caption{The bimodal distribution of metallicity of the halo candidates, which are selected from Cloud B subdwarfs with $v_{tot}\geq$220 km s$^{-1}$. The mean values and standard deviations of each Gaussian fit are also shown on the diagram.
\label{fig:5-bimod}}
\end{figure}

(c) As shown in Figure \ref{fig:5-bimod}, the halo candidates in Cloud B subdwarfs exhibit an inconspicuous bimodal distribution of metallicity, with two sub-populations, one around [Fe/H] = $-$1 dex and the other around $-1.5$ dex. 

(d) The mean values and dispersions of the $U$, $V$, and $W$ components agree quite well with the dwarf sample belonging to the thin disk, with a velocity lag of $V$ = $-$10 km s$^{-1}$. The green-marked and Cloud A objects have more velocity dispersion suggesting possible membership of the thick disk, but their mean $V$ value significantly deviates from the usually adopted $-$60 km s$^{-1}$. Cloud B objects are obviously dominated by halo members, with a mean $V$ largely retrograde and considerable dispersion. 

(e) The quartiles of the $|$z$|$ distributions show that the dwarf sample is the one which is the most concentrated around the Galactic plane, while the Cloud B contains the most diffuse population out of this plane. These histograms also show that the LAMOST+SDSS surveys sample the M dwarf population inside scale height of the thin disk above the Galactic plane (600 pc) \citep{2008ApJ...673..864J}, while the subdwarf one is sampled only up to less than 1 kpc (one scale height of the thick disk or less) \citep{1983MNRAS.202.1025G,2008A&A...480..753V,2010ApJ...712..692C}. 

(f) The striking similarity of all the indicators (except for a marginal difference in log $g$ ) between the green-marked objects and the Cloud A ones imply that they are members of the same stellar population as defined kinematically, only differing by the CaH1 and CaOH indices. In the following, these two groups will be considered as a unique subsample which will be designated as Extended Cloud A.
    
\subsubsection{The Toomre diagram} \label{subsubsec:Toomre}
Trying to push forward the analysis, we have plotted in Figure \ref{fig:5-Toomre} the Toomre diagram for all our stars.

The dwarf cloud appears slightly asymetric, with an extension towards negative $V$ velocity component. This is not unexpected since 2 or 3 - component disk (thin, thick and intermediate) model populations (\citealt{2009MNRAS.399.1145S} and references herein) predict such an asymmetry. The Extended Cloud A group seems to be centered more or less coincident with the dwarf cloud, while Cloud B objects are scattered across a large area of the diagram. 

\begin{enumerate}

\item Halo objects.

To select halo objects, a stringent condition on their total velocity \citep{2017ApJ...845..101B} which would avoid pollution by possibly numerous thick disk stragglers \citep{2009MNRAS.399.1223S}, is: 
\begin{equation}
|V - V_\text{LSR}| \geq 220\ \text{km s}^{-1}
\end{equation}
which translates, which the adopted velocity components of the LSR, ($U$ = 0, $V$ = 220, $W$ = 0 km s$^{-1}$) into:
\begin{equation}
v_\text{tot} = \sqrt{U^2 + V^2 + W^2} \geq 220\ \text{km s}^{-1}
\end{equation}
where $v_\text{tot}$ is the total space velocity with respect to the LSR.

\begin{figure*}[htb!]
\center
\includegraphics[width=155mm]{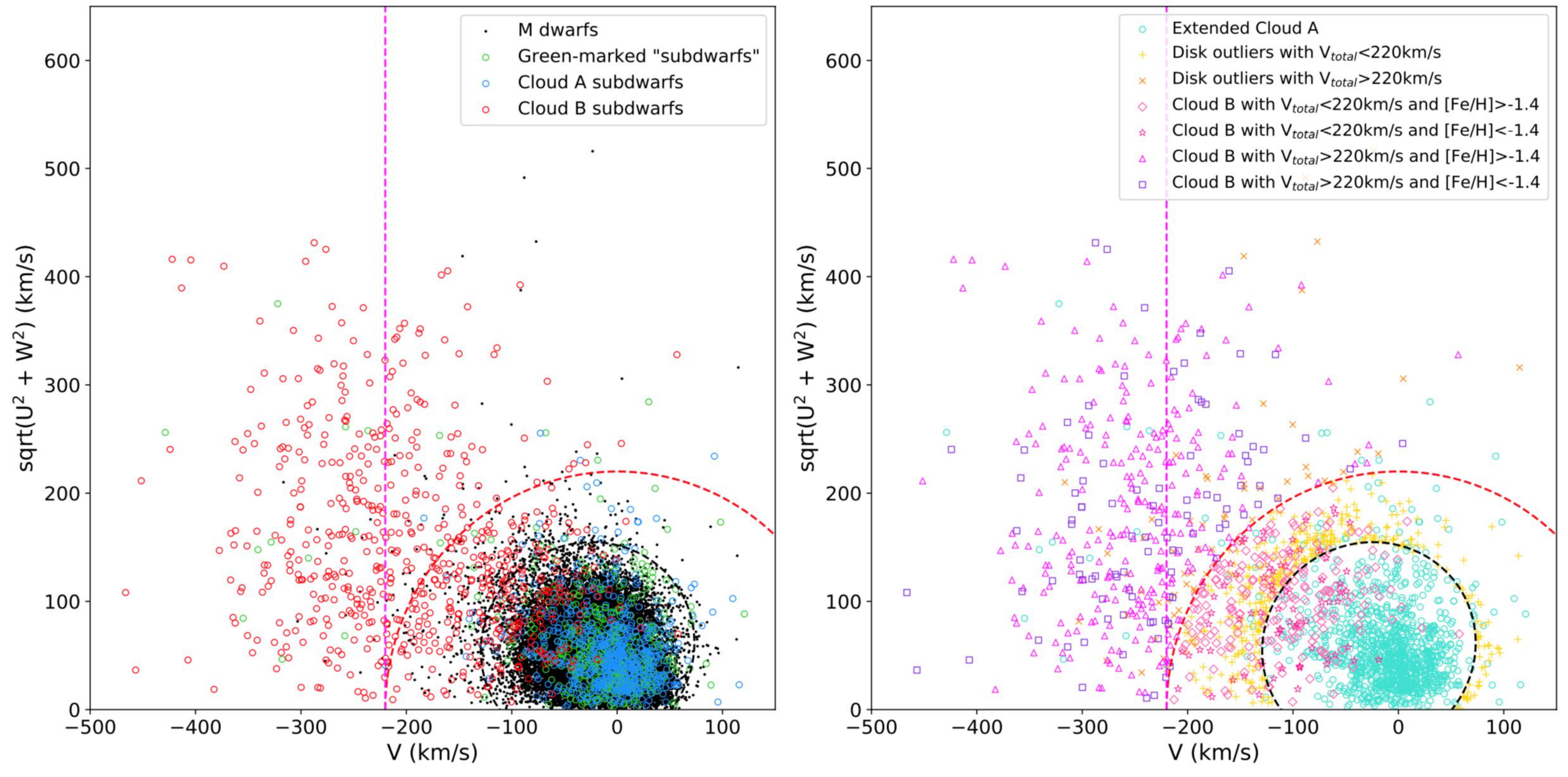}
\caption{Left panel: Toomre diagram for all stars in the present study. Black points: all spectroscopically identified M dwarfs, green circles: the green-marked ``subdwarfs'', blue circles: Cloud A subdwarfs, red circles: Cloud B subdwarfs. The black dashed curve encloses 99 \% of the M dwarf population: all black points outside this curve are considered as representative of ``outliers''. The dashed red circle is our adopted limit $v_\text{tot}$ = 220 km s$^{-1}$ for halo population assignment. The dashed magenta vertical line separates retrograde orbits on its left from prograde orbits on its right. \\ 
Right panel: the M dwarfs with thin disk kinematics have been removed. The green-marked ``subdwarfs'' and Cloud A subdwarfs are merged into a single group: Extended Cloud A. The M dwarfs outliers have been separated based on their total space velocity $v_\text{tot}$ w.r.t. the LSR. The Cloud B are color-coded in four categories based on their metallicity and $v_\text{tot}$. 
\label{fig:5-Toomre}}
\end{figure*}

This selection provides a kinematic sample of the local halo haunting the Solar Neighbourhood, which contains essentially stars from Cloud B, (220 objects), with a handful from Extended Cloud A, (9 objects) accompanied by targets considered as outliers from the dwarf cloud (71 objects)(see below). Its kinematic characteristics are in fair agreement with those of already identified halo population stars selected across widely more extended volumes and using more luminous targets \citep{2009MNRAS.399.1223S,2010ApJ...716....1B,2013AJ....145...40B}:
$<$$U$$>$ = 5 km s$^{-1}$ and $\sigma_{U}$ = 173 km s$^{-1}$, $<$$V$$>$ = $-$248 km s$^{-1}$ and $\sigma_{V}$ = 77 km s$^{-1}$, and $<$$W$$>$ = 6 km s$^{-1}$ and $\sigma_{W}$ = 89 km s$^{-1}$. With the adopted value of the LSR velocity, the fiducial line at $V$ = $-$220 km s$^{-1}$ separates the stars which have retrograde orbits from those who have prograde ones. A striking fact is the quite high fraction of prograde objects (258 stars or 40\%) observed in the present local halo sample, thus the strongly negative value of $<$$V$$>$ quoted above does not account completely for a complex orbital distribution.

The metallicity of this halo sample, in spite of the kinematical cut applied, remains bimodal. Previous studies, mostly based on F-G-K stars in which metallicity is derived from atomic lines have already found a substantial metal-rich component in the halo population, (and the metallicity distribution of globular clusters is one historic and well-documented example (see e.g. \citealt{1985ApJ...293..424Z,1996AJ....112.1487H}) but the metal-rich peak is generally found around [Fe/H] = $-$0.6 dex: \citet{2017ApJ...845..101B}. The prograde objects are also reputedly more metal-rich than the retrograde ones, implying that they could belong to different subpopulations, the metal-rich one containing inner halo objects while the metal-poor one is made of outer halo objects, these two subpopulations having different formation histories and different space distributions. In the present local halo sample, we do not find such a segregation, the number of prograde metal-poor objects remaining important while many higher metallicity stars are on obviously retrograde orbits, as shown in the right panel of Figure \ref{fig:5-Toomre}. 

\item Thin disk dwarfs:

Since some spectroscopically identified dwarfs are found to kinematically belong to the local halo, we could expect some continuity in the dwarf high velocity distribution tail, and therefore a number of them would also be kinematically mixed with the Extended Cloud A population and need to be disentangled.

To select thin disk dwarfs, a condition of type $v_\text{tot}$ = constant is not completely satisfactory because of the asymmetry of the dwarf cloud in the Toomre diagram. A correct way to proceed would be to compute for each star its probability of membership to one of the three populations - halo, thick disk and thin disk as proposed e.g. by \citet{2003A&A...410..527B} but we prefer to postpone such a calculation because the selection effects in the present sample are not correctly assessed. We choose a simplified and empirical procedure as follows: we assume that the distribution of the dwarf stars in the Toomre diagram is consistent with an incomplete ellipse, and that the isodensity contours of this ellipse trace constant probability level of a star to belong to its parent population. We select the isodensity contour containing 99\% of the dwarf cloud and fit an ellipse to this contour. All stars found outside are considered as ``outliers'' from the thin disk dwarf population. 

The kinematically ``purified'' dwarf sample shows typical velocity component values of the thin disk population \citep{1993A&A...275..101E,2003MNRAS.340..304R}:
$<$$U$$>$ = 0 km s$^{-1}$ and $\sigma_{U}$ = 44 km s$^{-1}$, $<$$V$$>$ = $-$13 km s$^{-1}$ and $\sigma_{V}$ = 30 km s$^{-1}$, and $<$$W$$>$ = 0 km s$^{-1}$ and $\sigma_{W}$ = 26 km s$^{-1}$, and its stars have exclusively prograde orbits as expected. The metallicity, as outlined above, is in average slightly deficient, the histogram showing three peaks, one solar or slightly super-solar, the principal one around $-$0.4 dex and the third one around $-$1 dex. Since the selection effects are not under control, we doubt that this three-component would be significant. 

\item Thick disk population and its elusive limits

The Extended Cloud A occupies a large area, somewhat fan-shaped vertically, in the Toomre diagram, broadly covering the dwarf cloud and merging beyond it until the halo region, with practically all orbits prograde. The shape of the $|$Z$|$ histogram, with its peak displaced from that of the thin disk dwarfs and its quartiles systematically larger shows that this population is clearly thicker in $z$. The average metal content is poorer than that of the thin disk dwarfs at [Fe/H] = $-$0.74 dex with a [Fe/H] distribution having a long metal-poor tail extending to $-$2.5 dex. We suspect that the Extended Cloud A population is an intricate mix of objects really belonging to a thick disk component with pollution by an important contribution of thin disk ``mild'' subdwarfs very difficult to disentangle, with unknown selection effects playing in background. A possible way to separate these contributions is the detailed computation of individual orbits and study of the $z_\text{max}$ and eccentricities, but such a work is far beyond the scope of the present study. 

In the Toomre diagram, the Extended Cloud A population is mixed with a fraction of the Cloud B population, corresponding to stars that preferentially belong to the high metallicity peak of the bimodal [Fe/H] distribution of Cloud B in Figure \ref{fig:5-histos}. These stars likely belong to the halo for a part, because our halo selection kinematic cut is very stringent, and to the thick disk for another part, but their respective assessment to the two components would probably also need a detailed orbital calculation. 

The [Fe/H] distribution of the outliers from the dwarf cloud show that they span the same broad metallicity range as the principal dwarf population, peaking at $-$0.6 dex, only slightly more metal-poor than the ordinary disk dwarfs. Their distribution in $|$Z$|$ is clearly more extended, making them probable members of the thick disk. Their velocity dispersions $\sigma_{U}$ = 119 km s$^{-1}$, $\sigma_{V}$ = 90 km s$^{-1}$, $\sigma_{W}$ = 75 km s$^{-1}$ suggest that they contain a majority of objects belonging to the thick disk with some pollution by an inner halo fraction.
 
\end{enumerate}

To conclude, the combined subdwarf sample extracted from LAMOST DR7 and SDSS DR7 contains a mix of objects belonging to various components: clear halo members with an implicit bimodal metallicity and considerable orbital mixing between retrograde and prograde cases, and an important thick disk fraction merging both into halo and thin disk. The spectroscopically selected M dwarfs themselves, whose quasi-totality belong to the standard thin disk, contain a small fraction (less than 1\% of the total) of outliers belonging both to the thick disk and the halo. It is quite unclear if the present situation is dependent or not on the selection effects, especially the favoured directions of investigation towards the Galactic anticenter in the LAMOST survey.
 
\section{Summary and Conclusion} \label{sec:con}

In the most widely used classification system, M-type main sequence stars are classified into dM/sdM/esdM/usdM subclasses corresponding to different metal abundances according to the CaH2, CaH3 and TiO5 molecular bands in the low-resolution spectrum. The metal-poor subclasses are named as subdwarfs, which have typical kinematic characteristics of thick disk and halo population. However, in the present work, we suggest the ``genuine'' subdwarfs belonging to population \Rmnum{2} be classified as luminosity class \Rmnum{6} on the H-R diagram. For these stellar objects, having larger surface gravity is a necessary condition as important as ``being metal deficient'', and the three spectral indices in classical classification system can not be fully competent for selecting these subdwarfs. This conclusion is drawn by a series of experiments including the comparison of synthetic spectra with the templates, measuring atmospheric parameters for a subdwarf sample and combine with kinematics analysis, etc). According to the spectral index CaOH-CaH1 relationship and the empirical relationship on RPM diagram, we divide a spectroscopically identified subdwarfs into multiple groups and analyze the multi-population properties in this sample from their multi-dimensional parameters.

The sample including subdwarfs from both LAMOST and SDSS combined with a control group of dwarfs selected from LAMOST, and estimate their atmospheric parameters fitting with the PHOENIX grid. The technical details of the whole fitting procedure of individual spectra is developed, including the use of different convolution kernels to smooth the synthetic spectra and the pseudocontinuum flux correction specifically implemented for the LAMOST data. The residuals of the fitting procedure can be used to determine the precision which depend on the S/N of the observed spectra. The accuracy of the process is evaluated from 167 M-type stars with measured atmospheric parameters from literature which have LAMOST spectra: and the bias and scatter of the comparison of parameters are 165$\pm$170 K for $T_{\text{eff}}$, 0.03$\pm$0.32 dex for gravity, and -0.17$\pm$0.41 dex for metallicity. A systematic bias in effective temperature exists for $T_{\text{eff}}$ lower than 3700 K while the gravity and metallicity both appear in fair agreement. The effective temperature scale produced by these detailed fits is reasonably consistent with those published by \citet{2014A&A...564A..90R} for the subdwarfs and \cite{2018A&A...620A.180R} for the dwarfs. The following analysis are mainly dependent on the latter two parameters.

For the sample of subdwarfs and dwarfs, apparent magnitudes, broad-band colors, and proper-motions are obtained from cross-matching these two sets with Gaia DR2, and the distances are from \citet{2018AJ....156...58B}. The objects are selected so as to have acceptable quality flags in photometry, distance and accuracy. The spectroscopic ``subdwarfs'' do not entirely lie below the main-sequence but in an extended region which overlaps that occupied by the M dwarfs in the color-absolute magnitude diagram, while the objects that lie away (and below) from the main sequence are the ones with both larger gravity and lower metal abundance. They are the ``genuine'' subdwarfs, for which the luminosity class \Rmnum{6} (\citep{2008AJ....136..840J}) is justified. Alternatively, some ``subdwarfs'' with gravity similar as that of the dwarfs or only slightly higher also behave like dwarfs in many aspects although they are medium metal-deficient through the definition using the CaH and TiO spectral indices, we have been able to sort out a significant subsample of objects which, in spite of average low metallicity, appear intricately mixed with the ordinary main-sequence dwarfs and also sharing their location in a reduced proper motion diagram. These ``subdwarfs'' should perhaps better be classified as ``metal-poor dwarfs''. 

A provisional kinematic analysis of these samples show that the sdM population lying below the main-sequence cloud is dominated by halo stars, in which an implicit bimodal metallicity distribution appears, with a peak around $-$1.0 dex and a second peak around $-$1.5 dex. This group, which samples the galactic halo only in a very local volume (typically less than 1 kpc radius), contains a mix of objects of prograde and retrograde orbits, as revealed by their location in the Toomre diagram, without clear trend with the metallicity. The other subdwarfs, including those which we prefer to consider as ``metal-poor dwarfs'' appear as a mixed population of thick disk and thin disk members. If ``lying below the main sequence'' and ``associated with halo kinematics'' (large negative $V$ component and large dispersions in $U$, $V$ and $W$) define intrinsic features of the genuine subdwarfs belonging to Population \Rmnum{2} stars, then both large gravity (average log $g$ = 5.2 to 5.4) $and$ low metal abundance should be essential attributes of them. The fact that the two properties must exist simultaneously needs to be underlined: we have only found a handful of halo objects among the other subdwarfs, even those which have a metallicity lower than $-$1.0 dex. 

The ordinary dwarfs populate in their vast majority the Galactic thin disk, as expected, with an average metallicity slightly lower than solar. A small fraction (less than 1\%) form a set of kinematic outliers, halo and thick disk probable members. Finally, CaH1 index is a good indicator of gravity experimentally, which behaves better in separating subdwarfs from dwarfs than the composite index CaH2+CaH3. 

Though the contamination by low ratio of multiplicity (12.5$\pm$1.9\%: \citealt{2015ApJ...804...30Z}) and the parameter coupling effect would influence the accuracy of parameters measurement, a preliminary indicative conclusion can still be drawn from the analysis results above. Besides, accurate discrimination of subdwarfs has to be revisited with more information such as high-resolution spectra and multi-band data, and the estimates of parameters also need a more complete model grid (an independent alpha enrichment such as the analysis in \citealt{2020AJ....159...30H}, more complete opacity and line lists, etc). The sub-population membership assessment of the subdwarfs found in numbers in the large area spectroscopic surveys, in turn needs careful selection effects control and detailed orbital calculation.

\acknowledgments
ZS thanks Dr. Yan. H.L. for the thoughtful discussion about smoothing the synthetic spectra, and Dr. Allard, F. for her helpful private communication about PHOENIX models. This work is supported by National Key R\&D Program of China(No. 2019YFA0405502), National Science Foundation of China (No. U1931209). Guoshoujing Telescope (the Large Sky Area Multi-Object Fiber Spectroscopic Telescope, LAMOST) is a National Major Scientific Project built by the Chinese Academy of Sciences. Funding for the project has been provided by the National Development and Reform Commission. LAMOST is operated and managed by the National Astronomical Observatories, Chinese Academy of Sciences.

\software{astropy \citep{2013A&A...558A..33A}, 
		matplotlib \citep{2005ASPC..347...91B},
		pandas\citep{mckinney2010data},
		numpy \citep{4160250},
		scipy \citep{2019arXiv190710121V}
        }
        
 \bibliography{manuscript_vr}{}

\begin{thebibliography}{}
\expandafter\ifx\csname natexlab\endcsname\relax\def\natexlab#1{#1}\fi

\bibitem[{{Ake} \& {Greenstein}(1980)}]{1980ApJ...240..859A}
{Ake}, T.~B., \& {Greenstein}, J.~L. 1980, \apj, 240, 859

\bibitem[{{Allard}(2016)}]{2016sf2a.conf..223A}
{Allard}, F. 2016, in SF2A-2016: Proceedings of the Annual meeting of the
  French Society of Astronomy and Astrophysics, ed. C.~{Reyl{\'e}},
  J.~{Richard}, L.~{Cambr{\'e}sy}, M.~{Deleuil}, E.~{P{\'e}contal},
  L.~{Tresse}, \& I.~{Vauglin}, 223--227

\bibitem[{{Allard} \& {Hauschildt}(1995)}]{1995ApJ...445..433A}
{Allard}, F., \& {Hauschildt}, P.~H. 1995, ApJ, 445, 433

\bibitem[{{Allard} {et~al.}(1997){Allard}, {Hauschildt}, {Alexander}, \&
  {Starrfield}}]{1997ARA&A..35..137A}
{Allard}, F., {Hauschildt}, P.~H., {Alexander}, D.~R., \& {Starrfield}, S.
  1997, Annual Review of Astronomy and Astrophysics, 35, 137

\bibitem[{{Allard} {et~al.}(2001){Allard}, {Hauschildt}, {Alexander},
  {Tamanai}, \& {Schweitzer}}]{2001ApJ...556..357A}
{Allard}, F., {Hauschildt}, P.~H., {Alexander}, D.~R., {Tamanai}, A., \&
  {Schweitzer}, A. 2001, \apj, 556, 357

\bibitem[{{Allard} {et~al.}(2011){Allard}, {Homeier}, \&
  {Freytag}}]{2011ASPC..448...91A}
{Allard}, F., {Homeier}, D., \& {Freytag}, B. 2011, Astronomical Society of the
  Pacific Conference Series, Vol. 448, {Model Atmospheres From Very Low Mass
  Stars to Brown Dwarfs}, ed. C.~{Johns-Krull}, M.~K. {Browning}, \& A.~A.
  {West}, 91

\bibitem[{{Allard} {et~al.}(2012){Allard}, {Homeier}, \&
  {Freytag}}]{2012RSPTA.370.2765A}
---. 2012, Philosophical Transactions of the Royal Society of London Series A,
  370, 2765

\bibitem[{{Allard} {et~al.}(2013){Allard}, {Homeier}, \&
  {Freytag}}]{2013MmSAI..84.1053A}
---. 2013, \memsai, 84, 1053

\bibitem[{{Asplund} {et~al.}(2009){Asplund}, {Grevesse}, {Sauval}, \&
  {Scott}}]{2009ARA&A..47..481A}
{Asplund}, M., {Grevesse}, N., {Sauval}, A.~J., \& {Scott}, P. 2009, ARA\&A,
  47, 481

\bibitem[{{Astropy Collaboration} {et~al.}(2013){Astropy Collaboration},
  {Robitaille}, {Tollerud}, {Greenfield}, {Droettboom}, {Bray}, {Aldcroft},
  {Davis}, {Ginsburg}, {Price-Whelan}, {Kerzendorf}, {Conley}, {Crighton},
  {Barbary}, {Muna}, {Ferguson}, {Grollier}, {Parikh}, {Nair}, {Unther},
  {Deil}, {Woillez}, {Conseil}, {Kramer}, {Turner}, {Singer}, {Fox}, {Weaver},
  {Zabalza}, {Edwards}, {Azalee Bostroem}, {Burke}, {Casey}, {Crawford},
  {Dencheva}, {Ely}, {Jenness}, {Labrie}, {Lim}, {Pierfederici}, {Pontzen},
  {Ptak}, {Refsdal}, {Servillat}, \& {Streicher}}]{2013A&A...558A..33A}
{Astropy Collaboration}, {Robitaille}, T.~P., {Tollerud}, E.~J., {et~al.} 2013,
  \aap, 558, A33

\bibitem[{{Bai} {et~al.}(2016){Bai}, {Luo}, {Comte}, {Zhao}, {Yang}, {Guo},
  {Wang}, {Li}, {Du}, {Hou}, {Kong}, {Yi}, {Song}, {Bai}, {Zhang}, {Wang},
  {Chen}, {Chen}, {Wu}, {Zuo}, {Wu}, {Cao}, {Hou}, {Wang}, \&
  {Zhang}}]{2016RAA....16..107B}
{Bai}, Y., {Luo}, A.~L., {Comte}, G., {et~al.} 2016, Research in Astronomy and
  Astrophysics, 16, 107

\bibitem[{{Bailer-Jones} {et~al.}(2018){Bailer-Jones}, {Rybizki}, {Fouesneau},
  {Mantelet}, \& {Andrae}}]{2018AJ....156...58B}
{Bailer-Jones}, C.~A.~L., {Rybizki}, J., {Fouesneau}, M., {Mantelet}, G., \&
  {Andrae}, R. 2018, \aj, 156, 58

\bibitem[{{Baraffe} {et~al.}(2015){Baraffe}, {Homeier}, {Allard}, \&
  {Chabrier}}]{2015A&A...577A..42B}
{Baraffe}, I., {Homeier}, D., {Allard}, F., \& {Chabrier}, G. 2015, A\&A, 577,
  A42

\bibitem[{{Barrett} {et~al.}(2005){Barrett}, {Hunter}, {Miller}, {Hsu}, \&
  {Greenfield}}]{2005ASPC..347...91B}
{Barrett}, P., {Hunter}, J., {Miller}, J.~T., {Hsu}, J.~C., \& {Greenfield}, P.
  2005, Astronomical Society of the Pacific Conference Series, Vol. 347,
  {matplotlib -- A Portable Python Plotting Package}, ed. P.~{Shopbell},
  M.~{Britton}, \& R.~{Ebert}, 91

\bibitem[{{Bensby} {et~al.}(2003){Bensby}, {Feltzing}, \&
  {Lundstr{\"o}m}}]{2003A&A...410..527B}
{Bensby}, T., {Feltzing}, S., \& {Lundstr{\"o}m}, I. 2003, \aap, 410, 527

\bibitem[{{Bochanski} {et~al.}(2013){Bochanski}, {Savcheva}, {West}, \&
  {Hawley}}]{2013AJ....145...40B}
{Bochanski}, J.~J., {Savcheva}, A., {West}, A.~A., \& {Hawley}, S.~L. 2013,
  \aj, 145, 40

\bibitem[{{Bonaca} {et~al.}(2017){Bonaca}, {Conroy}, {Wetzel}, {Hopkins}, \&
  {Kere{\v{s}}}}]{2017ApJ...845..101B}
{Bonaca}, A., {Conroy}, C., {Wetzel}, A., {Hopkins}, P.~F., \& {Kere{\v{s}}},
  D. 2017, \apj, 845, 101

\bibitem[{{Bond} {et~al.}(2010){Bond}, {Ivezi{\'c}}, {Sesar}, {Juri{\'c}},
  {Munn}, {Kowalski}, {Loebman}, {Ro{\v{s}}kar}, {Beers}, {Dalcanton},
  {Rockosi}, {Yanny}, {Newberg}, {Allende Prieto}, {Wilhelm}, {Lee},
  {Sivarani}, {Majewski}, {Norris}, {Bailer-Jones}, {Re Fiorentin}, {Schlegel},
  {Uomoto}, {Lupton}, {Knapp}, {Gunn}, {Covey}, {Allyn Smith}, {Miknaitis},
  {Doi}, {Tanaka}, {Fukugita}, {Kent}, {Finkbeiner}, {Quinn}, {Hawley},
  {Anderson}, {Kiuchi}, {Chen}, {Bushong}, {Sohi}, {Haggard}, {Kimball},
  {McGurk}, {Barentine}, {Brewington}, {Harvanek}, {Kleinman}, {Krzesinski},
  {Long}, {Nitta}, {Snedden}, {Lee}, {Pier}, {Harris}, {Brinkmann}, \&
  {Schneider}}]{2010ApJ...716....1B}
{Bond}, N.~A., {Ivezi{\'c}}, {\v{Z}}., {Sesar}, B., {et~al.} 2010, \apj, 716, 1

\bibitem[{{Burgasser} \& {Kirkpatrick}(2006)}]{2006ApJ...645.1485B}
{Burgasser}, A.~J., \& {Kirkpatrick}, J.~D. 2006, \apj, 645, 1485

\bibitem[{{Caffau} {et~al.}(2011){Caffau}, {Ludwig}, {Steffen}, {Freytag}, \&
  {Bonifacio}}]{2011SoPh..268..255C}
{Caffau}, E., {Ludwig}, H.~G., {Steffen}, M., {Freytag}, B., \& {Bonifacio}, P.
  2011, Solar Physics, 268, 255

\bibitem[{{Carollo} {et~al.}(2010){Carollo}, {Beers}, {Chiba}, {Norris},
  {Freeman}, {Lee}, {Ivezi{\'c}}, {Rockosi}, \& {Yanny}}]{2010ApJ...712..692C}
{Carollo}, D., {Beers}, T.~C., {Chiba}, M., {et~al.} 2010, \apj, 712, 692

\bibitem[{{Castelli} \& {Kurucz}(2004)}]{2004A&A...419..725C}
{Castelli}, F., \& {Kurucz}, R.~L. 2004, A\&A, 419, 725

\bibitem[{{Cui} {et~al.}(2012){Cui}, {Zhao}, {Chu}, {Li}, {Li}, {Zhang}, {Su},
  {Yao}, {Wang}, {Xing}, {Li}, {Zhu}, {Wang}, {Gu}, {Luo}, {Xu}, {Zhang},
  {Liu}, {Zhang}, {Yang}, {Cao}, {Chen}, {Chen}, {Chen}, {Chen}, {Chu}, {Feng},
  {Gong}, {Hou}, {Hu}, {Hu}, {Hu}, {Jia}, {Jiang}, {Jiang}, {Jiang}, {Jin},
  {Li}, {Li}, {Li}, {Liu}, {Liu}, {Lu}, {Mao}, {Men}, {Qi}, {Qi}, {Shi},
  {Tang}, {Tao}, {Wang}, {Wang}, {Wang}, {Wang}, {Wang}, {Wang}, {Wang},
  {Wang}, {Wang}, {Wang}, {Wang}, {Wang}, {Xu}, {Xu}, {Yang}, {Yu}, {Yuan},
  {Yuan}, {Zhai}, {Zhang}, {Zhang}, {Zhang}, {Zhao}, {Zhou}, {Zhou}, {Zhu}, \&
  {Zou}}]{2012RAA....12.1197C}
{Cui}, X.-Q., {Zhao}, Y.-H., {Chu}, Y.-Q., {et~al.} 2012, Research in Astronomy
  and Astrophysics, 12, 1197

\bibitem[{{Dahn} {et~al.}(1995){Dahn}, {Liebert}, {Harris}, \&
  {Guetter}}]{1995bmsb.conf..239D}
{Dahn}, C.~C., {Liebert}, J., {Harris}, H.~C., \& {Guetter}, H.~H. 1995, in The
  Bottom of the Main Sequence - and Beyond, ed. C.~G. {Tinney}, 239

\bibitem[{{Dhital} {et~al.}(2012){Dhital}, {West}, {Stassun}, {Bochanski},
  {Massey}, \& {Bastien}}]{2012AJ....143...67D}
{Dhital}, S., {West}, A.~A., {Stassun}, K.~G., {et~al.} 2012, AJ, 143, 67

\bibitem[{{Du} {et~al.}(2016){Du}, {Luo}, {Kong}, {Zhang}, {Guo}, {Cook},
  {Hou}, {Yang}, {Li}, {Song}, {Chen}, {Zuo}, {Wu}, {Wang}, {Wu}, {Wang}, \&
  {Zhao}}]{2016ApJS..227...27D}
{Du}, B., {Luo}, A.~L., {Kong}, X., {et~al.} 2016, \apjs, 227, 27

\bibitem[{{Edvardsson} {et~al.}(1993){Edvardsson}, {Andersen}, {Gustafsson},
  {Lambert}, {Nissen}, \& {Tomkin}}]{1993A&A...275..101E}
{Edvardsson}, B., {Andersen}, J., {Gustafsson}, B., {et~al.} 1993, \aap, 500,
  391

\bibitem[{{Faherty} {et~al.}(2009){Faherty}, {Burgasser}, {Cruz}, {Shara},
  {Walter}, \& {Gelino}}]{2009AJ....137....1F}
{Faherty}, J.~K., {Burgasser}, A.~J., {Cruz}, K.~L., {et~al.} 2009, \aj, 137, 1

\bibitem[{{Fuchs} {et~al.}(2009){Fuchs}, {Dettbarn}, {Rix}, {Beers}, {Bizyaev},
  {Brewington}, {Jahrei{\ss}}, {Klement}, {Malanushenko}, {Malanushenko},
  {Oravetz}, {Pan}, {Simmons}, \& {Snedden}}]{2009AJ....137.4149F}
{Fuchs}, B., {Dettbarn}, C., {Rix}, H.-W., {et~al.} 2009, \aj, 137, 4149

\bibitem[{{Gaia Collaboration} {et~al.}(2018{\natexlab{a}}){Gaia
  Collaboration}, {Babusiaux}, {van Leeuwen}, {Barstow}, {Jordi}, {Vallenari},
  {Bossini}, {Bressan}, {Cantat-Gaudin}, {van Leeuwen}, {Brown}, {Prusti}, {de
  Bruijne}, {Bailer-Jones}, {Biermann}, {Evans}, {Eyer}, {Jansen}, {Klioner},
  {Lammers}, {Lindegren}, {Luri}, {Mignard}, {Panem}, {Pourbaix}, {Randich},
  {Sartoretti}, {Siddiqui}, {Soubiran}, {Walton}, {Arenou}, {Bastian},
  {Cropper}, {Drimmel}, {Katz}, {Lattanzi}, {Bakker}, {Cacciari},
  {Casta{\~n}eda}, {Chaoul}, {Cheek}, {De Angeli}, {Fabricius}, {Guerra},
  {Holl}, {Masana}, {Messineo}, {Mowlavi}, {Nienartowicz}, {Panuzzo},
  {Portell}, {Riello}, {Seabroke}, {Tanga}, {Th{\'e}venin}, {Gracia-Abril},
  {Comoretto}, {Garcia-Reinaldos}, {Teyssier}, {Altmann}, {Andrae}, {Audard},
  {Bellas-Velidis}, {Benson}, {Berthier}, {Blomme}, {Burgess}, {Busso},
  {Carry}, {Cellino}, {Clementini}, {Clotet}, {Creevey}, {Davidson}, {De
  Ridder}, {Delchambre}, {Dell'Oro}, {Ducourant},
  {Fern{\'a}ndez-Hern{\'a}ndez}, {Fouesneau}, {Fr{\'e}mat}, {Galluccio},
  {Garc{\'\i}a-Torres}, {Gonz{\'a}lez-N{\'u}{\~n}ez}, {Gonz{\'a}lez-Vidal},
  {Gosset}, {Guy}, {Halbwachs}, {Hambly}, {Harrison}, {Hern{\'a}ndez},
  {Hestroffer}, {Hodgkin}, {Hutton}, {Jasniewicz}, {Jean-Antoine-Piccolo},
  {Jordan}, {Korn}, {Krone-Martins}, {Lanzafame}, {Lebzelter}, {L{\"o}ffler},
  {Manteiga}, {Marrese}, {Mart{\'\i}n-Fleitas}, {Moitinho}, {Mora}, {Muinonen},
  {Osinde}, {Pancino}, {Pauwels}, {Petit}, {Recio-Blanco}, {Richards},
  {Rimoldini}, {Robin}, {Sarro}, {Siopis}, {Smith}, {Sozzetti}, {S{\"u}veges},
  {Torra}, {van Reeven}, {Abbas}, {Abreu Aramburu}, {Accart}, {Aerts},
  {Altavilla}, {{\'A}lvarez}, {Alvarez}, {Alves}, {Anderson}, {Andrei},
  {Anglada Varela}, {Antiche}, {Antoja}, {Arcay}, {Astraatmadja}, {Bach},
  {Baker}, {Balaguer-N{\'u}{\~n}ez}, {Balm}, {Barache}, {Barata}, {Barbato},
  {Barblan}, {Barklem}, {Barrado}, {Barros}, {Bartholom{\'e} Mu{\~n}oz},
  {Bassilana}, {Becciani}, {Bellazzini}, {Berihuete}, {Bertone}, {Bianchi},
  {Bienaym{\'e}}, {Blanco-Cuaresma}, {Boch}, {Boeche}, {Bombrun}, {Borrachero},
  {Bouquillon}, {Bourda}, {Bragaglia}, {Bramante}, {Breddels}, {Brouillet},
  {Br{\"u}semeister}, {Brugaletta}, {Bucciarelli}, {Burlacu}, {Busonero},
  {Butkevich}, {Buzzi}, {Caffau}, {Cancelliere}, {Cannizzaro}, {Carballo},
  {Carlucci}, {Carrasco}, {Casamiquela}, {Castellani}, {Castro-Ginard},
  {Charlot}, {Chemin}, {Chiavassa}, {Cocozza}, {Costigan}, {Cowell}, {Crifo},
  {Crosta}, {Crowley}, {Cuypers}, {Dafonte}, {Damerdji}, {Dapergolas}, {David},
  {David}, {de Laverny}, {De Luise}, {De March}, {de Martino}, {de Souza}, {de
  Torres}, {Debosscher}, {del Pozo}, {Delbo}, {Delgado}, {Delgado}, {Diakite},
  {Diener}, {Distefano}, {Dolding}, {Drazinos}, {Dur{\'a}n}, {Edvardsson},
  {Enke}, {Eriksson}, {Esquej}, {Eynard Bontemps}, {Fabre}, {Fabrizio},
  {Faigler}, {Falc{\~a}o}, {Farr{\`a}s Casas}, {Federici}, {Fedorets},
  {Fernique}, {Figueras}, {Filippi}, {Findeisen}, {Fonti}, {Fraile}, {Fraser},
  {Fr{\'e}zouls}, {Gai}, {Galleti}, {Garabato}, {Garc{\'\i}a-Sedano},
  {Garofalo}, {Garralda}, {Gavel}, {Gavras}, {Gerssen}, {Geyer}, {Giacobbe},
  {Gilmore}, {Girona}, {Giuffrida}, {Glass}, {Gomes}, {Granvik}, {Gueguen},
  {Guerrier}, {Guiraud}, {Guti{\'e}}, {Haigron}, {Hatzidimitriou}, {Hauser},
  {Haywood}, {Heiter}, {Helmi}, {Heu}, {Hilger}, {Hobbs}, {Hofmann}, {Holland},
  {Huckle}, {Hypki}, {Icardi}, {Jan{\ss}en}, {Jevardat de Fombelle}, {Jonker},
  {Juh{\'a}sz}, {Julbe}, {Karampelas}, {Kewley}, {Klar}, {Kochoska}, {Kohley},
  {Kolenberg}, {Kontizas}, {Kontizas}, {Koposov}, {Kordopatis},
  {Kostrzewa-Rutkowska}, {Koubsky}, {Lambert}, {Lanza}, {Lasne}, {Lavigne}, {Le
  Fustec}, {Le Poncin-Lafitte}, {Lebreton}, {Leccia}, {Leclerc},
  {Lecoeur-Taibi}, {Lenhardt}, {Leroux}, {Liao}, {Licata}, {Lindstr{\o}m},
  {Lister}, {Livanou}, {Lobel}, {L{\'o}pez}, {Managau}, {Mann}, {Mantelet},
  {Marchal}, {Marchant}, {Marconi}, {Marinoni}, {Marschalk{\'o}}, {Marshall},
  {Martino}, {Marton}, {Mary}, {Massari}, {Matijevi{\v{c}}}, {Mazeh},
  {McMillan}, {Messina}, {Michalik}, {Millar}, {Molina}, {Molinaro},
  {Moln{\'a}r}, {Montegriffo}, {Mor}, {Morbidelli}, {Morel}, {Morris},
  {Mulone}, {Muraveva}, {Musella}, {Nelemans}, {Nicastro}, {Noval},
  {O'Mullane}, {Ord{\'e}novic}, {Ord{\'o}{\~n}ez-Blanco}, {Osborne}, {Pagani},
  {Pagano}, {Pailler}, {Palacin}, {Palaversa}, {Panahi}, {Pawlak},
  {Piersimoni}, {Pineau}, {Plachy}, {Plum}, {Poggio}, {Poujoulet},
  {Pr{\v{s}}a}, {Pulone}, {Racero}, {Ragaini}, {Rambaux}, {Ramos-Lerate},
  {Regibo}, {Reyl{\'e}}, {Riclet}, {Ripepi}, {Riva}, {Rivard}, {Rixon},
  {Roegiers}, {Roelens}, {Romero-G{\'o}mez}, {Rowell}, {Royer}, {Ruiz-Dern},
  {Sadowski}, {Sagrist{\`a} Sell{\'e}s}, {Sahlmann}, {Salgado}, {Salguero},
  {Sanna}, {Santana-Ros}, {Sarasso}, {Savietto}, {Schultheis}, {Sciacca},
  {Segol}, {Segovia}, {S{\'e}gransan}, {Shih}, {Siltala}, {Silva}, {Smart},
  {Smith}, {Solano}, {Solitro}, {Sordo}, {Soria Nieto}, {Souchay}, {Spagna},
  {Spoto}, {Stampa}, {Steele}, {Steidelm{\"u}ller}, {Stephenson}, {Stoev},
  {Suess}, {Surdej}, {Szabados}, {Szegedi-Elek}, {Tapiador}, {Taris}, {Tauran},
  {Taylor}, {Teixeira}, {Terrett}, {Teyssand ier}, {Thuillot}, {Titarenko},
  {Torra Clotet}, {Turon}, {Ulla}, {Utrilla}, {Uzzi}, {Vaillant}, {Valentini},
  {Valette}, {van Elteren}, {Van Hemelryck}, {Vaschetto}, {Vecchiato},
  {Veljanoski}, {Viala}, {Vicente}, {Vogt}, {von Essen}, {Voss}, {Votruba},
  {Voutsinas}, {Walmsley}, {Weiler}, {Wertz}, {Wevers}, {Wyrzykowski},
  {Yoldas}, {{\v{Z}}erjal}, {Ziaeepour}, {Zorec}, {Zschocke}, {Zucker},
  {Zurbach}, \& {Zwitter}}]{2018A&A...616A..10G}
{Gaia Collaboration}, {Babusiaux}, C., {van Leeuwen}, F., {et~al.}
  2018{\natexlab{a}}, \aap, 616, A10

\bibitem[{{Gaia Collaboration} {et~al.}(2018{\natexlab{b}}){Gaia
  Collaboration}, {Brown}, {Vallenari}, {Prusti}, {de Bruijne}, {Babusiaux},
  {Bailer-Jones}, {Biermann}, {Evans}, {Eyer}, {Jansen}, {Jordi}, {Klioner},
  {Lammers}, {Lindegren}, {Luri}, {Mignard}, {Panem}, {Pourbaix}, {Randich},
  {Sartoretti}, {Siddiqui}, {Soubiran}, {van Leeuwen}, {Walton}, {Arenou},
  {Bastian}, {Cropper}, {Drimmel}, {Katz}, {Lattanzi}, {Bakker}, {Cacciari},
  {Casta{\~n}eda}, {Chaoul}, {Cheek}, {De Angeli}, {Fabricius}, {Guerra},
  {Holl}, {Masana}, {Messineo}, {Mowlavi}, {Nienartowicz}, {Panuzzo},
  {Portell}, {Riello}, {Seabroke}, {Tanga}, {Th{\'e}venin}, {Gracia-Abril},
  {Comoretto}, {Garcia-Reinaldos}, {Teyssier}, {Altmann}, {Andrae}, {Audard},
  {Bellas-Velidis}, {Benson}, {Berthier}, {Blomme}, {Burgess}, {Busso},
  {Carry}, {Cellino}, {Clementini}, {Clotet}, {Creevey}, {Davidson}, {De
  Ridder}, {Delchambre}, {Dell'Oro}, {Ducourant},
  {Fern{\'a}ndez-Hern{\'a}ndez}, {Fouesneau}, {Fr{\'e}mat}, {Galluccio},
  {Garc{\'\i}a-Torres}, {Gonz{\'a}lez-N{\'u}{\~n}ez}, {Gonz{\'a}lez-Vidal},
  {Gosset}, {Guy}, {Halbwachs}, {Hambly}, {Harrison}, {Hern{\'a}ndez},
  {Hestroffer}, {Hodgkin}, {Hutton}, {Jasniewicz}, {Jean-Antoine-Piccolo},
  {Jordan}, {Korn}, {Krone-Martins}, {Lanzafame}, {Lebzelter}, {L{\"o}ffler},
  {Manteiga}, {Marrese}, {Mart{\'\i}n-Fleitas}, {Moitinho}, {Mora}, {Muinonen},
  {Osinde}, {Pancino}, {Pauwels}, {Petit}, {Recio-Blanco}, {Richards},
  {Rimoldini}, {Robin}, {Sarro}, {Siopis}, {Smith}, {Sozzetti}, {S{\"u}veges},
  {Torra}, {van Reeven}, {Abbas}, {Abreu Aramburu}, {Accart}, {Aerts},
  {Altavilla}, {{\'A}lvarez}, {Alvarez}, {Alves}, {Anderson}, {Andrei},
  {Anglada Varela}, {Antiche}, {Antoja}, {Arcay}, {Astraatmadja}, {Bach},
  {Baker}, {Balaguer-N{\'u}{\~n}ez}, {Balm}, {Barache}, {Barata}, {Barbato},
  {Barblan}, {Barklem}, {Barrado}, {Barros}, {Barstow}, {Bartholom{\'e}
  Mu{\~n}oz}, {Bassilana}, {Becciani}, {Bellazzini}, {Berihuete}, {Bertone},
  {Bianchi}, {Bienaym{\'e}}, {Blanco-Cuaresma}, {Boch}, {Boeche}, {Bombrun},
  {Borrachero}, {Bossini}, {Bouquillon}, {Bourda}, {Bragaglia}, {Bramante},
  {Breddels}, {Bressan}, {Brouillet}, {Br{\"u}semeister}, {Brugaletta},
  {Bucciarelli}, {Burlacu}, {Busonero}, {Butkevich}, {Buzzi}, {Caffau},
  {Cancelliere}, {Cannizzaro}, {Cantat-Gaudin}, {Carballo}, {Carlucci},
  {Carrasco}, {Casamiquela}, {Castellani}, {Castro-Ginard}, {Charlot},
  {Chemin}, {Chiavassa}, {Cocozza}, {Costigan}, {Cowell}, {Crifo}, {Crosta},
  {Crowley}, {Cuypers}, {Dafonte}, {Damerdji}, {Dapergolas}, {David}, {David},
  {de Laverny}, {De Luise}, {De March}, {de Martino}, {de Souza}, {de Torres},
  {Debosscher}, {del Pozo}, {Delbo}, {Delgado}, {Delgado}, {Di Matteo},
  {Diakite}, {Diener}, {Distefano}, {Dolding}, {Drazinos}, {Dur{\'a}n},
  {Edvardsson}, {Enke}, {Eriksson}, {Esquej}, {Eynard Bontemps}, {Fabre},
  {Fabrizio}, {Faigler}, {Falc{\~a}o}, {Farr{\`a}s Casas}, {Federici},
  {Fedorets}, {Fernique}, {Figueras}, {Filippi}, {Findeisen}, {Fonti},
  {Fraile}, {Fraser}, {Fr{\'e}zouls}, {Gai}, {Galleti}, {Garabato},
  {Garc{\'\i}a-Sedano}, {Garofalo}, {Garralda}, {Gavel}, {Gavras}, {Gerssen},
  {Geyer}, {Giacobbe}, {Gilmore}, {Girona}, {Giuffrida}, {Glass}, {Gomes},
  {Granvik}, {Gueguen}, {Guerrier}, {Guiraud}, {Guti{\'e}rrez-S{\'a}nchez},
  {Haigron}, {Hatzidimitriou}, {Hauser}, {Haywood}, {Heiter}, {Helmi}, {Heu},
  {Hilger}, {Hobbs}, {Hofmann}, {Holland}, {Huckle}, {Hypki}, {Icardi},
  {Jan{\ss}en}, {Jevardat de Fombelle}, {Jonker}, {Juh{\'a}sz}, {Julbe},
  {Karampelas}, {Kewley}, {Klar}, {Kochoska}, {Kohley}, {Kolenberg},
  {Kontizas}, {Kontizas}, {Koposov}, {Kordopatis}, {Kostrzewa-Rutkowska},
  {Koubsky}, {Lambert}, {Lanza}, {Lasne}, {Lavigne}, {Le Fustec}, {Le
  Poncin-Lafitte}, {Lebreton}, {Leccia}, {Leclerc}, {Lecoeur-Taibi},
  {Lenhardt}, {Leroux}, {Liao}, {Licata}, {Lindstr{\o}m}, {Lister}, {Livanou},
  {Lobel}, {L{\'o}pez}, {Managau}, {Mann}, {Mantelet}, {Marchal}, {Marchant},
  {Marconi}, {Marinoni}, {Marschalk{\'o}}, {Marshall}, {Martino}, {Marton},
  {Mary}, {Massari}, {Matijevi{\v{c}}}, {Mazeh}, {McMillan}, {Messina},
  {Michalik}, {Millar}, {Molina}, {Molinaro}, {Moln{\'a}r}, {Montegriffo},
  {Mor}, {Morbidelli}, {Morel}, {Morris}, {Mulone}, {Muraveva}, {Musella},
  {Nelemans}, {Nicastro}, {Noval}, {O'Mullane}, {Ord{\'e}novic},
  {Ord{\'o}{\~n}ez-Blanco}, {Osborne}, {Pagani}, {Pagano}, {Pailler},
  {Palacin}, {Palaversa}, {Panahi}, {Pawlak}, {Piersimoni}, {Pineau}, {Plachy},
  {Plum}, {Poggio}, {Poujoulet}, {Pr{\v{s}}a}, {Pulone}, {Racero}, {Ragaini},
  {Rambaux}, {Ramos-Lerate}, {Regibo}, {Reyl{\'e}}, {Riclet}, {Ripepi}, {Riva},
  {Rivard}, {Rixon}, {Roegiers}, {Roelens}, {Romero-G{\'o}mez}, {Rowell},
  {Royer}, {Ruiz-Dern}, {Sadowski}, {Sagrist{\`a} Sell{\'e}s}, {Sahlmann},
  {Salgado}, {Salguero}, {Sanna}, {Santana-Ros}, {Sarasso}, {Savietto},
  {Schultheis}, {Sciacca}, {Segol}, {Segovia}, {S{\'e}gransan}, {Shih},
  {Siltala}, {Silva}, {Smart}, {Smith}, {Solano}, {Solitro}, {Sordo}, {Soria
  Nieto}, {Souchay}, {Spagna}, {Spoto}, {Stampa}, {Steele},
  {Steidelm{\"u}ller}, {Stephenson}, {Stoev}, {Suess}, {Surdej}, {Szabados},
  {Szegedi-Elek}, {Tapiador}, {Taris}, {Tauran}, {Taylor}, {Teixeira},
  {Terrett}, {Teyssand ier}, {Thuillot}, {Titarenko}, {Torra Clotet}, {Turon},
  {Ulla}, {Utrilla}, {Uzzi}, {Vaillant}, {Valentini}, {Valette}, {van Elteren},
  {Van Hemelryck}, {van Leeuwen}, {Vaschetto}, {Vecchiato}, {Veljanoski},
  {Viala}, {Vicente}, {Vogt}, {von Essen}, {Voss}, {Votruba}, {Voutsinas},
  {Walmsley}, {Weiler}, {Wertz}, {Wevers}, {Wyrzykowski}, {Yoldas},
  {{\v{Z}}erjal}, {Ziaeepour}, {Zorec}, {Zschocke}, {Zucker}, {Zurbach}, \&
  {Zwitter}}]{2018A&A...616A...1G}
{Gaia Collaboration}, {Brown}, A.~G.~A., {Vallenari}, A., {et~al.}
  2018{\natexlab{b}}, \aap, 616, A1

\bibitem[{{Gilmore} \& {Reid}(1983)}]{1983MNRAS.202.1025G}
{Gilmore}, G., \& {Reid}, N. 1983, \mnras, 202, 1025

\bibitem[{{Gizis}(1997)}]{1997AJ....113..806G}
{Gizis}, J.~E. 1997, \aj, 113, 806

\bibitem[{{Gunn} {et~al.}(1998){Gunn}, {Carr}, {Rockosi}, {Sekiguchi}, {Berry},
  {Elms}, {de Haas}, {Ivezi{\'c}}, {Knapp}, {Lupton}, {Pauls}, {Simcoe},
  {Hirsch}, {Sanford}, {Wang}, {York}, {Harris}, {Annis}, {Bartozek},
  {Boroski}, {Bakken}, {Haldeman}, {Kent}, {Holm}, {Holmgren}, {Petravick},
  {Prosapio}, {Rechenmacher}, {Doi}, {Fukugita}, {Shimasaku}, {Okada}, {Hull},
  {Siegmund}, {Mannery}, {Blouke}, {Heidtman}, {Schneider}, {Lucinio}, \&
  {Brinkman}}]{1998AJ....116.3040G}
{Gunn}, J.~E., {Carr}, M., {Rockosi}, C., {et~al.} 1998, \aj, 116, 3040

\bibitem[{{Gunn} {et~al.}(2006){Gunn}, {Siegmund}, {Mannery}, {Owen}, {Hull},
  {Leger}, {Carey}, {Knapp}, {York}, {Boroski}, {Kent}, {Lupton}, {Rockosi},
  {Evans}, {Waddell}, {Anderson}, {Annis}, {Barentine}, {Bartoszek}, {Bastian},
  {Bracker}, {Brewington}, {Briegel}, {Brinkmann}, {Brown}, {Carr},
  {Czarapata}, {Drennan}, {Dombeck}, {Federwitz}, {Gillespie}, {Gonzales},
  {Hansen}, {Harvanek}, {Hayes}, {Jordan}, {Kinney}, {Klaene}, {Kleinman},
  {Kron}, {Kresinski}, {Lee}, {Limmongkol}, {Lindenmeyer}, {Long}, {Loomis},
  {McGehee}, {Mantsch}, {Neilsen}, {Neswold}, {Newman}, {Nitta}, {Peoples},
  {Pier}, {Prieto}, {Prosapio}, {Rivetta}, {Schneider}, {Snedden}, \&
  {Wang}}]{2006AJ....131.2332G}
{Gunn}, J.~E., {Siegmund}, W.~A., {Mannery}, E.~J., {et~al.} 2006, \aj, 131,
  2332

\bibitem[{{Gustafsson} {et~al.}(1975){Gustafsson}, {Bell}, {Eriksson}, \&
  {Nordlund}}]{1975A&A....42..407G}
{Gustafsson}, B., {Bell}, R.~A., {Eriksson}, K., \& {Nordlund}, A. 1975, A\&A,
  500, 67

\bibitem[{{Gustafsson} {et~al.}(2008){Gustafsson}, {Edvardsson}, {Eriksson},
  {J{\o}rgensen}, {Nordlund}, \& {Plez}}]{2008A&A...486..951G}
{Gustafsson}, B., {Edvardsson}, B., {Eriksson}, K., {et~al.} 2008, A\&A, 486,
  951

\bibitem[{{Harris}(1996)}]{1996AJ....112.1487H}
{Harris}, W.~E. 1996, \aj, 112, 1487

\bibitem[{{Hartwick} {et~al.}(1984){Hartwick}, {Cowley}, \&
  {Mould}}]{1984ApJ...286..269H}
{Hartwick}, F.~D.~A., {Cowley}, A.~P., \& {Mould}, J.~R. 1984, \apj, 286, 269

\bibitem[{{Hauschildt} {et~al.}(1999){Hauschildt}, {Allard}, \&
  {Baron}}]{1999ApJ...512..377H}
{Hauschildt}, P.~H., {Allard}, F., \& {Baron}, E. 1999, \apj, 512, 377

\bibitem[{{Hejazi} {et~al.}(2020){Hejazi}, {L{\'e}pine}, {Homeier}, {Rich}, \&
  {Shara}}]{2020AJ....159...30H}
{Hejazi}, N., {L{\'e}pine}, S., {Homeier}, D., {Rich}, R.~M., \& {Shara}, M.~M.
  2020, \aj, 159, 30

\bibitem[{{Jao} {et~al.}(2008){Jao}, {Henry}, {Beaulieu}, \&
  {Subasavage}}]{2008AJ....136..840J}
{Jao}, W.-C., {Henry}, T.~J., {Beaulieu}, T.~D., \& {Subasavage}, J.~P. 2008,
  \aj, 136, 840

\bibitem[{{Jao} {et~al.}(2011){Jao}, {Henry}, {Subasavage}, {Winters},
  {Riedel}, \& {Ianna}}]{2011AJ....141..117J}
{Jao}, W.-C., {Henry}, T.~J., {Subasavage}, J.~P., {et~al.} 2011, \aj, 141, 117

\bibitem[{{Jones}(1972)}]{1972ApJ...173..671J}
{Jones}, E.~M. 1972, \apj, 173, 671

\bibitem[{{Jones} {et~al.}(1996){Jones}, {Longmore}, {Allard}, \&
  {Hauschildt}}]{1996MNRAS.280...77J}
{Jones}, H. R.~A., {Longmore}, A.~J., {Allard}, F., \& {Hauschildt}, P.~H.
  1996, \mnras, 280, 77

\bibitem[{{Juri{\'c}} {et~al.}(2008){Juri{\'c}}, {Ivezi{\'c}}, {Brooks},
  {Lupton}, {Schlegel}, {Finkbeiner}, {Padmanabhan}, {Bond}, {Sesar},
  {Rockosi}, {Knapp}, {Gunn}, {Sumi}, {Schneider}, {Barentine}, {Brewington},
  {Brinkmann}, {Fukugita}, {Harvanek}, {Kleinman}, {Krzesinski}, {Long},
  {Neilsen}, {Nitta}, {Snedden}, \& {York}}]{2008ApJ...673..864J}
{Juri{\'c}}, M., {Ivezi{\'c}}, {\v{Z}}., {Brooks}, A., {et~al.} 2008, \apj,
  673, 864

\bibitem[{{Katz} {et~al.}(2019){Katz}, {Sartoretti}, {Cropper}, {Panuzzo},
  {Seabroke}, {Viala}, {Benson}, {Blomme}, {Jasniewicz}, {Jean-Antoine},
  {Huckle}, {Smith}, {Baker}, {Crifo}, {Damerdji}, {David}, {Dolding},
  {Fr{\'e}mat}, {Gosset}, {Guerrier}, {Guy}, {Haigron}, {Jan{\ss}en},
  {Marchal}, {Plum}, {Soubiran}, {Th{\'e}venin}, {Ajaj}, {Allende Prieto},
  {Babusiaux}, {Boudreault}, {Chemin}, {Delle Luche}, {Fabre}, {Gueguen},
  {Hambly}, {Lasne}, {Meynadier}, {Pailler}, {Panem}, {Royer}, {Tauran},
  {Zurbach}, {Zwitter}, {Arenou}, {Bossini}, {Gerssen}, {G{\'o}mez},
  {Lemaitre}, {Leclerc}, {Morel}, {Munari}, {Turon}, {Vallenari}, \&
  {{\v{Z}}erjal}}]{2019A&A...622A.205K}
{Katz}, D., {Sartoretti}, P., {Cropper}, M., {et~al.} 2019, \aap, 622, A205

\bibitem[{{Kesseli} {et~al.}(2019){Kesseli}, {Kirkpatrick}, {Fajardo-Acosta},
  {Penny}, {Gaudi}, {Veyette}, {Boeshaar}, {Henderson}, {Cushing},
  {Calchi-Novati}, {Shvartzvald}, \& {Muirhead}}]{2019AJ....157...63K}
{Kesseli}, A.~Y., {Kirkpatrick}, J.~D., {Fajardo-Acosta}, S.~B., {et~al.} 2019,
  \aj, 157, 63

\bibitem[{{Kurucz}(1973)}]{1973PhDT.........4K}
{Kurucz}, R.~L. 1973, PhD thesis, HARVARD UNIVERSITY.

\bibitem[{{Leggett} {et~al.}(1996){Leggett}, {Allard}, {Berriman}, {Dahn}, \&
  {Hauschildt}}]{1996ApJS..104..117L}
{Leggett}, S.~K., {Allard}, F., {Berriman}, G., {Dahn}, C.~C., \& {Hauschildt},
  P.~H. 1996, \apjs, 104, 117

\bibitem[{{Leggett} {et~al.}(2000){Leggett}, {Allard}, {Dahn}, {Hauschildt},
  {Kerr}, \& {Rayner}}]{2000ApJ...535..965L}
{Leggett}, S.~K., {Allard}, F., {Dahn}, C., {et~al.} 2000, \apj, 535, 965

\bibitem[{{L{\'e}pine} {et~al.}(2013){L{\'e}pine}, {Hilton}, {Mann}, {Wilde},
  {Rojas-Ayala}, {Cruz}, \& {Gaidos}}]{2013AJ....145..102L}
{L{\'e}pine}, S., {Hilton}, E.~J., {Mann}, A.~W., {et~al.} 2013, \aj, 145, 102

\bibitem[{{L{\'e}pine} {et~al.}(2003){L{\'e}pine}, {Rich}, \&
  {Shara}}]{2003AJ....125.1598L}
{L{\'e}pine}, S., {Rich}, R.~M., \& {Shara}, M.~M. 2003, \aj, 125, 1598

\bibitem[{{L{\'e}pine} {et~al.}(2007){L{\'e}pine}, {Rich}, \&
  {Shara}}]{2007ApJ...669.1235L}
---. 2007, ApJ, 669, 1235

\bibitem[{{L{\'e}pine} \& {Scholz}(2008)}]{2008ApJ...681L..33L}
{L{\'e}pine}, S., \& {Scholz}, R.-D. 2008, \apjl, 681, L33

\bibitem[{{L{\'e}pine} {et~al.}(2004){L{\'e}pine}, {Shara}, \&
  {Rich}}]{2004ApJ...602L.125L}
{L{\'e}pine}, S., {Shara}, M.~M., \& {Rich}, R.~M. 2004, \apjl, 602, L125

\bibitem[{{Lindegren} {et~al.}(2018){Lindegren}, {Hern{\'a}ndez}, {Bombrun},
  {Klioner}, {Bastian}, {Ramos-Lerate}, {de Torres}, {Steidelm{\"u}ller},
  {Stephenson}, {Hobbs}, {Lammers}, {Biermann}, {Geyer}, {Hilger}, {Michalik},
  {Stampa}, {McMillan}, {Casta{\~n}eda}, {Clotet}, {Comoretto}, {Davidson},
  {Fabricius}, {Gracia}, {Hambly}, {Hutton}, {Mora}, {Portell}, {van Leeuwen},
  {Abbas}, {Abreu}, {Altmann}, {Andrei}, {Anglada}, {Balaguer-N{\'u}{\~n}ez},
  {Barache}, {Becciani}, {Bertone}, {Bianchi}, {Bouquillon}, {Bourda},
  {Br{\"u}semeister}, {Bucciarelli}, {Busonero}, {Buzzi}, {Cancelliere},
  {Carlucci}, {Charlot}, {Cheek}, {Crosta}, {Crowley}, {de Bruijne}, {de
  Felice}, {Drimmel}, {Esquej}, {Fienga}, {Fraile}, {Gai}, {Garralda},
  {Gonz{\'a}lez-Vidal}, {Guerra}, {Hauser}, {Hofmann}, {Holl}, {Jordan},
  {Lattanzi}, {Lenhardt}, {Liao}, {Licata}, {Lister}, {L{\"o}ffler},
  {Marchant}, {Martin-Fleitas}, {Messineo}, {Mignard}, {Morbidelli}, {Poggio},
  {Riva}, {Rowell}, {Salguero}, {Sarasso}, {Sciacca}, {Siddiqui}, {Smart},
  {Spagna}, {Steele}, {Taris}, {Torra}, {van Elteren}, {van Reeven}, \&
  {Vecchiato}}]{2018A&A...616A...2L}
{Lindegren}, L., {Hern{\'a}ndez}, J., {Bombrun}, A., {et~al.} 2018, \aap, 616,
  A2

\bibitem[{{Liu} {et~al.}(2020){Liu}, {Fu}, {Shi}, {Wu}, {Han}, {Chen}, {Dong},
  {Zhao}, {Chen}, {Zhang}, {Bai}, {Chen}, {Cui}, {Du}, {Hsia}, {Jiang}, {Hou},
  {Hou}, {Li}, {Li}, {Li}, {Liu}, {Liu}, {Luo}, {Ren}, {Tian}, {Tian}, {Wang},
  {Wu}, {Xie}, {Yan}, {Yang}, {Yu}, {Zhang}, {Zhang}, {Zhang}, {Zhang}, {Zhao},
  {Zhong}, {Zong}, \& {Zuo}}]{2020arXiv200507210L}
{Liu}, C., {Fu}, J., {Shi}, J., {et~al.} 2020, arXiv e-prints, arXiv:2005.07210

\bibitem[{{Luo} {et~al.}(2015){Luo}, {Zhao}, {Zhao}, {Deng}, {Liu}, {Jing},
  {Wang}, {Zhang}, {Shi}, {Cui}, {Chu}, {Li}, {Bai}, {Wu}, {Cai}, {Cao}, {Cao},
  {Carlin}, {Chen}, {Chen}, {Chen}, {Chen}, {Chen}, {Chen}, {Chen},
  {Christlieb}, {Chu}, {Cui}, {Dong}, {Du}, {Fan}, {Feng}, {Fu}, {Gao}, {Gong},
  {Gu}, {Guo}, {Han}, {He}, {Hou}, {Hou}, {Hou}, {Hu}, {Hu}, {Hu}, {Huo},
  {Jia}, {Jiang}, {Jiang}, {Jiang}, {Jin}, {Kong}, {Kong}, {Lei}, {Li}, {Li},
  {Li}, {Li}, {Li}, {Li}, {Li}, {Li}, {Li}, {Li}, {Li}, {Li}, {Liang}, {Lin},
  {Liu}, {Liu}, {Liu}, {Liu}, {Lu}, {Luo}, {Mao}, {Newberg}, {Ni}, {Qi}, {Qi},
  {Shen}, {Shi}, {Song}, {Song}, {Su}, {Su}, {Tang}, {Tao}, {Tian}, {Wang},
  {Wang}, {Wang}, {Wang}, {Wang}, {Wang}, {Wang}, {Wang}, {Wang}, {Wang},
  {Wang}, {Wang}, {Wang}, {Wang}, {Wang}, {Wang}, {Wang}, {Wang}, {Wang},
  {Wang}, {Wei}, {Wei}, {Wu}, {Wu}, {Wu}, {Wu}, {Xing}, {Xu}, {Xu}, {Xu},
  {Yan}, {Yang}, {Yang}, {Yang}, {Yang}, {Yao}, {Yu}, {Yuan}, {Yuan}, {Yuan},
  {Yuan}, {Zhai}, {Zhang}, {Zhang}, {Zhang}, {Zhang}, {Zhang}, {Zhang},
  {Zhang}, {Zhang}, {Zhao}, {Zhou}, {Zhou}, {Zhu}, {Zhu}, {Zou}, \&
  {Zuo}}]{2015RAA....15.1095L}
{Luo}, A.-L., {Zhao}, Y.-H., {Zhao}, G., {et~al.} 2015, Research in Astronomy
  and Astrophysics, 15, 1095

\bibitem[{{Luri} {et~al.}(2018){Luri}, {Brown}, {Sarro}, {Arenou},
  {Bailer-Jones}, {Castro-Ginard}, {de Bruijne}, {Prusti}, {Babusiaux}, \&
  {Delgado}}]{2018A&A...616A...9L}
{Luri}, X., {Brown}, A.~G.~A., {Sarro}, L.~M., {et~al.} 2018, \aap, 616, A9

\bibitem[{McKinney {et~al.}(2010)}]{mckinney2010data}
McKinney, W., {et~al.} 2010, in Proceedings of the 9th Python in Science
  Conference, Vol. 445, Austin, TX, 51--56

\bibitem[{{Mould}(1976)}]{1976ApJ...210..402M}
{Mould}, J.~R. 1976, \apj, 210, 402

\bibitem[{{Mould} \& {McElroy}(1978)}]{1978ApJ...220..935M}
{Mould}, J.~R., \& {McElroy}, D.~B. 1978, \apj, 220, 935

\bibitem[{{Oliphant}(2007)}]{4160250}
{Oliphant}, T.~E. 2007, Computing in Science Engineering, 9, 10

\bibitem[{{Rajpurohit} {et~al.}(2018{\natexlab{a}}){Rajpurohit}, {Allard},
  {Rajpurohit}, {Sharma}, {Teixeira}, {Mousis}, \&
  {Kamlesh}}]{2018A&A...620A.180R}
{Rajpurohit}, A.~S., {Allard}, F., {Rajpurohit}, S., {et~al.}
  2018{\natexlab{a}}, \aap, 620, A180

\bibitem[{{Rajpurohit} {et~al.}(2018{\natexlab{b}}){Rajpurohit}, {Allard},
  {Teixeira}, {Homeier}, {Rajpurohit}, \& {Mousis}}]{2018A&A...610A..19R}
{Rajpurohit}, A.~S., {Allard}, F., {Teixeira}, G.~D.~C., {et~al.}
  2018{\natexlab{b}}, \aap, 610, A19

\bibitem[{{Rajpurohit} {et~al.}(2016){Rajpurohit}, {Reyl{\'e}}, {Allard},
  {Homeier}, {Bayo}, {Mousis}, {Rajpurohit}, \&
  {Fern{\'a}ndez-Trincado}}]{2016A&A...596A..33R}
{Rajpurohit}, A.~S., {Reyl{\'e}}, C., {Allard}, F., {et~al.} 2016, \aap, 596,
  A33

\bibitem[{{Rajpurohit} {et~al.}(2014){Rajpurohit}, {Reyl{\'e}}, {Allard},
  {Scholz}, {Homeier}, {Schultheis}, \& {Bayo}}]{2014A&A...564A..90R}
---. 2014, \aap, 564, A90

\bibitem[{{Reddy} {et~al.}(2003){Reddy}, {Tomkin}, {Lambert}, \& {Allende
  Prieto}}]{2003MNRAS.340..304R}
{Reddy}, B.~E., {Tomkin}, J., {Lambert}, D.~L., \& {Allende Prieto}, C. 2003,
  \mnras, 340, 304

\bibitem[{{Reid} {et~al.}(1995){Reid}, {Hawley}, \&
  {Gizis}}]{1995AJ....110.1838R}
{Reid}, I.~N., {Hawley}, S.~L., \& {Gizis}, J.~E. 1995, \aj, 110, 1838

\bibitem[{{Salim} \& {Gould}(2002)}]{2002ApJ...575L..83S}
{Salim}, S., \& {Gould}, A. 2002, \apjl, 575, L83

\bibitem[{{Sandage}(1969)}]{1969ApJ...158.1115S}
{Sandage}, A. 1969, \apj, 158, 1115

\bibitem[{{Sandage} \& {Kowal}(1986)}]{1986AJ.....91.1140S}
{Sandage}, A., \& {Kowal}, C. 1986, \aj, 91, 1140

\bibitem[{{Savcheva} {et~al.}(2014){Savcheva}, {West}, \&
  {Bochanski}}]{2014ApJ...794..145S}
{Savcheva}, A.~S., {West}, A.~A., \& {Bochanski}, J.~J. 2014, \apj, 794, 145

\bibitem[{{Sch{\"o}nrich} \& {Binney}(2009)}]{2009MNRAS.399.1145S}
{Sch{\"o}nrich}, R., \& {Binney}, J. 2009, \mnras, 399, 1145

\bibitem[{{Sch{\"o}nrich} {et~al.}(2010){Sch{\"o}nrich}, {Binney}, \&
  {Dehnen}}]{2010MNRAS.403.1829S}
{Sch{\"o}nrich}, R., {Binney}, J., \& {Dehnen}, W. 2010, \mnras, 403, 1829

\bibitem[{{Smith} {et~al.}(2009){Smith}, {Evans}, {Belokurov}, {Hewett},
  {Bramich}, {Gilmore}, {Irwin}, {Vidrih}, \& {Zucker}}]{2009MNRAS.399.1223S}
{Smith}, M.~C., {Evans}, N.~W., {Belokurov}, V., {et~al.} 2009, \mnras, 399,
  1223

\bibitem[{{Subasavage} {et~al.}(2005){Subasavage}, {Henry}, {Hambly}, {Brown},
  {Jao}, \& {Finch}}]{2005AJ....130.1658S}
{Subasavage}, J.~P., {Henry}, T.~J., {Hambly}, N.~C., {et~al.} 2005, \aj, 130,
  1658

\bibitem[{{Veltz} {et~al.}(2008){Veltz}, {Bienaym{\'e}}, {Freeman}, {Binney},
  {Bland-Hawthorn}, {Gibson}, {Gilmore}, {Grebel}, {Helmi}, {Munari},
  {Navarro}, {Parker}, {Seabroke}, {Siebert}, {Steinmetz}, {Watson},
  {Williams}, {Wyse}, \& {Zwitter}}]{2008A&A...480..753V}
{Veltz}, L., {Bienaym{\'e}}, O., {Freeman}, K.~C., {et~al.} 2008, \aap, 480,
  753

\bibitem[{{Virtanen} {et~al.}(2019){Virtanen}, {Gommers}, {Oliphant},
  {Haberland}, {Reddy}, {Cournapeau}, {Burovski}, {Peterson}, {Weckesser},
  {Bright}, {van der Walt}, {Brett}, {Wilson}, {Jarrod Millman}, {Mayorov},
  {Nelson}, {Jones}, {Kern}, {Larson}, {Carey}, {Polat}, {Feng}, {Moore}, {Vand
  erPlas}, {Laxalde}, {Perktold}, {Cimrman}, {Henriksen}, {Quintero}, {Harris},
  {Archibald}, {Ribeiro}, {Pedregosa}, {van Mulbregt}, \&
  {Contributors}}]{2019arXiv190710121V}
{Virtanen}, P., {Gommers}, R., {Oliphant}, T.~E., {et~al.} 2019, arXiv
  e-prints, arXiv:1907.10121

\bibitem[{{West} {et~al.}(2004){West}, {Hawley}, {Walkowicz}, {Covey},
  {Silvestri}, {Raymond}, {Harris}, {Munn}, {McGehee}, {Ivezi{\'c}}, \&
  {Brinkmann}}]{2004AJ....128..426W}
{West}, A.~A., {Hawley}, S.~L., {Walkowicz}, L.~M., {et~al.} 2004, \aj, 128,
  426

\bibitem[{{Woolf} \& {Wallerstein}(2005)}]{2005MNRAS.356..963W}
{Woolf}, V.~M., \& {Wallerstein}, G. 2005, \mnras, 356, 963

\bibitem[{{Yee} {et~al.}(2017){Yee}, {Petigura}, \& {von
  Braun}}]{2017ApJ...836...77Y}
{Yee}, S.~W., {Petigura}, E.~A., \& {von Braun}, K. 2017, \apj, 836, 77

\bibitem[{{Yong} \& {Lambert}(2003{\natexlab{a}})}]{2003PASP..115...22Y}
{Yong}, D., \& {Lambert}, D.~L. 2003{\natexlab{a}}, \pasp, 115, 22

\bibitem[{{Yong} \& {Lambert}(2003{\natexlab{b}})}]{2003PASP..115..796Y}
---. 2003{\natexlab{b}}, \pasp, 115, 796

\bibitem[{{York} {et~al.}(2000){York}, {Adelman}, {Anderson}, {Anderson},
  {Annis}, {Bahcall}, {Bakken}, {Barkhouser}, {Bastian}, {Berman}, {Boroski},
  {Bracker}, {Briegel}, {Briggs}, {Brinkmann}, {Brunner}, {Burles}, {Carey},
  {Carr}, {Castander}, {Chen}, {Colestock}, {Connolly}, {Crocker}, {Csabai},
  {Czarapata}, {Davis}, {Doi}, {Dombeck}, {Eisenstein}, {Ellman}, {Elms},
  {Evans}, {Fan}, {Federwitz}, {Fiscelli}, {Friedman}, {Frieman}, {Fukugita},
  {Gillespie}, {Gunn}, {Gurbani}, {de Haas}, {Haldeman}, {Harris}, {Hayes},
  {Heckman}, {Hennessy}, {Hindsley}, {Holm}, {Holmgren}, {Huang}, {Hull},
  {Husby}, {Ichikawa}, {Ichikawa}, {Ivezi{\'c}}, {Kent}, {Kim}, {Kinney},
  {Klaene}, {Kleinman}, {Kleinman}, {Knapp}, {Korienek}, {Kron}, {Kunszt},
  {Lamb}, {Lee}, {Leger}, {Limmongkol}, {Lindenmeyer}, {Long}, {Loomis},
  {Loveday}, {Lucinio}, {Lupton}, {MacKinnon}, {Mannery}, {Mantsch}, {Margon},
  {McGehee}, {McKay}, {Meiksin}, {Merelli}, {Monet}, {Munn}, {Narayanan},
  {Nash}, {Neilsen}, {Neswold}, {Newberg}, {Nichol}, {Nicinski}, {Nonino},
  {Okada}, {Okamura}, {Ostriker}, {Owen}, {Pauls}, {Peoples}, {Peterson},
  {Petravick}, {Pier}, {Pope}, {Pordes}, {Prosapio}, {Rechenmacher}, {Quinn},
  {Richards}, {Richmond}, {Rivetta}, {Rockosi}, {Ruthmansdorfer}, {Sand ford},
  {Schlegel}, {Schneider}, {Sekiguchi}, {Sergey}, {Shimasaku}, {Siegmund},
  {Smee}, {Smith}, {Snedden}, {Stone}, {Stoughton}, {Strauss}, {Stubbs},
  {SubbaRao}, {Szalay}, {Szapudi}, {Szokoly}, {Thakar}, {Tremonti}, {Tucker},
  {Uomoto}, {Vanden Berk}, {Vogeley}, {Waddell}, {Wang}, {Watanabe},
  {Weinberg}, {Yanny}, {Yasuda}, \& {SDSS Collaboration}}]{2000AJ....120.1579Y}
{York}, D.~G., {Adelman}, J., {Anderson}, John~E., J., {et~al.} 2000, \aj, 120,
  1579

\bibitem[{{Zhang} {et~al.}(2016){Zhang}, {Pi}, {Han}, {Shi}, {Wang}, {Luo},
  {Zhang}, {Hou}, \& {Wang}}]{2016NewA...44...66Z}
{Zhang}, L., {Pi}, Q., {Han}, X.~L., {et~al.} 2016, \na, 44, 66

\bibitem[{{Zhang} {et~al.}(2019){Zhang}, {Luo}, {Comte}, {Gizis}, {Wang}, {Li},
  {Qin}, {Kong}, {Bai}, \& {Yi}}]{2019ApJS..240...31Z}
{Zhang}, S., {Luo}, A.~L., {Comte}, G., {et~al.} 2019, \apjs, 240, 31

\bibitem[{{Zhang}(2019)}]{2019MNRAS.489.1423Z}
{Zhang}, Z. 2019, \mnras, 489, 1423

\bibitem[{{Zhang} {et~al.}(2017{\natexlab{a}}){Zhang}, {Homeier}, {Pinfield},
  {Lodieu}, {Jones}, {Allard}, \& {Pavlenko}}]{2017MNRAS.468..261Z}
{Zhang}, Z.~H., {Homeier}, D., {Pinfield}, D.~J., {et~al.} 2017{\natexlab{a}},
  \mnras, 468, 261

\bibitem[{{Zhang} {et~al.}(2017{\natexlab{b}}){Zhang}, {Pinfield},
  {G{\'a}lvez-Ortiz}, {Burningham}, {Lodieu}, {Marocco}, {Burgasser},
  {Day-Jones}, {Allard}, {Jones}, {Homeier}, {Gomes}, \&
  {Smart}}]{2017MNRAS.464.3040Z}
{Zhang}, Z.~H., {Pinfield}, D.~J., {G{\'a}lvez-Ortiz}, M.~C., {et~al.}
  2017{\natexlab{b}}, \mnras, 464, 3040

\bibitem[{{Zhang} {et~al.}(2018){Zhang}, {Galvez-Ortiz}, {Pinfield},
  {Burgasser}, {Lodieu}, {Jones}, {Mart{\'\i}n}, {Burningham}, {Homeier},
  {Allard}, {Zapatero Osorio}, {Smith}, {Smart}, {L{\'o}pez Mart{\'\i}},
  {Marocco}, \& {Rebolo}}]{2018MNRAS.480.5447Z}
{Zhang}, Z.~H., {Galvez-Ortiz}, M.~C., {Pinfield}, D.~J., {et~al.} 2018,
  \mnras, 480, 5447

\bibitem[{{Zhong} {et~al.}(2015){Zhong}, {L{\'e}pine}, {Hou}, {Shen}, {Yuan},
  {Huo}, {Zhang}, {Xiang}, {Zhang}, \& {Liu}}]{2015AJ....150...42Z}
{Zhong}, J., {L{\'e}pine}, S., {Hou}, J., {et~al.} 2015, \aj, 150, 42

\bibitem[{{Ziegler} {et~al.}(2015){Ziegler}, {Law}, {Baranec}, {Riddle}, \&
  {Fuchs}}]{2015ApJ...804...30Z}
{Ziegler}, C., {Law}, N.~M., {Baranec}, C., {Riddle}, R.~L., \& {Fuchs}, J.~T.
  2015, \apj, 804, 30

\bibitem[{{Zinn}(1985)}]{1985ApJ...293..424Z}
{Zinn}, R. 1985, \apj, 293, 424

\end{thebibliography}
\bibliographystyle{aasjournal}
         
\appendix
\begin{deluxetable}{llll}[ht!]
\scriptsize
\tablecaption{Data description of the subdwarf catalog. \label{tab:desrcription}}
\tablewidth{0pt}
\tablenum{4}
\tablehead{
\colhead{Column Name\qquad\qquad\qquad} & \colhead{Unit\qquad \qquad\qquad\qquad} & \colhead{Description\qquad \qquad \qquad \qquad \qquad \qquad\qquad\qquad\qquad\qquad\qquad\qquad\qquad\qquad\qquad\qquad} }
\startdata
ID	& - & SpecObjID for SDSS objects and obsid for LAMOST objects \\
specname & - & Name of spectrum\\
ra & degrees & Right Ascension (J2000) \\
dec & degrees & Declination (J2000) \\
l & degrees & Galactic longitude\\
b & degrees & Galactic latitude \\
rv & km s$^{-1}$ & Heliocentric radial velocity  \\
Teff & K & Estimated effective temperature  \\
logg & dex & Estimated surface gravity  \\
FeH & dex & Estimated overall metallicity  \\
snri & - & Average spectrum signal to noise ratio in i band \\
spt & - & Spectral subtype \\
CaOH & - & Spectral index of CaOH band \\
CaH1 & - & Spectral index of CaH1 band  \\
CaH2 & - & Spectral index of CaH2 band  \\
CaH3 & - & Spectral index of CaH3 band  \\
TiO5 & - & Spectral index of TiO5 band  \\
ZetaL07 & - & The value of $\zeta$ parameter calibrated by \citet{2007ApJ...669.1235L}  \\
ZetaD12 & - & The value of $\zeta$ parameter calibrated by \citet{2012AJ....143...67D}  \\
ZetaL13 & - & The value of $\zeta$ parameter calibrated by \citet{2013AJ....145..102L}  \\
ZetaZ19 & - & The value of $\zeta$ parameter calibrated by \citet{2019ApJS..240...31Z}  \\
source & - & The source of spectrum (lamost/sdss)  \\
parallax & mas & Gaia DR2 parallax  \\
parallax\_error & mas & Standard error of Gaia DR2 parallax \\
pmra & mas yr$^{-1}$ & Gaia DR2 proper motion in right ascension direction\\
pmra\_error & mas yr$^{-1}$  & Standard error of Gaia DR2 proper motion in right ascension direction\\
pmdec & mas yr$^{-1}$ & Gaia DR2 proper motion in declination direction\\
pmdec\_error & mas yr$^{-1}$ & Standard error of Gaia DR2 proper motion in declination direction\\
astrometric\_gof\_al & - & Goodness of fit statistic of model wrt along-scan observations \\
astrometric\_excess\_noise\_sig & - & Significance of excess noise \\
phot\_g\_mean\_flux\_over\_error & -  & G-band mean flux divided by its error \\
phot\_g\_mean\_mag & mag & Gaia DR2 G-band mean magnitude \\
phot\_bp\_mean\_flux\_over\_error & - & Integrated BP mean flux divided by its error \\
phot\_bp\_mean\_mag & mag & Gaia DR2 integrated BP mean magnitude \\
phot\_rp\_mean\_flux\_over\_error & - & Integrated RP mean flux divided by its error\\
phot\_rp\_mean\_mag & mag & Gaia DR2 integrated RP mean magnitude\\
bp\_rp & mag & Gaia DR2 BP-RP colour. \\
ruwe & - & Renormalized Unit Weight Error associated to each Gaia source \\
rest & pc & Estimated distance from \citet{2018AJ....156...58B}, based on  Gaia DR2 \\
ResFlag & - & The Result Flag from \citet{2018AJ....156...58B} \\
ModFlag & -  & The Modality Flag from \citet{2018AJ....156...58B} \\
flag & - & Set as 1 for the final sample used in kinematic part, see table \ref{tab:subsamples} for details\\
$V_{total}$ & km s$^{-1}$ & Total spacial velocity \\
$Z$ & pc & Vertical distance away from Galactic plane \\
$U$ & km s$^{-1}$ & Velocity component points towards the galactic center \\
$V$ & km s$^{-1}$ & The tangential velocity component rotating around galactic center\\
$W$ & km s$^{-1}$ & Velocity component points towards the north galactic pole \\
\enddata
\tablecomments{The complete catalog can be accessed online\footnote{\url{http://paperdata.china-vo.org/work/subdwarf-catalog.csv}}.}
\end{deluxetable}

\end{document}